\documentclass[11pt]{article}

\pdfoutput=1

\usepackage[textwidth=15.5cm,textheight=22.5cm]{geometry}
\usepackage{bm}
\usepackage{amsmath,amssymb,color,graphicx,amscd,amsfonts}
\usepackage{latexsym}
\usepackage{graphicx}
\usepackage{cite}
\usepackage{subfig}
\usepackage{bbm}
\usepackage{ulem}

\def\ul#1{{\underline{#1}}}

\def\a{\alpha}

\def\e{\epsilon}

\def\s{\sigma}

\def\th{\theta}

\def\tA{\totalarea}
\def\totalarea{[\sigma]}

\allowdisplaybreaks[4]

\numberwithin{equation}{section}

\newcommand{\be}{\begin{equation}}
\newcommand{\ee}{\end{equation}}
\newcommand{\ndt}{\noindent}
\def\bea{\begin{eqnarray}}
\def\eea{\end{eqnarray}}
\def\beas{\begin{eqnarray*}}
\def\eeas{\end{eqnarray*}}
\def\sla{\raise.15ex\hbox{$/$}\kern-.57em}

\newcommand\fr[1]{\frac{1}{#1}}

\newcommand {\nn} {\nonumber}

\renewcommand{\th}{\theta}

\allowdisplaybreaks[1]

\newcommand{\er}{{\rm e}}
\newcommand{\dr}{{\rm d}}
\newcommand{\tr}{{\rm tr}}
\newcommand{\R}{\mathbbm{R}}
\newcommand{\Z}{\mathbbm{Z}}
\newcommand{\C}{\mathbbm{C}}

\newcommand{\Id}{\mathbbm{1}}

\newcommand{\del}{\partial}
\newcommand{\ba}{\begin{eqnarray}}
\newcommand{\ea}{\end{eqnarray}}
\newcommand{\bdm}{\begin{displaymath}}
\newcommand{\edm}{\end{displaymath}}

\def\a{\alpha}

\def\s{\sigma}

\def\veps{\varepsilon}

\def\v{\varphi}

\newcommand{\ie}{{\it i.e.\ }}
\newcommand{\eg}{{\it e.g.\ }}

\newcommand{\calM}{{\mathcal M}}
\newcommand{\calN}{{\mathcal N}}

\DeclareMathAlphabet{\mathpzc}{OT1}{pzc}{m}{it}

\newcommand\vsp[1]{\vspace*{#1 cm}}

\makeatletter
\newif\if@borderstar
\def\bordermatrix{\@ifnextchar*{%
 \@borderstartrue\@bordermatrix@i}{\@borderstarfalse\@bordermatrix@i*}%
}
\def\@bordermatrix@i*{\@ifnextchar[{\@bordermatrix@ii}{\@bordermatrix@ii[()]}}
\def\@bordermatrix@ii[#1]#2{%
\begingroup
 \m@th\@tempdima8.75\p@\setbox\z@\vbox{%
 \def\cr{\crcr\noalign{\kern 2\p@\global\let\cr\endline }}%
 \ialign {$##$\hfil\kern 2\p@\kern\@tempdima & \thinspace %
  \hfil $##$\hfil && \quad\hfil $##$\hfil\crcr\omit\strut %
  \hfil\crcr\noalign{\kern -\baselineskip}#2\crcr\omit %
  \strut\cr}}%
 \setbox\tw@\vbox{\unvcopy\z@\global\setbox\@ne\lastbox}%
 \setbox\tw@\hbox{\unhbox\@ne\unskip\global\setbox\@ne\lastbox}%
 \setbox\tw@\hbox{%
  $\kern\wd\@ne\kern -\@tempdima\left\@firstoftwo#1%
  \if@borderstar\kern 2pt\else\kern -\wd\@ne\fi%
 \global\setbox\@ne\vbox{\box\@ne\if@borderstar\else\kern 2\p@\fi}%
 \vcenter{\if@borderstar\else\kern -\ht\@ne\fi%
  \unvbox\z@\kern -\if@borderstar2\fi\baselineskip}%
 \if@borderstar\kern-2\@tempdima\kern2\p@\else\,\fi\right\@secondoftwo#1 $%
 }\null \;\vbox{\kern\ht\@ne\box\tw@}%
\endgroup
}
\makeatother

\begin{document}

\thispagestyle{empty}

\setcounter{page}{0}

\begin{flushright} 
DIAS-STP-15-05, WITS-MITP-018, OIQP-15-4

\end{flushright} 

\vspace{0.1cm}

\begin{center}
{\LARGE
On membrane interactions 
and a three-dimensional analog of Riemann surfaces
\rule{0pt}{20pt}  }
\end{center}

\vspace*{0.2cm}

\renewcommand{\thefootnote}{\alph{footnote}}

\begin{center}

Stefano K{\sc ovacs}\,$^{\ddag\,\sharp\,}$\footnote
{
E-mail address: skovacs@stp.dias.ie },
Yuki S{\sc ato}\,$^{\flat\,}$\footnote
{
E-mail address: Yuki.Sato@wits.ac.za
}
and
Hidehiko S{\sc himada}\,$^{\S\,}$\footnote
         {
E-mail address: shimada.hidehiko@googlemail.com}

\vspace{0.3cm}

${}^{\ddag}$ {\it Dublin Institute for Advanced Studies \\
10 Burlington Road, Dublin 4, Ireland }\\[0.2cm]
${}^\sharp$ {\it ICTP South American Institute for Fundamental Research \\
IFT-UNESP, S\~ao Paulo, SP Brazil  01440-070 } \\ [0.2cm]
${}^{\flat}$ {\it National Institute for Theoretical Physics, }\\
{\it School of Physics and Mandelstam Institute for Theoretical Physics,} \\
{\it University of the Witwartersrand, Wits 2050, South Africa}\\[0.2cm]
${}^{\S}$ {\it Okayama Institute for Quantum Physics, Okayama, Japan}\\

\end{center}

\vspace{1cm}

\begin{abstract}
Membranes in M-theory are expected to interact via splitting and joining 
processes. We study these effects in the pp-wave matrix model, in
which they are associated with transitions between states in sectors built 
on vacua with different numbers of membranes. Transition amplitudes 
between such states receive contributions from BPS instanton 
configurations interpolating between the different vacua.
Various properties of the moduli space of BPS instantons are known,
but there are very few known examples of explicit solutions. We present
a new approach to the construction of instanton solutions 
interpolating between states containing arbitrary numbers of membranes, 
based on a continuum approximation valid for matrices of large size.
The proposed scheme uses functions on a two-dimensional 
space to approximate matrices and it relies on the same ideas behind the 
matrix regularisation of membrane degrees of freedom in M-theory. We 
show that the BPS instanton equations have a continuum counterpart 
which can be mapped to the three-dimensional Laplace equation through
a sequence of changes of variables. A description of configurations 
corresponding to membrane splitting/joining processes can be given in
terms of solutions to the Laplace equation in a three-dimensional analog
of a Riemann surface, consisting of multiple copies of $\R^3$ connected 
via a generalisation of branch cuts. We discuss various general features
of our proposal and we also present explicit analytic solutions.

\end{abstract}


\newpage

\setcounter{footnote}{0}

\renewcommand{\thefootnote}{\arabic{footnote}}

\section{Introduction}
\label{RSIntro}

The best candidate for a microscopic description of M-theory is the matrix 
model originally proposed in~\cite{RBdWHN} as a regularisation of the 
supermembrane theory and subsequently conjectured 
in~\cite{RBBFSS} to describe the full dynamics of the theory in light-front 
quantisation when the size of the matrices is sent to infinity. The matrix 
model provides a formulation of M-theory which is capable of describing 
multi-membrane configurations, arising as block-diagonal matrices.
Membranes in M-theory should interact via splitting and joining processes 
and therefore the matrix model should capture such effects. However, no 
concrete proposal for the description of these splitting/joining interactions
in the matrix model has been formulated. 
More generally there is no quantitative prescription for the 
study of this type of membrane interactions, which would make it 
possible to evaluate the associated transition amplitudes. This makes it 
difficult to test any results for these amplitudes that may be obtained from 
the matrix model. This is of course a general difficulty with all predictions of 
the matrix model, which so far have been mostly tested at the level of the 
low energy supergravity approximation. 

The AdS/CFT correspondence -- and specifically the duality proposed
in~\cite{RBABJM}, relating M-theory in AdS$_4\times S^7/\Z_k$ to an
$\calN=6$ Chern--Simons theory -- provides in principle new means of 
testing the matrix model predictions by comparing them to a dual CFT. 
In the context of the duality of~\cite{RBABJM} it was recently shown 
in~\cite{RBKSS1} that the matrix model description of M-theory in 
AdS$_4\times S^7/\Z_k$ can be quantitatively compared to the dual
gauge theory without relying on the supergravity approximation or 
compactification to type IIA string theory in ten dimensions. 
The crucial observation of~\cite{RBKSS1} was that, focussing on large
angular momentum states in M-theory and the dual CFT sector involving
monopole operators, natural approximation schemes arise on the two sides
of the duality, so that a systematic, quantitative comparison is possible.

On the gravity side, M-theory states with large angular momentum, $J$, along 
a great circle in $S^7$ can be studied using the pp-wave approximation. The 
associated matrix model uses $J\times J$ matrices and its action was 
constructed in~\cite{RBBMN}. We will use basic properties of the model 
which were further studied in~\cite{RBDSVR1}. 
Multi-membrane states in the pp-wave 
matrix model consist of concentric membranes and their fluctuations. More
precisely the vacua of the theory consist of spherical membranes which 
extend in AdS$_4$ directions and are point-like in $S^7$. They are 
classified by a set of integers, $J_i$, $i=1,\ldots,n$, corresponding to a 
partition of the total angular momentum among $n$ membranes.  The 
fluctuations of the spherical membranes described by the
pp-wave matrix model are associated, in the dual gauge theory, with certain 
monopole operators. The latter are characterised by their integer GNO
charges, which are in one-to-one correspondence with the angular 
momenta, $J_i$, of the membranes. The sector of monopole operators with 
large GNO charges can be reliably studied using a weakly-coupled effective 
low-energy approximation.

The AdS$_4$/CFT$_3$ duality in this M-theoretic regime relates 
correlation functions of monopole operators in the ABJM theory to 
processes on the gravity side in which the dual states interact in the bulk 
and propagate to the boundary. The simplest such process involves a single 
membrane splitting into two -- with the associated three states propagating to 
the boundary -- and it corresponds to a three-point correlation 
function of monopole operators. Similar processes have been studied in 
the case of the AdS$_5$/CFT$_4$ correspondence, in which three-point 
functions of gauge-invariant operators are related to cubic interaction 
vertices in the bulk associated with the splitting of 
strings in the pp-wave approximation.
A method to compare the cubic vertices in 
string theory to the corresponding conformal three-point functions in the CFT
was discussed in~\cite{RBShimada3}, following a spacetime interpretation
first proposed in~\cite{RBDobashiShimadaYoneya}. 
Studies using other approaches can be found in  
~\cite{RBDobashiYoneya1, RBDobashiYoneya2, RBDobashi, 
RBGrignaniZayakin1,RBGrignaniZayakin2,RBSchulginZayakin}.
Similar aspects have been extensively  studied beyond the pp-wave 
approximation taking advantage of the integrability appearing in the analysis 
of the system. For recent interesting developments and up-to-date 
references, 
see~\cite{RBBassoKomatsuVieira, RBKazamaKomatsuNishimura}.

The three-point correlation functions of monopole operators, 
in particular those with large charge $J$, which are
relevant in the present case, are well defined in the ABJM theory and one 
expects that they should be computable within the CFT framework.
Although in this paper we focus on the gravity side of the 
correspondence, it is important that in principle the results of the matrix 
model calculation can be verified by independent means. Successfully 
comparing the physical transition amplitudes associated with a membrane 
splitting or joining process to the gauge theory results for 
three-point functions would provide such an independent test. The 
agreement would not only represent a highly non-trivial  
test of the AdS$_4$/CFT$_3$ correspondence, but it would also provide 
strong evidence indicating that the matrix model captures the aspects 
of the dynamics of membranes related to splitting/joining interactions.

In the context of the pp-wave matrix model, 
processes of joining or splitting of membranes correspond to transitions between states built on 
vacua with different numbers of membranes. A semi-classical description of 
these transitions can be given in terms of tunnelling amplitudes associated 
with instanton configurations which interpolate between states containing 
different numbers of membranes. In this paper we initiate a study of these 
interpolating configurations. The relevant instantons are 
BPS solutions to the Euclidean equations of motion of the pp-wave 
matrix model~\cite{RBYeeYi}, with boundary conditions specifying the
single or multiple membrane states between which they interpolate. 
The equations obeyed by these BPS instantons, with the same boundary conditions, 
were studied in a different context in~\cite{RBBHP}, where they were found to arise as the 
equations describing domain walls interpolating between distinct isolated 
vacua of the so-called $\calN=1^*$ SYM theory. 
A number of properties of these instanton equations
and the associated moduli spaces were studied 
in~\cite{RBBHP}. An important ingredient in the analysis presented 
in~\cite{RBBHP} is the observation that the relevant instanton equations 
can be mapped to the Nahm equations~\cite{RBNahm} which arise in
the construction of BPS monopoles, although with different boundary 
conditions.

The BPS instantons of the pp-wave matrix model are saddle points 
representing the dominant contribution to the tunnelling processes. Physical
transition amplitudes associated with the splitting/joining processes can
be computed using a semi-classical approximation around the instanton
solutions. This involves evaluating the determinants arising from the 
non-zero mode fluctuations as well as the integration over the collective
coordinates associated with zero modes. Although there may be alternative 
approaches to the computation of the splitting/joining transition amplitudes, 
in order to carry out the standard semi-classical calculations it is necessary  
to obtain the explicit form of the classical instanton solutions. However, 
although various general
features of the instanton equations -- including conditions 
for the existence of solutions and properties
of their moduli spaces -- were studied in~\cite{RBBHP}, 
little is known about explicit solutions. In particular, 
no solution is known for the most elementary process we are interested in, 
\ie the splitting of a single membrane into two.

In this paper we show that an efficient approach to the construction of
solutions to the instanton equations as well as to the study of their moduli 
spaces can be developed using a continuum approximation. Such an 
approximation scheme is valid in the case of large matrices and, 
therefore, it is applicable in the context of the M-theoretic regime of the 
AdS$_4$/CFT$_3$ duality considered in~\cite{RBKSS1}. In this 
approximation the original instanton equations can be mapped to a 
continuum version of the Nahm equations, which in turn can be shown to be 
equivalent to the three-dimensional Laplace equation, using results 
in~\cite{RBWard,RBHoppeLaplaceEq, RBBordemannHoppe}. 
In this formulation equipotential 
surfaces for the solution to the Laplace equation directly represent the 
profile of the membranes at different times. Using this approach we present 
explicit analytic solutions describing the splitting of membranes.

A remarkable feature of our construction is that, in order to 
obtain a solution representing the splitting of one membrane into two,
we are led to consider the Laplace equation not in the ordinary 
three-dimensional Euclidean space, but rather in a three-dimensional 
generalisation of a two-sheeted Riemann surface, which we refer to as a 
Riemann space following~\cite{RBSommerfeld}. The emergence of Riemann 
spaces, which can be naturally motivated in the context of the construction 
of our solutions, is quite intriguing. The possibility that in the description of 
membrane interactions Riemann spaces may play a central role, similar to 
that played by Riemann surfaces in string perturbation theory, is very 
interesting and deserves further investigation. 

This paper is organised as follows. In section~\ref{RSBPSInstantons} we 
describe the instanton equations and we recall some general properties of 
their solutions and of the associated moduli space which were obtained 
in~\cite{RBBHP}. In section~\ref{RSApprox} we discuss the continuum 
approximation scheme used in our analysis and the mapping of the 
instanton equations to the three-dimensional Laplace equation. In 
section~\ref{RSSolutions} we present explicit solutions to the Laplace 
equation which are relevant for the pp-wave matrix model 
and we describe the corresponding solutions to the BPS instanton
equation within our continuum approximation. The most interesting solution 
-- which utilises a two-sheeted Riemann space to describe the splitting of a 
single membrane into two -- is considered in 
section~\ref{RSSSolutionHobson}. We conclude with a discussion of 
our results in section~\ref{RSConclusion}. Technical details are discussed in 
various appendices.

\section{BPS instantons}
\label{RSBPSInstantons}

In our study of configurations describing membrane splitting/joining 
processes we focus on the pp-wave matrix model~\cite{RBBMN}.
As discussed in~\cite{RBKSS1} this model provides a good approximation to 
the dynamics of M-theory in AdS$_4\times S^7/\Z_k$ in a sector containing 
membranes with large angular momentum, $J$. M-theory in this background 
is dual to the $\mathcal{N}=6$ Chern-Simons theory with gauge group 
U($N$)$\times$ U($N$) and level $k$ constructed in~\cite{RBABJM}. The 
region of applicability of the pp-wave approximation is defined by the 
conditions $1\ll J \ll (Nk)^{1/2}$. Moreover for $J\gg (Nk)^{1/3}$ the model is 
weakly coupled.  
 
In this section we review the BPS instanton equations of the pp-wave matrix 
model and recall some of their properties. These equations were first 
studied in a different context in~\cite{RBBHP}, where they were shown to 
describe supersymmetric domain walls interpolating between vacua of the 
$\calN=1^*$ SYM theory. Their relevance in the pp-wave matrix model was 
pointed out in~\cite{RBYeeYi}, where they were first shown to describe 
instanton configurations interpolating between different vacua.

\subsection{Instanton equations}
\label{RSBPSEq}

The Euclidean action describing the M-theory side of the AdS$_4$/CFT$_3$ 
duality in the large $J$ sector takes the 
form~\cite{RBBMN, RBDSVR1, RBKSS1} 
\begin{align}
S_E = \int  \dr t \; \text{tr}&  
\biggl\{ \frac{k}{2R} \left( \frac{D Y^i}{D t} \right)^2
+\frac{k}{2R} \left( \frac{D X^m}{D  t} \right)^2
+(2\pi T)^2 \frac{R}{4k}
\left((i[X^m, X^n])^2 + 2 (i[X^m,Y^i])^2
\right) \nonumber \\
& + \frac{k}{2R^3} (X^m)^2
+ (2\pi T)^2 \frac{R}{2k}
\left( \frac{i}{2} \epsilon_{ijk} [Y^j, Y^k]
+ \frac{1}{2\pi T} \frac{2 k }{R^2} Y^i
\right)^{\!\!2} \label{RFeuclideanaction} \\
&+\frac{1}{2} \Psi^T \frac{D \Psi}{D t} 
+ 2\pi T \frac{R}{k} \frac{1}{2} \left(\Psi^T \gamma^m[ X^m, \Psi]
+\Psi^T \gamma^i[ Y^i, \Psi] \right)
- \frac{3i}{4} \frac{1}{R} \, \Psi^T \gamma^{123} \Psi
\biggl\} , \nn
\end{align}
where $T=1/[(2\pi)^2 l_P^3]$ is the membrane tension,   
$R$ is the radius of $S^7$ (and twice the radius of AdS$_4$) and 
$\gamma^{\alpha}$ $(\alpha=1,2,\ldots,9)$ are SO($9$) gamma matrices, 
with $\gamma^{123}=\gamma^1\gamma^2\gamma^3$. Following the 
notation in~\cite{RBKSS1}, we have denoted by $Y^i$ $(i=1,2,3)$ and 
$X^m$ $(m=4, 5, \ldots ,9)$ matrices originating from the membrane 
coordinates in AdS$_4$ and $S^7$ respectively. $\Psi$ is a 16 
component matrix valued spinor. The $Y^i$'s, $X^m$'s and the 
components of $\Psi$ are $K \times K$ matrices, with $K=J/k$. The 
covariant derivative in~(\ref{RFeuclideanaction}) is defined by 
\begin{equation}
\frac{D X}{D t} = \frac{\text{d} X}{\text{d} t} - i [A_0, X]
\end{equation}
where $A_0$ is the gauge potential associated with the invariance of the
model under time dependent unitary transformations. In the following we will 
choose the gauge $A_0=0$, which is compatible with the boundary 
conditions relevant for the tunnelling processes we are interested in.

The matrix model defined by the action~(\ref{RFeuclideanaction}) is a 
special case of a family of pp-wave matrix models, which are usually 
parameterised by a mass scale $\mu$. In the following we use the 
model~(\ref{RFeuclideanaction}), which is the one relevant for the
large $J$ sector of the AdS/CFT duality of~\cite{RBABJM} and corresponds 
to a specific choice of $\mu$. To simplify notation and equations, we also 
focus on the $k=1$ case. The generalisation is straightforward and in 
appendix~\ref{RSAGeneralisedFormulae} we present the main formulae for 
the case of a general pp-wave background. Using the equations in the 
appendix it is easy to recover the results for arbitrary $k$.

The potential term in the action~(\ref{RFeuclideanaction}) is written as a 
sum of squares and thus it is manifestly non-negative. This makes it easy 
to identify all the vacua of the model, \ie the zero-energy minima 
of~(\ref{RFeuclideanaction}). They correspond to the 
configurations~\cite{RBBMN, RBDSVR1}
\begin{align}
& Y_0^i = \frac{2}{(2\pi T)R^2} L^i \, , \quad {\rm with} \quad 
[L^i,L^j] = i \epsilon^{ijk}L^k \, , 
\label{RFMMVacua} \\
& X_0^m = 0 \, , \qquad \Psi_0=0 \, 
\end{align}
and therefore they are classified by (generally reducible) $J$ dimensional 
representations of SU(2). The solution in which the $Y^i$'s are proportional 
to the $J$ dimensional irreducible representation corresponds to a single 
spherical membrane. It extends in the AdS$_4$ directions, with radius 
\begin{equation}
r = \frac{\sqrt{J^2-1}}{(2\pi T)R^2} \approx \frac{J}{(2\pi T) R^2} \, ,
\label{RFradius}
\end{equation}
and carries momentum $J/R$ along a great circle of $S^7$. Generic vacua 
are given by block diagonal $Y^i$ matrices corresponding to reducible SU(2) 
representations. The blocks have size $J_i$, $i=1,2,\ldots,n$, with 
$\sum_i J_i = J$ and they represent $n$ concentric membranes of radii
\begin{equation}
r_i \approx \frac{J_i}{(2\pi T)R^2} \, ,
\end{equation}
with angular momenta $J_i$ along the same great circle in $S^7$. 

We will denote the irreducible $J$ dimensional SU(2) representation by 
$\ul{J}$ and the reducible representation which is the direct sum of 
irreducible representations of dimension $J_1,J_2,\ldots,J_n$ by 
$\ul{J_1} \oplus\ul{J_2} \oplus\cdots\oplus\ul{J_n}$. 

We are interested in the tunnelling processes corresponding to 
classical solutions interpolating between a vacuum associated with 
the SU($2$) representation $L^i_{(-\infty)}$ in the infinite past 
($t=-\infty$) and another vacuum associated with $L^i_{(+\infty)}$ 
at $t=+\infty$. These processes are governed by the path integral with 
boundary conditions
\begin{align}
& Y^i(-\infty) = \frac{2}{(2\pi T)R^2} L^i_{(-\infty)} \nn \\
& Y^i(+\infty) = \frac{2}{(2\pi T)R^2} \,U L^i_{(+\infty)} U^{-1} \, .
\label{RFGenBoundaryCond}
\end{align}
In~(\ref{RFGenBoundaryCond}) we have chosen the generators 
$L^i_{(\pm\infty)}$ to correspond to the standard embedding of SU(2) 
consisting of block diagonal $J\times J$ matrices. $U$ denotes 
an arbitrary unitary matrix, which we need to include to take into account 
the gauge choice, $A_0=0$.

In order to obtain the equations obeyed by the classical interpolating 
configurations we rewrite the bosonic part of the Euclidean 
action (\ref{RFeuclideanaction}) in the form of a sum of squares plus a 
boundary term, 
\begin{align}
S_E =&  \frac{1}{2 R}  \int \dr t \: \text{tr}  \biggl[ 
\left( \frac{\text{d} X^m}{\text{d} t} \right)^2
+\frac{1}{R^2}\left( X^m \right)^2
+\frac{(2\pi T)^2R^2}{2} \left\{ 
(i[X^m,X^n])^2 +(i[X^m,Y^i])^2  \right\} 
\notag \\
&+ \left( \frac{\text{d} Y^i}{ \text{d}t} \pm \frac{2}{R} Y^i \pm i (2\pi T) 
\frac{R}{2} \epsilon^{ijk} [Y^j,Y^k]   \right)^2 
\mp \frac{\text{d}}{\text{d}t} \left(  \frac{2}{R}  Y^iY^i + i (2\pi T) 
\frac{R}{3} \epsilon^{ijk}Y^i[Y^j,Y^k]  \right) \nn \\
&+ {\rm fermions} \biggl] \, .
\label{RFactionsumofsquares}
\end{align}
From~(\ref{RFactionsumofsquares}) it follows that, on configurations 
interpolating between vacua at $t=\pm\infty$, the 
bosonic part of the Euclidean action 
obeys the bound
\begin{equation}
S_E \ge \mp \ \text{tr}\!\left( \frac{1}{R^2}  Y^i Y^i + i \frac{(2\pi T)}{6} 
\epsilon^{ijk}Y^i[Y^j,Y^k] \right)\!\biggl|^{+\infty}_{-\infty} \, .
\label{RFbound1}
\end{equation}
Using the explicit form of the vacuum configurations at $t=\pm\infty$ the 
bound~(\ref{RFbound1}) becomes
\begin{equation}
S_E \ge \left.\mp\fr{2R}\, W[Y]\right|_{-\infty}^{+\infty}
= \mp \frac{1}{3R^2} \left( \frac{2}{(2\pi T) R^2 } \right)^{\!2} 
\tr\!\left( L^i_{(+\infty)} L^i_{(+\infty)} - L^i_{(-\infty)} L^i_{(-\infty)} \right) \, ,
\label{RFbound}
\end{equation}
where following~\cite{RBBHP} we defined the functional
\begin{equation}
W[Y] = \tr\left(\frac{2}{R}  Y^iY^i + i (2\pi T) \frac{R}{3} 
\epsilon^{ijk}Y^i[Y^j,Y^k]\right)  \, .
\label{RFsuperpotential}
\end{equation}
The BPS (anti-)instantons are  the configurations~\cite{RBBHP,RBYeeYi}
which saturate the bound (\ref{RFbound}) and therefore satisfy the 
equations 
\begin{equation}
\frac{\text{d} Y^i}{ \text{d}t} \pm \frac{2}{R} Y^i 
\pm 
i (2\pi T) \frac{R}{2} \epsilon^{ijk} [Y^j,Y^k]  
=0,
\label{RFBPSEq}
\end{equation}
with $X^m=0$ and $\Psi=0$. In this paper, we refer to solutions 
of~(\ref{RFBPSEq}) in which one chooses the upper or lower signs as 
instantons or anti-instantons, respectively. We note that the Gauss law 
constraints are satisfied as a consequence of~(\ref{RFBPSEq}),
\begin{equation}
\left[ Y^i, \frac{\text{d} Y^i}{\text{d} t} \right]
=
\mp 
i (2\pi T) \frac{R}{2} \epsilon^{ijk} 
[Y^i, [Y^j,Y^k] ]=0.
\end{equation}
Since the (bosonic part of) the Euclidean action~(\ref{RFeuclideanaction}) 
is non-negative, so should be the expressions on the right hand side of
(\ref{RFbound1}) and (\ref{RFbound}), when the bounds are saturated.  
Therefore $W[Y]$ should decrease from $t=-\infty$ to $t=+\infty$ in the case 
of instantons (upper signs in~(\ref{RFbound1})-(\ref{RFBPSEq})) and 
increase in the case anti-instantons. 
This can be seen explicitly noticing that the instanton equations imply
\begin{equation}
\frac{\dr}{\dr t}W[Y] = \mp \fr{2}\, 
\!\left|\frac{\del W[Y]}{\del Y^i}\right|^2 \, ,
\label{RFWtimederivative}
\end{equation}
where the signs are correlated with those in~(\ref{RFBPSEq}). 

In the following we focus on instanton configurations and all the equations 
that we present correspond to the choice of upper signs in~(\ref{RFBPSEq}). 
Anti-instanton solutions can be simply obtained from the corresponding  
instanton by changing $t$ to $-t$. 

A detailed discussion of the general conditions for the existence of solutions 
to the instanton equations~(\ref{RFBPSEq}) can be found in~\cite{RBBHP}. 
We will revisit these issues in section~\ref{RSSModuliSpaces}.
Here we only recall a necessary condition. 
For a BPS solution connecting two 
vacua to exist, the vacuum with the larger value of $W[Y]$ should correspond 
to a representation which does not contain more irreducible blocks than the 
one with the smaller value of $W[Y]$. This allows us to identify instanton 
configurations as corresponding to membrane splitting processes, while 
anti-instantons describe joining processes. 

Evaluating $W[Y]$ on a vacuum configuration~(\ref{RFMMVacua}) 
containing $m$ membranes with angular momenta $J_i$, $i=1,\ldots,m$, 
one gets
\begin{equation}
W[Y_0] = \frac{2}{3R}\left( \frac{2}{(2\pi T)R^2} \right)^{\!2}  
\tr\!\left(L^iL^i\right) 
= \frac{2}{3R}\left( \frac{2}{(2\pi T)R^2} \right)^2 \frac{1}{4} 
\sum^{m}_{i=1} \tr \left[ \left( J^2_i -1 \right) \Id_{J_i \times J_i}  \right] \, .
\label{RFW0}
\end{equation}
Therefore on a generic instanton interpolating between the 
representations $\ul{J_1}\oplus\ul{J_2} \oplus\cdots\oplus\ul{J_m}$ 
and $\ul{J'_1}\oplus\ul{J'_2} \oplus\cdots\oplus\ul{J'_n}$ (with 
$\sum_{i=1}^m J_i = \sum_{i=1}^n J'_i = J$, $n\ge m$) the Euclidean action 
is 
\begin{align}
S_E &= \frac{-1}{2R}\big(\tr\left(W[Y_0(+\infty)]\right)
-\tr\left(W[Y_0(-\infty)]\right)\big) \label{RFinstaction} 
\\
&
= \frac{-4}{3(2\pi T)^2 R^6 } \left(\tr\!\left( L^i_{(+\infty)} L^i_{(+\infty)} 
- L^i_{(-\infty)} L^i_{(-\infty)} \right) \right)
= \fr{3(2\pi T)^2R^6} \left( \sum_{i=1}^m {J_i}^3 
- \sum_{i=1}^n {J'_i}^3 \right) \, \ge 0 \nn
\end{align}
In the following sections we will focus on the most elementary process in 
which a single membrane splits into two. This corresponds to an instanton 
solution in which $L^i_{(-\infty)}$ is taken to be the irreducible representation 
$\ul{J}$ and $L^i_{(+\infty)}$ is the representation 
$\ul{J_1}\oplus\ul{J_2}$, where $J_1+J_2=J$ as required by angular 
momentum conservation. For this process the Euclidean action is
\begin{align}
S_E = \frac{1}{4R^2} \left( \frac{2}{(2\pi T) R^2 } \right)^{\!2} J J_1J_2 
= \fr{8} \frac{JJ_1J_2}{N} \, ,
\label{RFlowest}
\end{align}
where in the final equality we have used the relations $T=1/[(2\pi)^2l_P^3]$ 
and $(R/l_P)^6=2^5\pi^2 N$.

The (anti-)instantons described above are local minima of the Euclidean 
action of the pp-wave matrix model. Their contribution to physical transition
amplitudes associated with membrane splitting/joining processes can be
evaluated using a standard semi-classical approximation. The latter involves
the integration over the bosonic and fermionic collective coordinates 
associated with zero modes in the instanton background. From the 
semi-classical calculation of transition amplitudes it should be possible to 
extract effective interaction vertices for membranes. In the present paper 
we focus on the construction of classical instanton solutions, leaving the 
semi-classical evaluation of the corresponding amplitudes for a future 
publication. Among the questions that we will not address is whether the 
BPS instantons discussed here give the dominant contributions to the 
amplitudes, or instead other saddle points associated with non-BPS 
configurations need to be taken into account. 

\subsection{Some properties of BPS instantons and their moduli spaces}
\label{RSSInstantonandmoduli}

A key observation used in the analysis of~\cite{RBBHP} is 
that~(\ref{RFBPSEq}) can be mapped to the Nahm equations arising in 
the construction of BPS multi-monopole configurations~\cite{RBNahm}. 
In order to reduce~(\ref{RFBPSEq}) to the Nahm equations one considers 
the change of variables~\cite{RBBHP}
\begin{equation}
Z^i = C \,\er^{2t/R} Y^i, \qquad
s =  \er^{-2t/R}, 
\label{RFBHPToNahm}
\end{equation}
where $C$ is a constant, which we take to be real, so that the $Z^i$'s are 
Hermitian matrices. Substituting into~(\ref{RFBPSEq}) gives 
\begin{equation}
\frac{\text{d} Z^i}{ \text{d} s} =  
i (2\pi T) \frac{R^2}{4C}  \epsilon^{ijk}[Z^j,Z^k] \, ,
\label{RFNahm}
\end{equation}
which by a suitable choice of the constant $C$ can be brought to the 
canonical form of the Nahm equations. 

In terms of the original variables, we are interested in solutions $Y^i(t)$ 
approaching constant configurations proportional to different SU(2)  
representations at $t=\pm\infty$. After the change of 
variables~(\ref{RFBHPToNahm}) we therefore need to consider Nahm
equations defined on a semi-infinite interval with boundary 
conditions~\cite{RBBHP}
\begin{equation}
Z^i(s) \sim \frac{L^i_{(-\infty)}}{s} + \cdots  \quad {\rm for}~s\to\infty \, ,
\qquad
Z^i(s) \sim \frac{L^i_{(+\infty)}}{s} + \cdots  \quad {\rm for}~ s\to 0 \, ,
\label{RFZboundaryconditions}
\end{equation}
where the ellipses indicate sub-leading terms.

The Nahm equations have been extensively studied for their central role in
the physics of monopoles and a great deal is known about their 
properties. The particular boundary 
conditions~(\ref{RFZboundaryconditions}) relevant in the present case 
are non-standard and less studied, although they have been considered in a
different context in~\cite{RBKronheimer}. In general, little is known about 
explicit solutions to~(\ref{RFNahm})-(\ref{RFZboundaryconditions}). The 
only known examples appear to be solutions presented in~\cite{RBBHP}.
These include a particular configuration interpolating between a vacuum 
associated with the representation $L^i_{(-\infty)}$ and the trivial vacuum 
corresponding to the $J$-dimensional representation 
$\ul{1}\oplus\ul{1}\oplus\cdots\oplus\ul{1}$. The explicit form of this 
solution, in terms of the original $Y^i(t)$ matrices, is
\begin{equation}
Y^i(t)= \frac{2}{(2\pi T)R^2}
\frac{1}{1+ \er^{2(t-t_0)/R}} \,L^i_{(-\infty)} \, .
\label{RFBHPSolution}
\end{equation}
In the following we will refer to~(\ref{RFBHPSolution}) as the 
Bachas-Hoppe-Pioline (BHP) solution. 
The only other known explicit solutions are slight variations of this one,
obtained by means of tensor products. They interpolate between different 
specific pairs of representations and were also presented in~\cite{RBBHP}. 
However, no systematic approach to the construction of solutions has been 
proposed and no concrete examples are known for the class of processes 
we are interested in, \ie those in which all the membranes involved in the 
splitting or joining transition carry large angular momenta. In the following 
sections we propose a strategy for the construction of instanton 
configurations in this large angular momentum sector. For the important 
case of the splitting of one membrane into two we also present an explicit 
solution, which represents an essential step in the calculation of the 
associated physical transition amplitude.

Various properties of the moduli space of BPS instantons were discussed 
in~\cite{RBBHP}, where, in particular, the dimension of the moduli space of 
a general instanton configuration was determined. For an instanton 
describing the splitting process with $m$ membranes at $t=-\infty$
and $n$ membranes at $t=+\infty$ -- \ie connecting the two representations 
$\ul{J_1}\oplus\ul{J_2}\oplus\cdots\oplus\ul{J_m}$ and 
$\ul{J'_1}\oplus\ul{J'_2}\oplus\cdots\oplus\ul{J'_n}$ (with 
$\sum_{i=1}^m J_i = \sum_{i=1}^n J'_i = J$ and $n\ge m$ and assuming that
this is an allowed process according to the criterion in~\cite{RBBHP}) -- the 
(complex) dimension of the moduli space, $\calM$, was found to be
\begin{equation}
\dim_\C\left(\calM\right) = \sum_{i=1}^n (2i-1) J'_i 
- \sum_{i=1}^m (2i-1) J_i \, ,
\label{RFDimModuliSpace}
\end{equation}
where in the sums the representations are ordered by decreasing 
dimension, \ie $J_1\ge J_2 \ge \cdots \ge J_m$ and $J'_1 \ge J'_2 \ge 
\cdots \ge J'_n$. For the splitting of one membrane into 
two~(\ref{RFDimModuliSpace}) gives a (real) dimension
\begin{equation}
\dim_\R\left(\calM\right) = 2(J'_1 + 3 J'_2 - J_1) = 4 J'_2 \, ,
\label{RFDimM12}
\end{equation}
where we used $J_1=J'_1+J'_2=J$. 

An interesting feature of the moduli space of BPS instantons is an 
additivity rule~\cite{RBBHP}. Given three vacua associated with the 
SU(2) representations $L_{(A)}$, $L_{(B)}$ and $L_{(C)}$, the moduli 
spaces $\calM_{(A,B)}$, $\calM_{(B,C)}$ and $\calM_{(A,C)}$ of 
configurations connecting $A$ and $B$, $B$ and $C$ and $A$ and $C$ 
respectively satisfy
\begin{equation}
\dim\left(\calM_{(A,B)}\right) + \dim\left(\calM_{(B,C)}\right) = 
\dim\left(\calM_{(A,C)}\right) \, .
\label{RFAdditivity}
\end{equation}
In section~\ref{RSSSolutionGeneral} we will comment on how this property 
is reflected in the description of the instanton moduli space arising from our
reformulation in terms of the Laplace equation. 

The instanton configurations discussed in the previous subsection 
interpolate between vacua of the pp-wave matrix model containing 
different numbers of membranes. In~\cite{RBDSVR2} it was shown, 
based on properties of the representations of the relevant 
supersymmetry algebra, that these vacua are 
non-perturbatively protected 
and their energies are exactly zero in the full quantum theory. This means 
that different vacua do not mix and tunnelling transitions should be 
possible only if excited states are involved. 
It is natural to expect that the 
mechanism forbidding transitions between vacua should be associated 
with selection rules induced by the integration over fermion zero modes 
in the instanton background. This was shown explicitly in~\cite{RBYeeYi}
in the case of the special solution~(\ref{RFBHPSolution}) and similar
selection rules should exist for more general tunnelling amplitudes such
as those that we study in this paper.

\section{Mapping to three-dimensional Laplace equation}
\label{RSApprox}

In this section we present our general strategy for the construction of 
instanton solutions describing tunnelling configurations interpolating between 
vacua containing large membranes. 

\subsection{Continuum approximation}

The central ingredient in the approach that we develop is a continuum 
approximation valid for large $J$, \ie for large matrix size. In this 
approximation scheme, the matrices, $Y^i(t)$, are replaced by functions, 
$y^i(t, \s^1,\s^2)$, of two spatial coordinates, $(\s^1, \s^2)$, and (Euclidean)
time, $t$. The mathematics underlying this replacement is essentially the 
same that is utilised in the derivation of the matrix model as a regularisation 
of the supermembrane theory~\cite{RBGoldstoneHoppe, RBdWHN}. In the 
matrix regularisation, the starting point is the continuum theory which is 
formulated in terms of functions on the membrane 2+1 dimensional 
world-volume. At fixed time, for any function, $f$, defined on the membrane 
world-space parametrised by the coordinates $(\s^1,\s^2)$, one introduces 
a corresponding matrix, $\rho(f)$. The map $\rho$  between functions and 
matrices satisfies the properties~\footnote{For a discussion of the matrix 
regularisation emphasising its interpretation as an approximation between 
discrete and continuum objects, see \cite{RBShimadaMR}. 
}
\begin{align}
\rho(f g) &\approx \frac{1}{2} \Big(\rho(f) \rho(g) + \rho(g) \rho(f)\Big),
\label{RFMRMultiplication} \\
\rho\left(\{f, g\}\right) &\approx \frac{2\pi J}{i \tA} 
\left[ \rho(f), \rho(g) \right],
\label{RFMRCommutator} \\
\frac{1}{\tA}\int f \, \dr^2 \s &\approx \frac{1}{J} \tr \big(\rho(f)\big) \, ,
\label{RFMRTr}
\end{align}
where the two sides are understood to be equal up to terms of higher order in 
$1/J$. The Lie bracket, $\{f, g\}$, between two functions, $f(\s^1,\s^2)$ and 
$g(\s^1,\s^2)$, is defined as 
\begin{equation}
\{f,g\} = \frac{\del f}{\del\s^1}\frac{\del g}{\del\s^2}
- \frac{\del f}{\del\s^2}\frac{\del g}{\del\s^1} 
\label{RFLieBracket}
\end{equation}
and the conventional constant $[\s]$ denotes the area of the base space 
\begin{equation}
[\s]=\int \text{d}^2\s.
\end{equation}
In the present case we use a substitution of this type in reverse order. The 
starting point are time-dependent matrices, which we approximate by 
functions depending on two auxiliary continuous coordinates. Such an 
approach, based on the approximation of discrete objects by functions of
continuous variables is of course quite standard. An elementary example of 
this type of approximation is the description of sound waves (phonons) in a 
solid, \ie a discrete lattice, by a continuum wave equation. 

Using the rules~(\ref{RFMRMultiplication})-(\ref{RFMRTr}),
the instanton equations~(\ref{RFBPSEq}) obeyed by the matrices $Y^i(t)$ 
get mapped to non-linear partial differential equations for a set of functions 
$y^i(t,\s^2,\s^2)$, 
\begin{equation}
\frac{\partial y^i}{\partial t} 
\pm \frac{2}{R} y^i 
\mp \frac{(2\pi T)R}{4\pi}\frac{[\sigma]}{J} \epsilon^{ijk} \{y^j,y^k \}  = 0 \, .
\label{RFContinuumBPSEq}
\end{equation}
The boundary conditions for the matrices $Y^i(t)$ at $t=\pm\infty$ 
determine the corresponding boundary conditions for the functions 
$y^i(t,\s^1,\s^2)$. The latter should  describe families of concentric 
spheres at $t=\pm\infty$. The approximate 
identities~(\ref{RFMRMultiplication})-(\ref{RFMRTr}) imply 
that, for large $J$, solutions to the original equations~(\ref{RFBPSEq}) are
well approximated by functions satisfying these equations with the 
appropriate boundary conditions.

The linear term in the continuum version~(\ref{RFContinuumBPSEq}) 
of the BPS equations can be eliminated by the change of variables
\begin{equation}
z^i = \frac{(2\pi T)R^2}{4\pi} \,\er^{2 t /R} y^i \, , \qquad
s =  \er^{-2 t/R}, 
\label{RFContiBPSToContiNahm}
\end{equation}
which is analogous to~(\ref{RFBHPToNahm}) that was used 
in~\cite{RBBHP}~\footnote{\label{RFootnoteChangeVariable}
This change of variables is similar to that used in \cite{RBShimada3}
to relate the Poincar\'e coordinates of the AdS space to the coordinates 
of the pp-wave space. It is also reminiscent of the coordinate transformation 
involved in the radial quantisation of the CFT on the boundary.}.
Substituting into~(\ref{RFContinuumBPSEq}) yields the equations 
\begin{equation}
\frac{\partial z^i}{ \partial s} = - \frac{[\sigma]}{2 J} 
\epsilon^{ijk} \{ z^j,z^k \} \, ,
\label{RFContinuumnahm}
\end{equation}
for the functions $z^i(s,\s^1,\s^2)$, $i=1,2,3$, where the variable $s$ is 
defined on a semi-infinite interval with $s=+\infty$ corresponding to 
$t=-\infty$ and $s=0$ corresponding to $t=+\infty$. As in the case of the 
$y^i$'s the boundary conditions for the $z^i$'s at $s=+\infty$ and $s=0$ 
follow from those required for the corresponding 
matrices~(\ref{RFZboundaryconditions}). We will return to the analysis of the
relevant boundary conditions for the equations~(\ref{RFContinuumnahm}) 
in the next section where we discuss various explicit solutions.

The equations~(\ref{RFContinuumnahm}), sometimes 
referred to as the SU($\infty$) Nahm equations (after an appropriate 
rescaling to normalise the coefficients), are known to be (locally) equivalent 
to the three-dimensional Laplace 
equation~\cite{RBWard, RBHoppeLaplaceEq}~\footnote{
The continuum Nahm equations were also considered in connection 
with membrane theory in \cite{RBFloratosLeontaris}.}.
In order to map~(\ref{RFContinuumnahm}) to the Laplace equation one
uses a so-called hodograph transformation -- a change of variables in
which the roles of dependent and independent variables are exchanged,
\begin{equation}
(z^1,z^2,z^3) ~ \longleftrightarrow ~ (s,\s^1,\s^2) \, .
\label{RFhodograph}
\end{equation}
A fundamental feature of such a transformation is that it allows one to 
map a non-linear differential equation to a linear one. 

The change of variables~(\ref{RFhodograph}) means, in particular, that one 
considers the originally independent variable $s$ as a function, $\phi(z^i)$, 
of the new independent variables $z^i$, $i=1,2,3$. One can then show that if 
the $z^i$'s satisfy the equations~(\ref{RFContinuumnahm}) as functions of 
$s, \s^1, \s^2$, then $s=\phi(z^i)$ obeys the Laplace equation
\begin{equation}
\bm{\nabla}^2 \phi=
\frac{\partial}{\partial z^i}\frac{\partial}{\partial z^i}\phi=0 \, .
\label{RFLaplaceEquation}
\end{equation}
A geometric derivation of this result is presented in  
appendix~\ref{RSAGeometricalDerivationCNahmtoLaplace}. Alternative
proofs can be found in~\cite{RBWard, RBHoppeLaplaceEq}.

This reformulation of the continuum version of the Nahm equations 
has a simple and very intuitive interpretation. The time variable $s$ 
becomes a `potential' function obeying the Laplace equation. The 
associated equipotential surfaces, $\phi(z^i)\!=\!{\rm const.}$, for different 
values of the potential represent implicit equations describing the membrane 
profile at the given time. Therefore a sequence of equipotential surfaces 
for values of $s$ ranging from $s=+\infty$ to $s=0$ directly captures the 
evolution of the membrane configurations in Euclidean time. 

\subsection{Classical membrane perspective}
\label{RSSClassicalMembranePerspective}

In the previous subsection we have obtained the 
equations~(\ref{RFContinuumBPSEq}) as an approximation, valid for
large matrices, to the original instanton equations~(\ref{RFBPSEq}). 
More specifically the approximation requires the dimension of all the 
irreducible representations contained in the vacua at $t=\pm\infty$ to be 
large. The same equations can also be understood as classical 
(Euclidean) equations of motion for the membrane theory in the 
AdS$_4\times S^7$ background in the pp-wave approximation.
In the notations of~\cite{RBKSS1} the Euclidean action for the pp-wave 
membrane theory (in the $A_0=0$ gauge) is
\begin{align}
S_E = \int & \text{d}t \,\text{d}^2 \s  
\biggl[ \frac{1}{2R}\frac{J}{[\sigma]} 
\left( \frac{\partial y^i}{\partial t}  \right)^2 
+ \frac{1}{2R}\frac{J}{[\sigma]} 
\left( \frac{\partial x^m}{\partial t} \right)^2 
+\frac{1}{2R^3} \frac{J}{[\sigma]}   (x^m)^2 
+ \frac{2}{R^3} \frac{J}{[\sigma]}  (y^i)^2  \notag \\[0.1cm]
& + \frac{RT^2}{4} \frac{[\sigma]}{J} \left( \{x^m,x^n \}^2 
+2\{x^m,y^i \}^2 + \{y^i,y^j \}^2  \right)
- \frac{T}{R}\epsilon^{ijk}y^i \{ y^j,y^k \} \notag \\[0.1cm]
&+\frac{1}{R} \frac{J}{[\sigma]} \theta^T \frac{ \partial \theta}{\partial t} 
+i T \left( \theta^T \gamma^i \{ y^i , \theta \} 
+ \theta^T \gamma^m \{ x^m , \theta \} \right) 
-\frac{3i}{2} \frac{J}{[\sigma]} 
\frac{1}{R^2} \theta^T \gamma^{123} \theta \biggl] \, .
\label{RFeuclideanmembraneaction1}
\end{align}
As in the case of the matrix model, the bosonic part of this action can be
rewritten as a sum of squares plus a boundary term, 
\begin{align}
\frac{1}{2R}\frac{J}{[\sigma]}
\int & \text{d}t \text{d}^2 \s 
\biggl[ \left(\frac{\del x^m}{\del t}\right)^{\!2} +\frac{1}{R^2}(x^m)^2 
+\frac{R^2T^2}{2} \frac{[\sigma]^2}{J^2} \left( \{x^m,x^n \}^2 
+2\{x^m,y^i \}^2  \right) 
\label{RFeuclideanmembraneaction2}
\\[0.1cm]
&+ \left( \frac{\partial y^i}{\partial t} \pm \frac{2}{R} y^i 
\mp \frac{RT}{2}\frac{[\sigma]}{J} \epsilon^{ijk} \{y^j,y^k \} \right)^{\!2} 
\mp \frac{\partial}{\partial t}
\left( \frac{2}{R} (y^i)^2 - \frac{RT}{3}\frac{[\sigma]}{J} 
\epsilon^{ijk} y^i \{ y^j,y^k \} \right) 
\biggl] \, . \nn
\end{align} 
This formula shows that~(\ref{RFContinuumBPSEq}) can be obtained 
minimising the membrane Euclidean action and therefore these equations
describe the BPS instanton configurations of the membrane theory.

This is of course consistent and not surprising. The matrix model contains
membrane degrees of freedom and the matrix configurations we focussed
on represent regularised membrane states. However, we prefer to emphasise 
the point of view presented in the previous subsection in which the continuum 
equations~(\ref{RFContinuumBPSEq}) are viewed as an approximation to 
the corresponding matrix model instanton equations. This is because the 
matrix model itself is more fundamental as a candidate for a microscopic 
formulation of quantum M-theory. Moreover the calculation of physical 
transition amplitudes in semi-classical approximation should be carried
out in the matrix model. This in particular should allow one to compute 
tunnelling amplitudes between states corresponding to non-Abelian degrees 
of freedom associated with excitations in off-diagonal blocks, which have no 
simple counterpart in the continuum. 

\section{Solutions}
\label{RSSolutions}

In this section we discuss solutions to the Laplace equation which 
correspond to membrane splitting processes. We will see that, in order
to construct solutions with the required properties, we need to introduce 
the concept of Riemann space. After explaining the general features
of our proposal, we present exact solutions to the Laplace 
equation~(\ref{RFLaplaceEquation}) with appropriate boundary conditions, 
which, based on the arguments in the previous section, provide approximate 
solutions to the BPS instanton equations~(\ref{RFBPSEq}). 

We first discuss the simplest solution corresponding to a static single 
spherical membrane, \ie  the simplest stable vacuum of the pp-wave matrix 
model. 
This case (which can be thought of as a solution in the zero instanton 
sector) allows us to illustrate the effects of the sequence of changes of 
variables that we use to map the original instanton equations to the Laplace 
equation. In section~\ref{RSSSolutionHobson} we then discuss the solution 
corresponding to the splitting of a single membrane into two membranes, 
which requires the introduction of the notion of Riemann space.  More general 
splitting processes, which involve more complex examples of Riemann 
spaces, are discussed in section~\ref{RSSSolutionGeneral}. A reformulation 
of the solution~(\ref{RFBHPSolution}) discussed in~\cite{RBBHP, RBYeeYi} in 
terms of the Laplace equation is presented, together with other simple new 
solutions, in appendix~\ref{RSAOtherSolutions}.

\subsection{Stable sphere}
\label{RSSSolutionCoulomb}

In order to understand the asymptotic behaviour of non-trivial instanton 
solutions and the corresponding required boundary conditions, it is 
instructive to first consider the simplest solution to the BPS instanton 
equations~(\ref{RFBPSEq}), namely the static configuration corresponding to
the irreducible representation $\ul{J}$,
\begin{equation}
Y^i(t) = \frac{2}{(2\pi T)R^2} L^i \, , \quad \forall \, t \in (-\infty,+\infty) \, ,
\label{RF1StableSphere}
\end{equation}
where $[L^i,L^j] = i \epsilon^{ijk}L^k$. The continuum counterpart of this 
solution is
\begin{equation}
y^i(t, \s^1, \s^2) = r \,n^i(\s^1,\s^2) \, ,
\label{RFcontinuuum1sphere}
\end{equation}
where $r=J/[(2\pi T)R^2]$ and $n^i(\s^1,\s^2)$ is a unit vector in the radial 
direction. Using the standard parametrisation of the sphere we can take
\begin{equation}
n^i(\s^1,\s^2) = (\sin\theta\cos\v,\,\sin\theta\sin\v,\,\cos\theta) \, , 
\quad {\rm with} \quad \s^1 = 1-\cos\theta \,, ~ \s^2 = \v \, .
\label{RFradialunitvec}
\end{equation} 
Equations~(\ref{RFcontinuuum1sphere})-(\ref{RFradialunitvec}) provide a 
static solution to the continuum version of the BPS instanton 
equations~(\ref{RFContinuumBPSEq}), as can be verified using the 
definition of the Lie bracket~(\ref{RFLieBracket}). In~(\ref{RFradialunitvec})
we have chosen a parameterisation suitable for the North pole patch of the
unit sphere. In general there is freedom in the choice of the $(\s^1,\s^2)$
variables associated with the invariance of the membrane theory under 
area preserving diffeomorphisms. The $(\s^1, \s^2)$ coordinates should be 
such that the area element $\dr\s^1 \dr\s^2$ coincides with the natural 
SO(3) symmetric area element on a round sphere.

Using the change of variables~(\ref{RFContiBPSToContiNahm})
we get a solution to the continuum Nahm 
equations~(\ref{RFContinuumnahm})
\begin{equation}
z^i(s, \s^1, \s^2)= \frac{J}{4\pi} \fr{s} \, n^i(\s^1,\s^2) \, .
\label{RFContNahm1Sphere}
\end{equation}
This represents a sphere with radius increasing as $s$ varies from
$+\infty$ to $0$. Inverting~(\ref{RFContNahm1Sphere}) to obtain 
$s=\phi(z^i)$ we immediately deduce that the potential solving the associated 
Laplace equation satisfies
\begin{equation}
\phi(\bm{z}) 
= \frac{J}{4\pi |\bm{z}|},
\label{RFpotential1}
\end{equation}
where $\bm{z}=(z^1,z^2,z^3)$. Therefore the solution to the Laplace equation
corresponding to a single static membrane configuration is simply the 
Coulomb potential generated by a positive point charge $J$ located at the 
origin.

We note that a static spherical membrane in the $(t, \bm{y})$ variables
becomes a time-dependent spherical membrane in the $(s, \bm{z})$ 
variables, with radius changing in time  as $1/s$.
Using the transformation~(\ref{RFContiBPSToContiNahm}) in reverse order, 
one finds indeed that~(\ref{RFpotential1}) corresponds to a membrane 
whose distance from the origin is constant in time $t$,
\begin{equation}
|\bm{y}| = \frac{J}{(2\pi T)R^2} \, , \quad {\rm with} \quad 
\bm{y}=(y^1,y^2,y^3) \, ,
\label{membranelocation1}
\end{equation}
which of course represents a static sphere of radius $J/(2\pi T R^2)$.

As a simple generalisation of~(\ref{RFpotential1}) we can consider the case
of a point charge located away from the origin. The potential is
\begin{equation}
\phi (\bm{z}) 
= \frac{J}{4\pi |\bm{z} - \bm{z}_{0}|} \, , \quad \bm{z}_0 \ne 0 \, .
\end{equation} 
In this case the equipotential surfaces are spheres centred at $\bm{z}_0$.
In terms of the original $\bm{y}$ variables we have
\begin{equation}
|\bm{y}| = \frac{J}{(2\pi T)R^2} \frac{|\bm{z}|}{|\bm{z}-\bm{z}_0|}.
\label{RFRunAway}
\end{equation}
This configuration corresponds to a spherical membrane moving in 
towards the origin from infinity. As (Euclidean) time evolves from $t=-\infty$,  
corresponding to $s=+\infty$ where $|\bm{z}-\bm{z}_0|\to 0$, to $t=+\infty$, 
corresponding to $s=0$, $|\bm{y}|$ decreases from $\infty$ to a constant.
This run-away behaviour of the solution in the infinite past makes it 
non-physical,
because the solution does not satisfy the required boundary 
conditions. This is a general result, which implies that the potentials relevant 
for the description of configurations interpolating between stable vacua of 
the pp-wave matrix model can only involve positive charges located at the
origin~\footnote{
The location of negative charges can be arbitrary.
In appendix \ref{RSAOtherSolutions} we will present explicit examples of 
physically acceptable solutions involving negative charges away from the
origin. 
}.

\subsection{Single membrane splitting and Riemann space}
\label{RSSSolutionHobson}

In this section we consider the most interesting solution, which describes a 
single  membrane with angular momentum $J$ splitting into two membranes 
with angular momenta $J_1$ and $J_2$ (with $J_1+J_2=J$). This 
corresponds to a configuration interpolating between a spherical membrane 
of radius $J/(2\pi T R^2)$ and two concentric membranes of radii 
$J_i/(2\pi T R^2)$, $i=1,2$. It is the most elementary example of a splitting 
process and it allows us to illustrate general properties which are common to 
all splitting/joining transitions.

\subsubsection{Qualitative picture: introduction of the concept of Riemann
space}

\begin{figure}[htb]
\centering
\includegraphics[width=\textwidth]{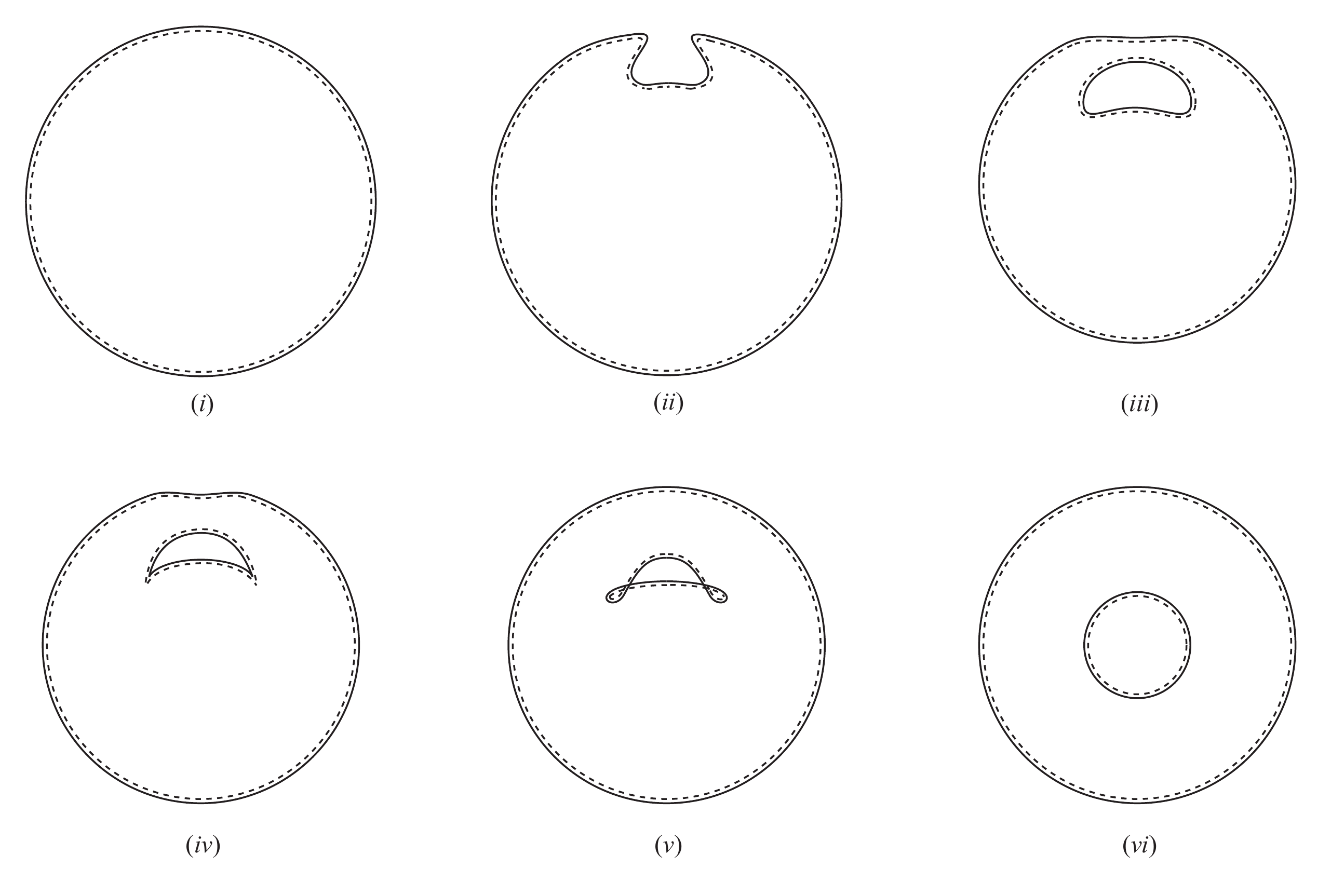}
\caption{Qualitative depiction of the splitting process.
The process necessarily involves both splitting and flipping.}
\label{RPQualitativeSplit}
\end{figure}

We will now argue that it is not possible to describe a membrane splitting 
process in terms of a solution to the Laplace equation formulated in a 
standard Euclidean space $\R^3$. Instead, we are led to introduce 
the concept of Riemann space explained in detail below.
In order to motivate our proposal, let us first discuss the qualitative features 
of a typical instanton solution describing a splitting process. The transition 
from a spherical membrane to two concentric membranes can be 
qualitatively described as involving the steps depicted in 
figure~\ref{RPQualitativeSplit}. Starting with a single membrane of radius 
$J/(2\pi T R^2)$ a small perturbation forms and grows until the membrane 
splits into two, resulting in the creation of a second disconnected membrane 
inside the first one, as shown in steps ($i$) to ($iii$) in the 
figure~\footnote{
The point where the splitting occurs is singular from the point of view of 
the partial differential equation governing the evolution of the membranes.
The nature of the singularity and the properties of the instanton
solution in the vicinity of a generic splitting point are universal and
can be studied using the description in terms of the Laplace equation, 
as discussed in appendix~\ref{RSASplittingPoint}.}.

To understand the subsequent steps it is important to notice that membranes 
are charged and thus oriented objects. In figure~\ref{RPQualitativeSplit} the 
orientation of the membranes is indicated by the relative position of the 
continuous and dashed contour lines: membranes have an outer continuous 
contour, whereas in anti-membranes the outer contour is dashed. In the 
pp-wave background a configuration of two concentric membranes with the 
same orientation, which is BPS, is stable as a result of the balance between 
the gravitational attraction and the force associated with the three-form flux. 
The two forces add up in the case of a membrane anti-membrane pair, 
resulting in an unstable configuration. 

As shown in figure~\ref{RPQualitativeSplit}, after the splitting takes place the 
internal membrane that is formed has opposite orientation compared to the 
initial one and the resulting configuration is not stable. To complete the 
transition to a stable vacuum consisting of two concentric membranes it is 
necessary for the internal membrane to flip its orientation. This process 
requires intermediate steps in which the membrane self-intersects after 
developing cusps, see ($iv$)-($vi$) in figure~\ref{RPQualitativeSplit}.

The process shown in the figure has axial symmetry, but this is of course
not a requirement. In general the details of the intermediate configurations 
can vary, however, an essential element, common to all splitting
processes, is the fact that a flipping of the orientation is inevitable.  
In the illustrative example shown in the figure the sequence of steps leading 
to the final two-membrane state involves a splitting resulting in the formation 
of a second internal anti-membrane followed by the flipping of the 
orientation of the latter. However, as we will see in explicit examples in the 
next subsection, the splitting and flipping steps can also take place in 
reverse order, or simultaneously. 
 
Applying the approach described in section~\ref{RSApprox}, our goal is to
construct a solution, $\phi(\bm{z})$, to the Laplace equation such that the 
associated equipotential surfaces reproduce the qualitative behaviour shown 
in figure~\ref{RPQualitativeSplit}. 

The crucial feature of any splitting process, which emerges from the 
qualitative arguments we presented above, is the fact that necessarily 
there exists a finite interval of values of the potential for which the 
equipotential surfaces are self-intersecting. Such a behaviour is not 
compatible with the potential $\phi(\bm{z})$ being a solution to the Laplace 
equation in the ordinary $\R^3$ space, because the associated gradient, 
$\bm{\nabla}\phi(\bm{z})$, would be ill-defined, since the normal direction 
differs on the two portions of a self-intersecting equipotential surface. 
In the case of the intersections occurring during the flipping of the 
membrane orientation, this issue arises not just at isolated singular 
points, but over a finite region corresponding to an interval of values of 
$\phi(\bm{z})$. This motivates us to consider multi-valued potential functions. 

More concretely the above observations lead us to propose that the potential 
relevant for the representation of a membrane splitting process should be a 
solution to the Laplace equation in a space, which we refer to as a Riemann 
space, that is a three-dimensional generalisation of a two-sheeted Riemann 
surface. Such a space consists of two copies of $\R^3$ connected by a 
branch surface, bounded by a branch curve or loop. For simplicity we will 
assume the surface connecting the two $\R^3$'s to always have the topology 
of a disk and in the following we will use the expression branch disk without 
necessarily implying a circular planar shape~\footnote{
In appendix~\ref{RSAConnectingConditionIntegralEq} we present a 
reformulation of the Laplace equation in a Riemann space as an
integral equation and we discuss in more detail the boundary conditions 
at the branch disk. To construct solutions relevant for the description  
of membrane splitting processes we require the potential to be finite at
the branch loop and also to decay sufficiently rapidly at infinity.}.
Deformations of the branch disk which leave the branch loop fixed 
yield physically equivalent Riemann spaces.
This is analogous to properties of branch cuts and branch 
points of Riemann surfaces.
A space defined in this way is locally equivalent to 
$\R^3$, but differs globally. The use of a Riemann space makes it possible to 
have a potential which locally solves the Laplace equation, while avoiding the 
problems caused by self-intersecting equipotential surfaces. The solution is 
such that different portions of self-intersecting surfaces live on different sheets 
of the Riemann space, so that no intersections take place within the same 
copy of $\R^3$.

The description of a stable spherical membrane in 
section~\ref{RSSSolutionCoulomb} provides an indication of what the 
asymptotic behaviour of the potential should be in the case of a splitting
process. The infinite past, $t\to-\infty$, corresponds to $s\to+\infty$, \ie 
large values of the potential. In this region the equipotential surfaces should 
be small spheres with radius proportional to $J$ and approaching zero. 
The infinite future, $t\to+\infty$, corresponds to $s\to 0$, \ie small values of 
the potential approaching zero from above. In this region the equipotential 
surfaces should be two large concentric spheres with radii proportional to
$J_1$ and $J_2$ and growing indefinitely. Using a Riemann space we
can construct a potential which provides a concrete realisation 
of this type of asymptotic behaviour and is also well-defined in the 
intermediate region where the membranes self-intersect.
More precisely, in 
order to account for the behaviour at large $\phi$ we consider a point 
charge $J$ at the origin of the first $\R^3$, so that close to the charge 
we have a Coulomb potential, whose equipotential surfaces are small 
spheres. The behaviour in the two $\R^3$'s for small $s$, on the other
hand, can be understood as resulting from the splitting of the initial flux 
$J$ of $\bm{\nabla}\phi$, with a fraction $J_1$ going to infinity in the first 
space and a fraction $J_2=J-J_1$ passing into the second space through 
the branch disk. 

\begin{figure}[p]
\centering
\includegraphics[width=0.99\textwidth]{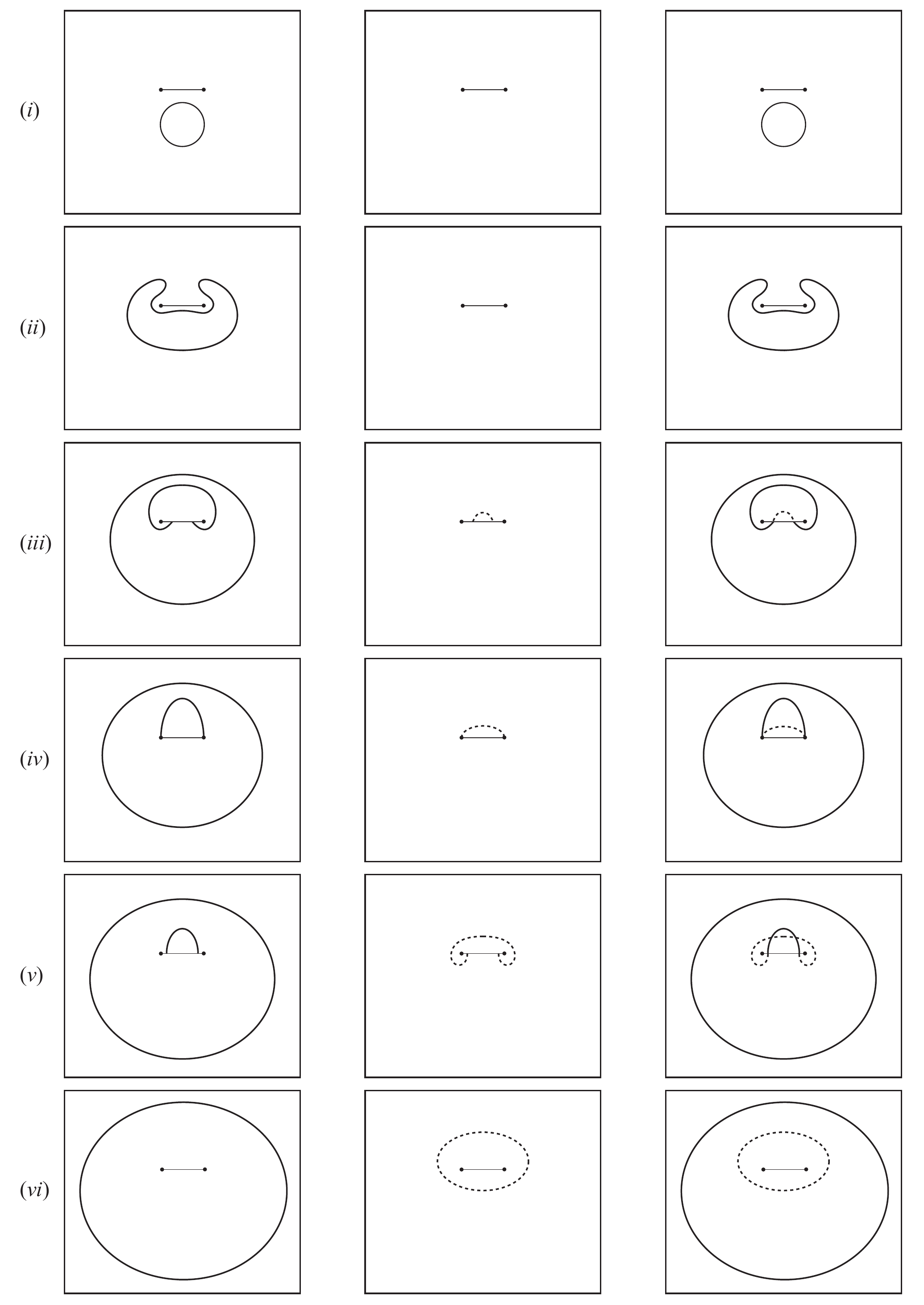}
\caption{Qualitative depiction of splitting in Riemann space.}
\label{RPQualitativeSplitRiemann}
\end{figure}

Figure~\ref{RPQualitativeSplitRiemann} illustrates qualitatively 
the splitting process as rendered in the Riemann space, with the six rows in 
the picture corresponding to the same stages of the splitting process shown in 
figure~\ref{RPQualitativeSplit}. The first two columns depict the two sheets of 
the Riemann space, with each copy of $\R^3$ represented as a rectangle. 
The horizontal slit corresponds to the branch disk, with the dots at the end 
points indicating the branch loop. The third column shows the two copies of 
$\R^3$ superposed. 
Membranes, or portions of membranes, are represented as 
continuous lines in the first $\R^3$ and as dashed lines in the second $\R^3$.
The same notation will be used in the figures throughout the paper. 
Re-analysing the splitting process as presented in the
figure allows us to clarify how the issues associated with the flipping of the
membrane orientation and the self-intersections in the equipotential surfaces
can be addressed using a Riemann space.  As the value of the potential 
decreases, away from the point charge, the equipotential surfaces intersect 
the branch disk and therefore extend partially into the second copy of $\R^3$. 
This is shown at step ($iii$), where the splitting has already taken place. 
Crucially the use of a Riemann space allows us to have self-intersecting 
surfaces which result from the superposition of portions of surfaces without 
self-intersections in each sheet. This is illustrated in row ($v$) in 
figure~\ref{RPQualitativeSplitRiemann}. Notice that in the third 
column intersections occur only between a continuous and a dashed line,
never between two continuous or two dashed lines.
The final configuration in the splitting 
process is represented by two membranes, each contained entirely in one 
copy of $\R^3$, with the superposition resulting in two concentric surfaces.

As already pointed out, a feature of the solution is that during 
the flipping process the equipotential surfaces develop cusps. These 
special points occur as the equipotential surfaces intersect the branch loop, 
as shown at step ($iv$) in figure~\ref{RPQualitativeSplitRiemann}. From the 
explicit solutions discussed in the next subsection one can verify that the 
potential itself is regular 
over the entire Riemann space (except for the origin 
of the first $\R^3$ where the point charge is located), including the branch 
loop where the cusps arise~\footnote{This is true in general and not only in 
the cases with axial symmetry for which we present plots.}. At these points, 
however, the gradient of the potential diverges.  This allows us to physically 
characterise the branch loop in the Riemann space in terms of the behaviour 
of the solution to the continuum Nahm equations. Recalling that the potential, 
$s=\phi(\bm{z})$, is obtained inverting the functions $z^i(s)$, we deduce that 
the branch loop, where $\bm{\nabla}\phi(\bm{z})$ diverges, corresponds to 
points where the velocity of the front of the membrane, as described by 
$z^i(s)$ in Euclidean time, vanishes. 

Various other properties of the instanton solutions corresponding to 
membrane splitting processes and their moduli spaces can be given an 
intuitive interpretation using a description in terms of Riemann spaces. We 
will discuss some of these aspects in section~\ref{RSSModuliSpaces}.

\subsubsection{Analytic solution}
\label{RSSAnalyticSolution}

In section~\ref{RSApprox} we have shown that the problem of solving the 
continuous version of the instanton equations can be mapped to that of 
finding solutions to the three-dimensional Laplace equation. Then in the 
previous subsection we have discussed the boundary conditions that we 
propose to consider in order to obtain a potential with the required properties.
We now present analytic examples showing explicitly that solutions obeying 
such boundary conditions can be constructed. 

Based on the examples of static solutions presented in 
section~\ref{RSSSolutionCoulomb}, we expect the potential for large $s$ 
(corresponding to $t\to -\infty$ in the original Euclidean time variable) to have 
a Coulomb-like singularity. The considerations in the previous subsection 
about the expected qualitative behaviour of the potential then lead us to 
consider the Laplace equation in a Riemann space made of two copies of 
$\R^3$, with boundary conditions associated with the presence 
of a single positive point charge at the origin of the first $\R^3$. 

The idea of studying the Laplace equation in a Riemann space has actually
been considered long ago by Sommerfeld in a 1896 
paper~\cite{RBSommerfeld}, which introduces the idea of Riemann spaces
to develop a generalisation of the standard method of images for the 
solution of electrostatics problems. The examples studied 
in~\cite{RBSommerfeld} involve multiple copies of $\R^3$ with branch 
curves consisting of straight lines. Following Sommerfeld's proposal, 
an analytic solution to the Laplace equation with boundary conditions 
precisely of the type we are interested in was constructed by Hobson 
in~\cite{RBHobson}. This paper considers a Riemann space consisting of two 
copies of $\R^3$ connected by a flat circular branch disk and computes
the potential generated by a point charge located in the first $\R^3$. 
More recently, Riemann spaces have been considered 
in~\cite{RBHeise,RBdMeSKKSF}.
It can be shown that Sommerfeld's solution with a straight
branch line and Hobson's solution with a circular branch loop
can be related by an inversion transformation, see 
appendix~\ref{RSASommerfeldHobson}. 

\begin{figure}[htb]
\centering
\includegraphics[width=\textwidth]{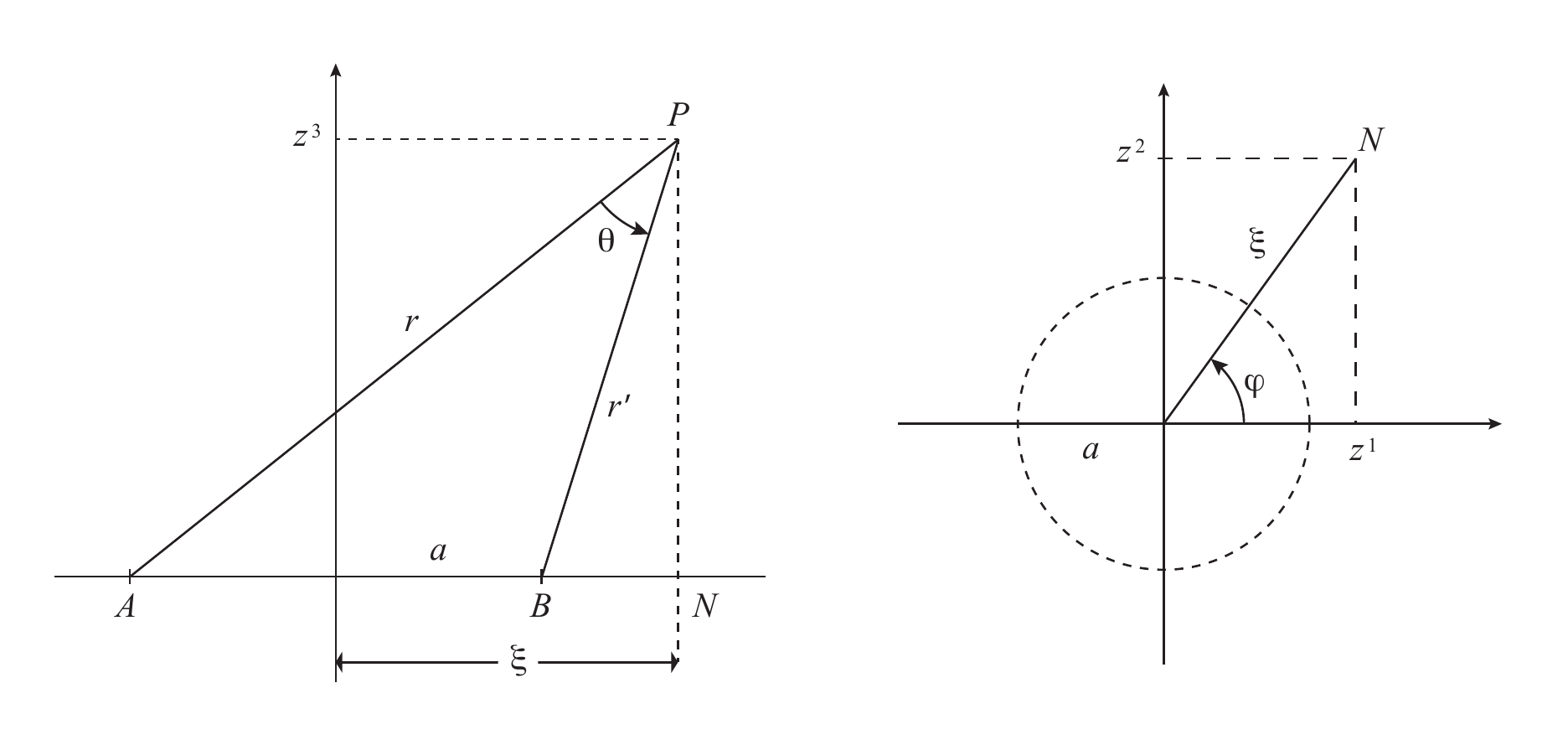}
\caption{Relation between peripolar and Cartesian coordinates}
\label{RPperipolar-cartesian}
\end{figure}

Hobson's solution is written in terms of so-called peripolar coordinates, which
are particularly suited to the specific geometry under consideration. The 
peripolar coordinates of a point $P$ are designated by $(\rho,\theta,\v)$. 
Their definition in $\R^3$ and their relation to the Cartesian coordinates, 
$(z^1,z^2,z^3)$, can be given as follows. One starts with a circle of radius 
$a$, which will later be identified with the branch loop and we conventionally 
take to lie in the $(z^1,z^2)$ plane. A plane containing the point $P$ and the 
axis of the circle, which we identify with the $z^3$ axis, intersects the circle at 
two diametrically opposite points, $A$ and $B$. Denoting the distances of 
$P$ from $A$ and $B$ respectively by $r$ and $r'$, the coordinate $\rho$ 
of $P$ is defined as 
\begin{equation}
\rho = \log\frac{r}{r'} \, ,
\label{RFperipolarrho}
\end{equation}
while $\theta$ is the $\widehat{APB}$ angle. The angle $\v$ is the 
standard polar angle for the projection, $N$, of $P$ onto the $(z^1,z^2)$ 
plane. Figure~\ref{RPperipolar-cartesian} illustrates the definition of $\rho$,
$\theta$ and $\v$ and their relation to the Cartesian coordinates. 

The angle $\v$ is taken to vary between $0$ and $2\pi$. From the definition
$\rho\in(-\infty,+\infty)$ and $\theta$ is defined to be in the interval 
$[-\pi,+\pi]$. The angle $\theta$ goes to $0$ when the distance of $P$ from 
the origin goes to $\infty$ and in the region of the $(z^1,z^2)$ plane outside 
the circle. It has a discontinuity as one passes through the interior of the disk
bounded by the circle of radius $a$. It approaches $+\pi$ or $-\pi$ if one 
approaches a point inside the disk from above ($z^3>0$) or below ($z^3<0$) 
respectively. Points on the $z^3$ axis correspond to $\rho=0$ and points on 
the circle of radius $a$ in the $(z^1,z^2)$ plane have $\rho=\pm\infty$.

Denoting by $\xi$ the distance of the point $N$ from the origin, the peripolar 
and Cartesian coordinates are related by
\begin{equation}
z^1= \xi \cos\v \, , \quad z^2 = \xi \sin\v \, , \quad 
z^3 = \frac{a\sin\theta}{\cosh\rho - \cos\theta} \, , 
\label{RFperipolar-cartesian}
\end{equation}
where
\begin{equation}
\xi = \frac{a\sinh\rho}{\cosh\rho-\cos\theta} \, .
\label{RFxiperipolar}
\end{equation}
To describe (in peripolar coordinates) a Riemann space consisting of two 
copies of $\R^3$ connected by a branch disk coinciding with the disk 
bounded by the circle of radius $a$ in the $(z^1,z^2)$ plane, one 
simply allows the range of $\theta$ to extend from $-\pi$ to $3\pi$. The 
intervals  $[-\pi,\pi]$ and $[\pi,3\pi]$ correspond to the first and second $\R^3$ 
respectively. Moving in the first $\R^3$ from infinity towards the origin along 
the positive $z^3$ axis (or any other curve in the $z^3>0$ region) $\theta$ is 
positive and increases from $0$ to $\pi$. Crossing the disk at $z^3=0$, one 
crosses into the second $\R^3$, while $\theta$ varies continuously. Moving 
along the negative $z^3$ away from the origin in the second space, $\theta$ 
continues to increase from $\pi$, reaching $2\pi$ as $z^3\to-\infty$. 
Approaching the origin along the positive $z^3$ axis in the second space the 
angle $\theta$ varies from $2\pi$ to $3\pi$, which is reached on the upper 
side of the disk. Passing through the disk one crosses back into the first 
space (with $z^3<0$) and $\theta$ goes back to $-\pi$. 

The Laplacian in peripolar coordinates takes the form
\begin{align}
\bm{\nabla}^2 = & \left( \frac{\cosh \rho - \cos \theta}{a^2} \right) 
\left[(\cosh \rho - \cos \theta) \frac{\del^2}{\del\rho^2} 
+ \frac{(1-\cos \theta \cosh \rho )}{\sinh\rho} \frac{\del}{\del\rho} 
\right. \nn \\
& \left.+(\cosh\rho - \cos\theta) \frac{\del^2}{\del \theta^2} 
- \sin\theta \frac{\del}{\del\theta} 
+ \frac{(\cosh\rho -\cos \theta)}{\sinh^2 \rho} 
\frac{\del^2}{\del\varphi^2} \right] \, .
\label{RFperipolarlaplacian}
\end{align}
In~\cite{RBHobson} Hobson computed the potential solving the Laplace 
equation in the two-sheeted Riemann space with a circular branch loop for
an arbitrary relative position of the point charge in the first space relative 
to the branch disk. Denoting by $\bm{z}_0$ the location of the point charge, 
with peripolar coordinates $(\rho_0,\theta_0,\v_0)$, the potential at a 
generic point $\bm{z}$ of coordinates $(\rho,\theta,\v)$ is 
\begin{equation}
\phi(\bm{z},\bm{z}_0) = \frac{J}{4\pi|\bm{z}-\bm{z}_0|}  
\left[ \frac{1}{2} + \frac{1}{\pi} \arcsin\!\left( 
\cos \left(\frac{\theta - \theta_0}{2}\right) 
\sqrt{\frac{2}{\cosh \alpha +1}}  \right) \right] \, ,
\label{RFHobsonspotential}
\end{equation}
where 
\begin{equation}
\cosh\alpha = \cosh \rho \cosh \rho_0 - 
\cos (\varphi-\varphi_0) \sinh \rho \sinh \rho_0 
\label{RFHobsonscoshalpha}
\end{equation}
and
\begin{equation}
|\bm{z}-\bm{z}_0| = a \sqrt{2} \sqrt{\frac{ \cosh \alpha - 
\cos (\theta - \theta_0)}{(\cosh \rho - \cos \theta)( \cosh \rho_0 
- \cos \theta_0)}} \, .
\label{RFHobsonsdistance}
\end{equation}
Notice that the potential~(\ref{RFHobsonspotential}) is periodic in $\theta$ 
with period $4\pi$, as appropriate for the two-sheeted Riemann space. 
One can verify that $\phi(\bm{z},\bm{z}_0)$ is well-defined and finite 
everywhere, including the branch loop, except for the location of the
point charge, \ie $\bm{z}=\bm{z}_0$ in the first space, 
where it has a Coulomb-like divergence.

As anticipated in the previous subsection, the flux of 
$\bm{\nabla}\phi$ provides a measure of the angular momentum carried by 
the different membranes, which is also proportional to their respective radius. 
In the case under consideration we have a two-sheeted Riemann space with 
one point charge in the first copy of $\R^3$. We will see that the equipotential 
surfaces for large values of the potential are small spheres centred at the 
location of the point charge. They represent the single membrane with 
angular momentum $J$ at large negative values of the original Euclidean time 
$t$. The flux of $\bm{\nabla}\phi$ through these surfaces is $J$. The 
asymptotic equipotential surfaces for small values of the potential 
approaching zero will be shown to be two (approximately) concentric spheres 
with diverging radii, one in each copy of $\R^3$. The flux of 
$\bm{\nabla}\phi$ through these spheres (coinciding with the flux at infinity in 
the respective $\R^3$) equals the angular momentum of the corresponding 
membrane, $J_i$, $i=1,2$. Conservation of the flux, which corresponds to 
conservation of angular momentum for the membranes, implies 
$J=J_1+J_2$. The way in which the flux $J$ is split between the two spaces  
is controlled by the size of the branch disk and by its position relative to the 
point charge. For a disk of given radius, if the point charge is located very far 
from the disk only a small fraction of the flux passes through the disk into the 
second space and thus we have $J_2\ll J_1$, corresponding to a final state 
with one membrane much larger than the other. If we reduce the distance 
between the point charge and the branch disk, the fraction of flux passing into 
the second space increases. When the distance of the charge from the disk 
tends to zero, we approach the case in which the flux is equally spit between 
the two spaces, which in turn corresponds to having two membranes of equal 
radius in the final state. 

These general considerations can be made more precise
by studying the 
asymptotic behaviour of $\phi(\bm{z},\bm{z}_0)$ at a large distance from both 
the point charge and the branch disk in each of the two spaces. For this 
purpose we analyse $\phi(\bm{z},\bm{z}_0)$ in the region defined by
$\rho \rightarrow 0$, $\th \rightarrow 0$ in the first space and 
$\rho \rightarrow 0$, $\th \rightarrow 2\pi$ in the second space.
In both cases we have $\cosh\a \approx \cosh\rho_0$.
The asymptotic behaviour at infinity in the first space is
\begin{equation}
\phi(\bm{z},\bm{z}_0)\approx\frac{J}{4\pi|\bm{z}-\bm{z}_0|}
\left[ \frac{1}{2} + \frac{1}{\pi} \arcsin\left(
\frac{\cos\left(\frac{\th_0}{2}\right)}
{\cosh\left(\frac{\rho_0}{2}\right)}\right) \right]
\label{RFphi1aympt}
\end{equation}
and in the second space it is
\begin{equation}
\phi(\bm{z},\bm{z}_0)\approx\frac{J}{4\pi|\bm{z}-\bm{z}_0|}
\left[ \frac{1}{2} - \frac{1}{\pi} \arcsin\left(
\frac{\cos\left(\frac{\th_0}{2}\right)}
{\cosh\left(\frac{\rho_0}{2}\right)}\right) \right].
\label{RFphi2aympt}
\end{equation}
The factors in square brackets in~(\ref{RFphi1aympt}) 
and~(\ref{RFphi2aympt}), multiplying the Coulomb potential 
$J/(4\pi|\bm{z}-\bm{z}_0|)$, represent the fractions of the total flux staying 
in the first space or escaping into the second space, respectively.
They determine the angular momenta $J_1$ and $J_2$ of the two 
membranes in the final state as fractions of the total angular momentum $J$.

In the axially symmetric case in which the point charge is located on the axis 
of the branch disk, corresponding to $\rho_0=0$, the asymptotic 
formulae~(\ref{RFphi1aympt}) 
and~(\ref{RFphi2aympt}) simplify and we get
\begin{equation}
\phi(\bm{z},\bm{z}_0)\approx\frac{J}{4\pi|\bm{z}-\bm{z}_0|}
\left(1-\frac{|\th_0|}{2\pi}\right),
\label{RFphi1aymptaxial}
\end{equation}
in the first space and 
\begin{equation}
\phi(\bm{z},\bm{z}_0)\approx\frac{J}{4\pi|\bm{z}-\bm{z}_0|} 
\frac{|\th_0|}{2\pi},
\label{RFphi2aymptaxial}
\end{equation}
in the second space. Notice that for $\th_0 \to \pm \pi$, \ie when the location 
of the point charge approaches the disk, the ratio goes to $1/2$ as it should 
be. 

In the following we set $a=1$ and we present plots of the equipotential 
surfaces of the potential~(\ref{RFHobsonspotential}) in the axially symmetric 
case, $\rho_0=0$. We can also set $\v_0=0$ without loss of generality, so
that the only remaining parameter in the solution is the angle 
$\theta_0$, whose value controls the distance of the branch loop from 
the point charge~\footnote{We note that solutions with different $\theta_0$ can be
related by a conformal transformation.}.
As will be shown in the explicit examples below, the value of 
$\theta_0$ also affects the order in which the splitting and flipping of the 
membrane orientation take place. Therefore the sequence in which these 
steps occur is correlated with the way the angular momentum gets divided 
between the two membranes in the final state.

Using the relations~(\ref{RFperipolar-cartesian})-(\ref{RFxiperipolar})
in~(\ref{RFHobsonspotential}), 
one can obtain the form of the potential in 
Cartesian coordinates, $\phi=\phi(z^1,z^2,z^3)$, which is the expression 
used to produce the plots presented below. In all the following 
figures we use a rescaled potential, $\hat\phi$, related 
to~(\ref{RFHobsonspotential}) by $\hat\phi=(4\pi/J)\phi$. 

\begin{figure}[htb]
\centering
\includegraphics[width=0.48\textwidth]{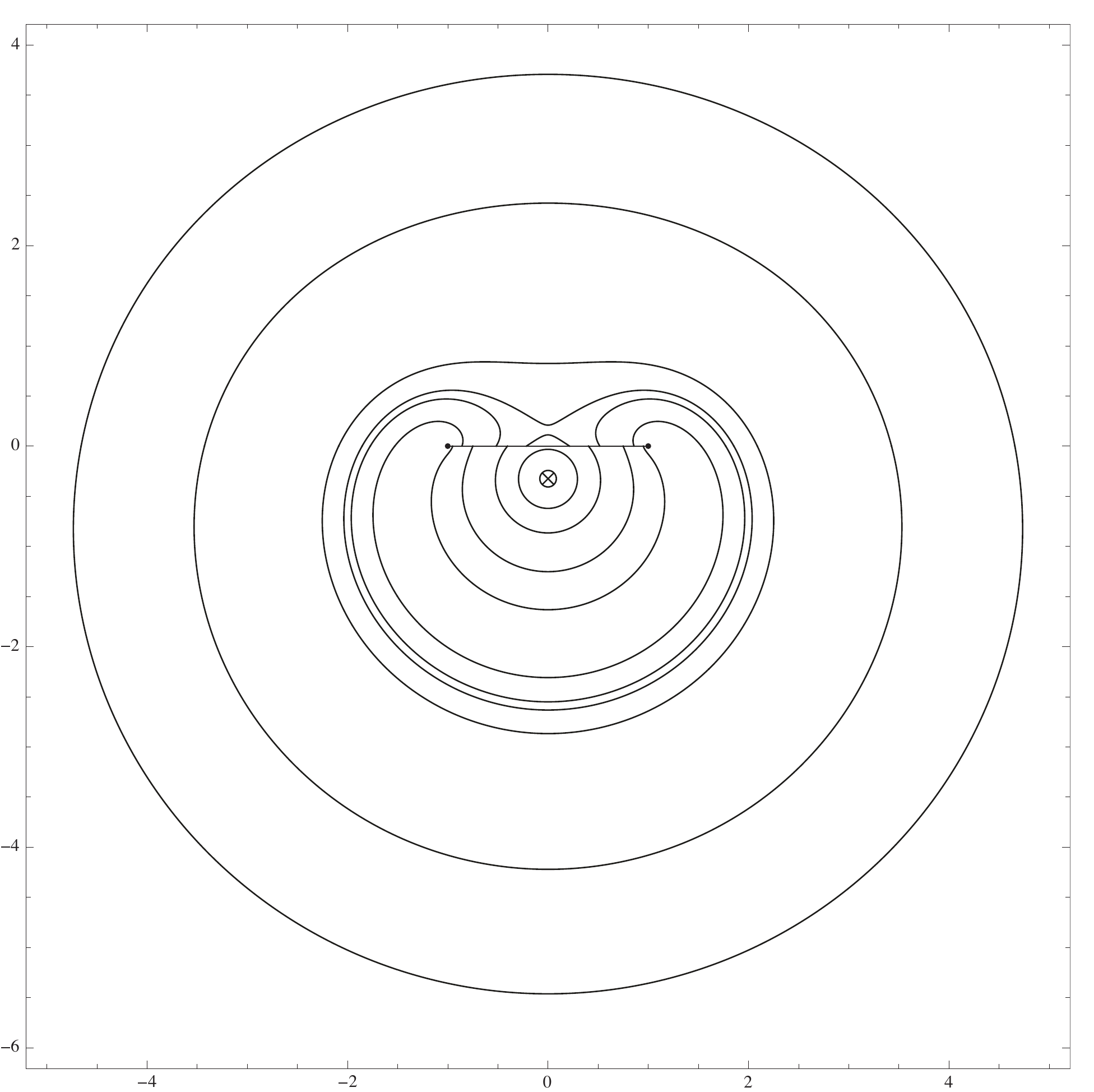} 
\hspace*{0.2cm} 
\includegraphics[width=0.48\textwidth]{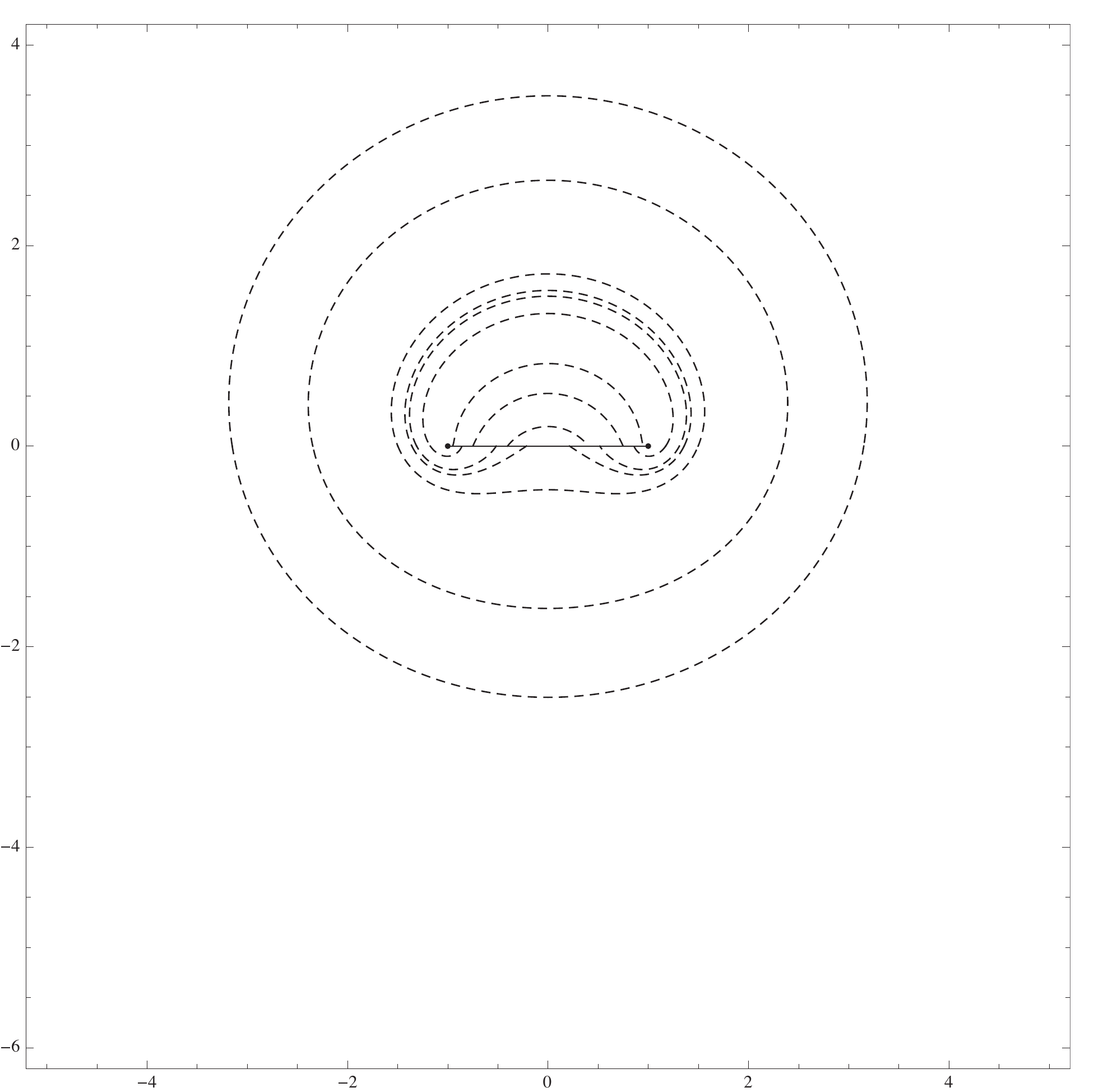}
\caption{Contour plots showing families of equipotential surfaces in each of 
the two copies of $\R^3$ which constitute the Riemann space. The contours
shown are for $\theta_0=-\frac{4\pi}{5}$ and 
correspond to unequally spaced values of $\hat{\phi}$. 
The values of the potential used here are $\hat{\phi}=
\{3.113,1.613,0.878,0.593,0.368,0.323,0.310,0.278,0.173,0.128\}$ and 
include those used in figure~\ref{RPmembranesplittingb}.}
\label{RPseparatecontourplots}
\end{figure}

In figure~\ref{RPseparatecontourplots} we show contour plots for the 
potential~(\ref{RFHobsonspotential}) with $\theta_0=-4\pi/5$. The figure
shows families of equipotential surfaces separately in the two copies of 
$\R^3$. Each $\R^3$ is represented by a square, with the vertical direction 
being the direction of the $z^3$ axis. The branch disk is indicated by a 
horizontal slit, the point charge (contained in the first $\R^3$, which is on
the left) is denoted by a small crossed circle below the disk. The contours 
are displayed as continuous lines in the first space and as dashed lines in the 
second space.

The following figures depict the evolution of the profile of the membranes
throughout the splitting process for different values of $\theta_0$. They 
show how the equipotential surfaces for the potential in the analytic 
solution~(\ref{RFHobsonspotential}) reproduce the steps that were 
qualitatively discussed in the previous subsection. In these 
figures the two copies of $\R^3$ are superposed. The (portions of) 
membranes living in the first or second space are depicted as continuous or 
dashed lines respectively. The figures display axially symmetric solutions,
therefore the three-dimensional shape of the membranes can be generated 
rotating the contours about the vertical axis. 

It is interesting to notice that the equipotential surfaces deviate significantly
from a spherical shape only for a rather small range of the 
potential around the value where the splitting takes place. This is the region
in which the equipotential surfaces cross the branch disk. The surfaces then 
quickly revert to Coulomb-like behaviour outside this region for both larger 
and smaller values of $\hat\phi$. 

\begin{figure}[htbp]
\centering
\includegraphics[height=0.312\textheight]{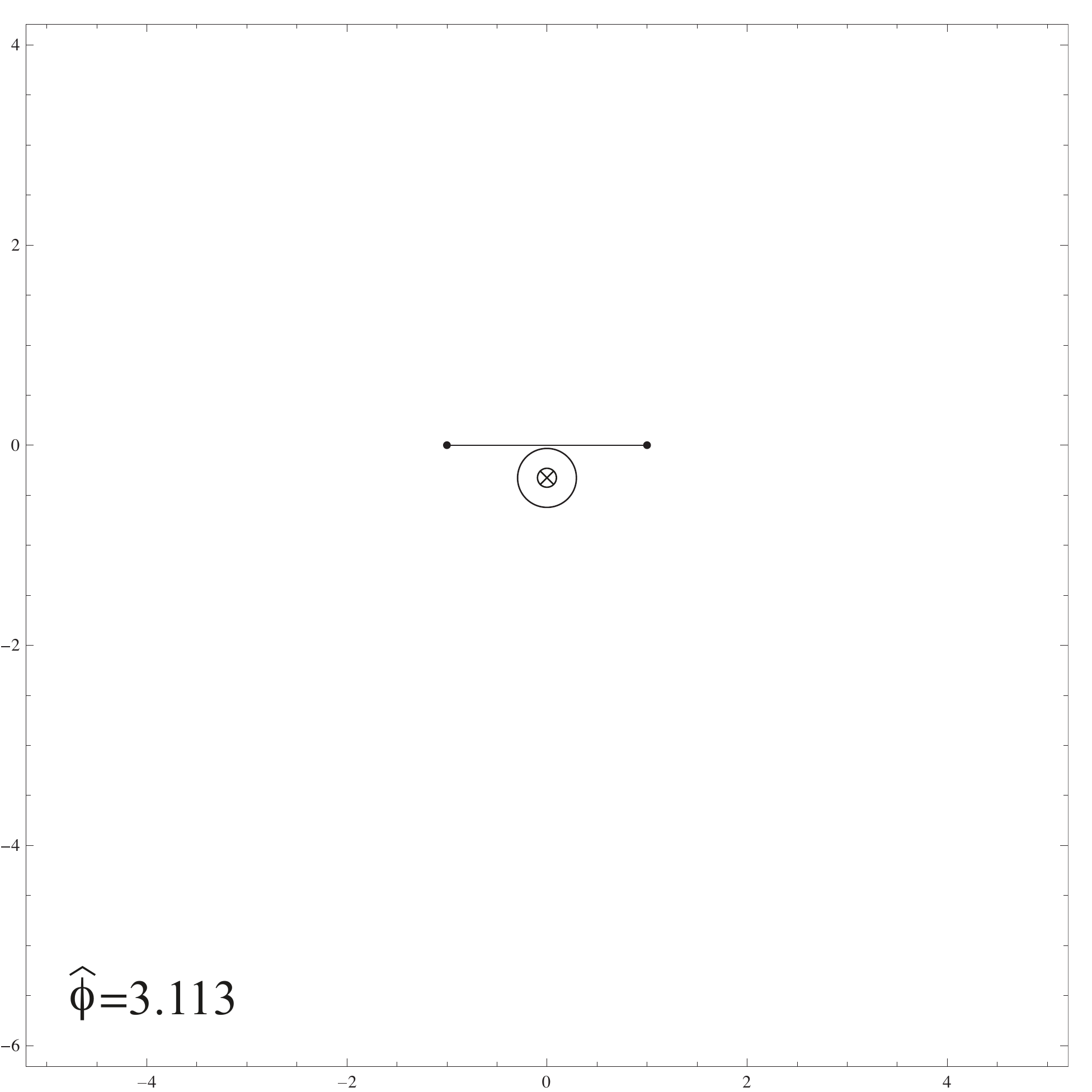} \hspace*{0.2cm}
\includegraphics[height=0.312\textheight]{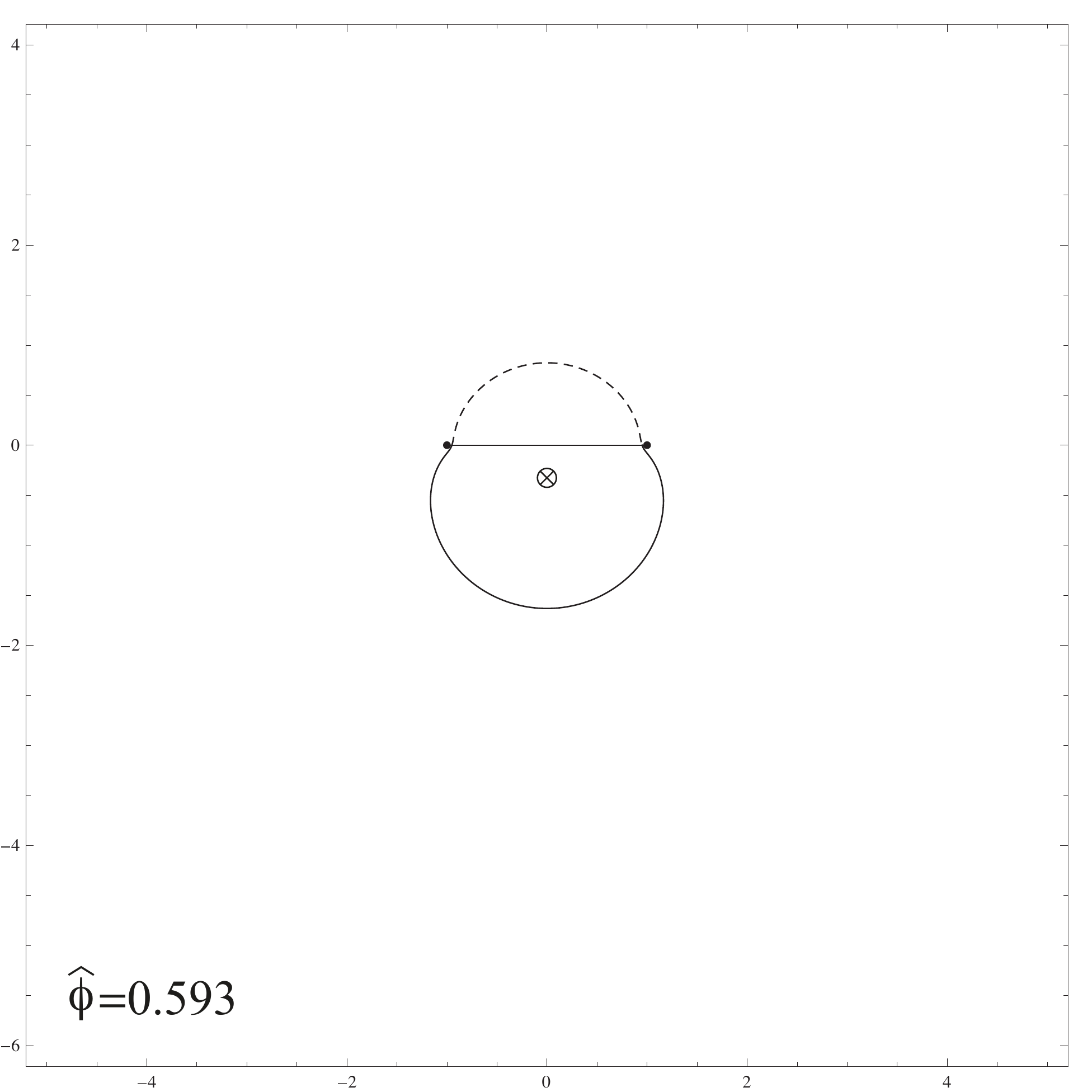} \\
\includegraphics[height=0.312\textheight]{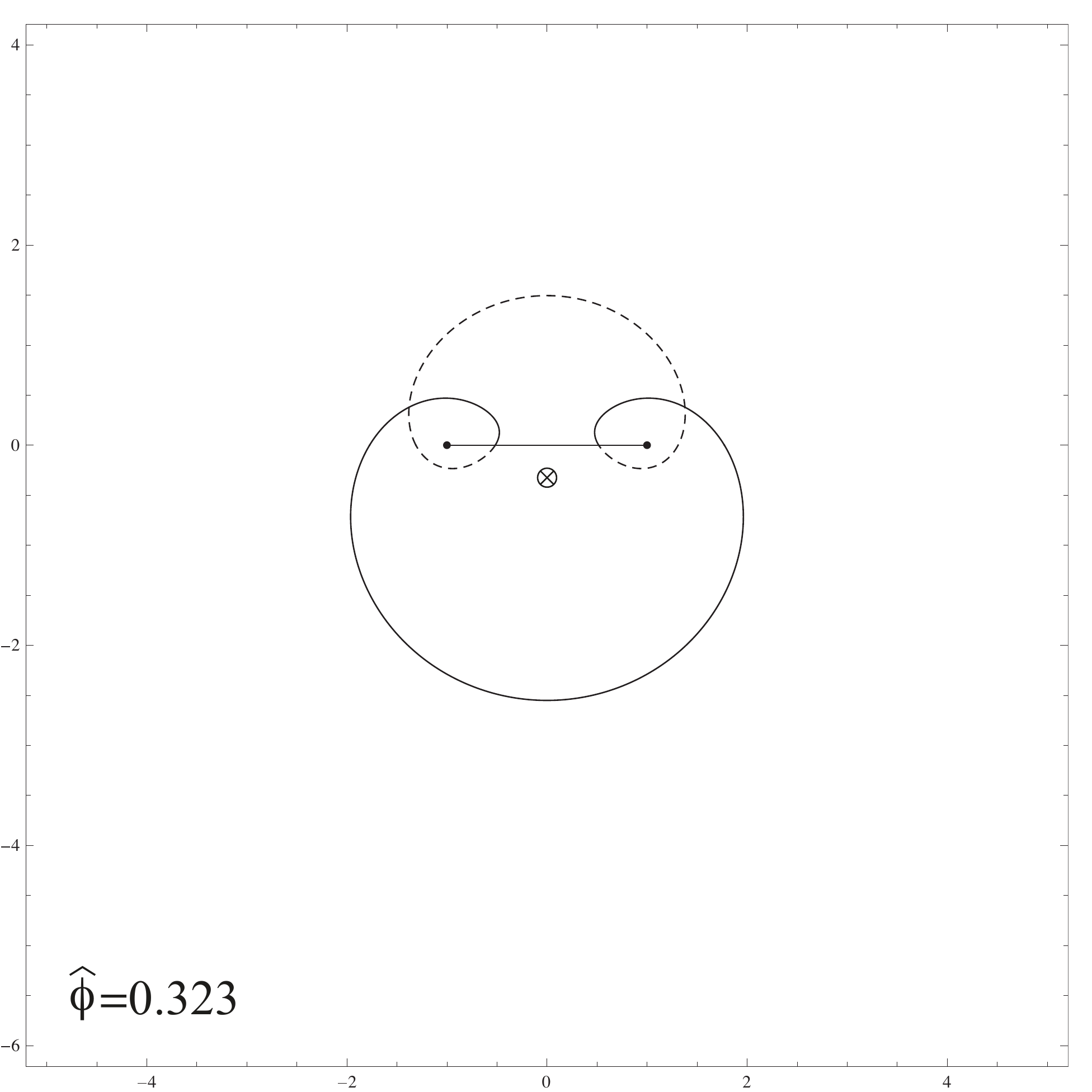} \hspace*{0.2cm}
\includegraphics[height=0.312\textheight]{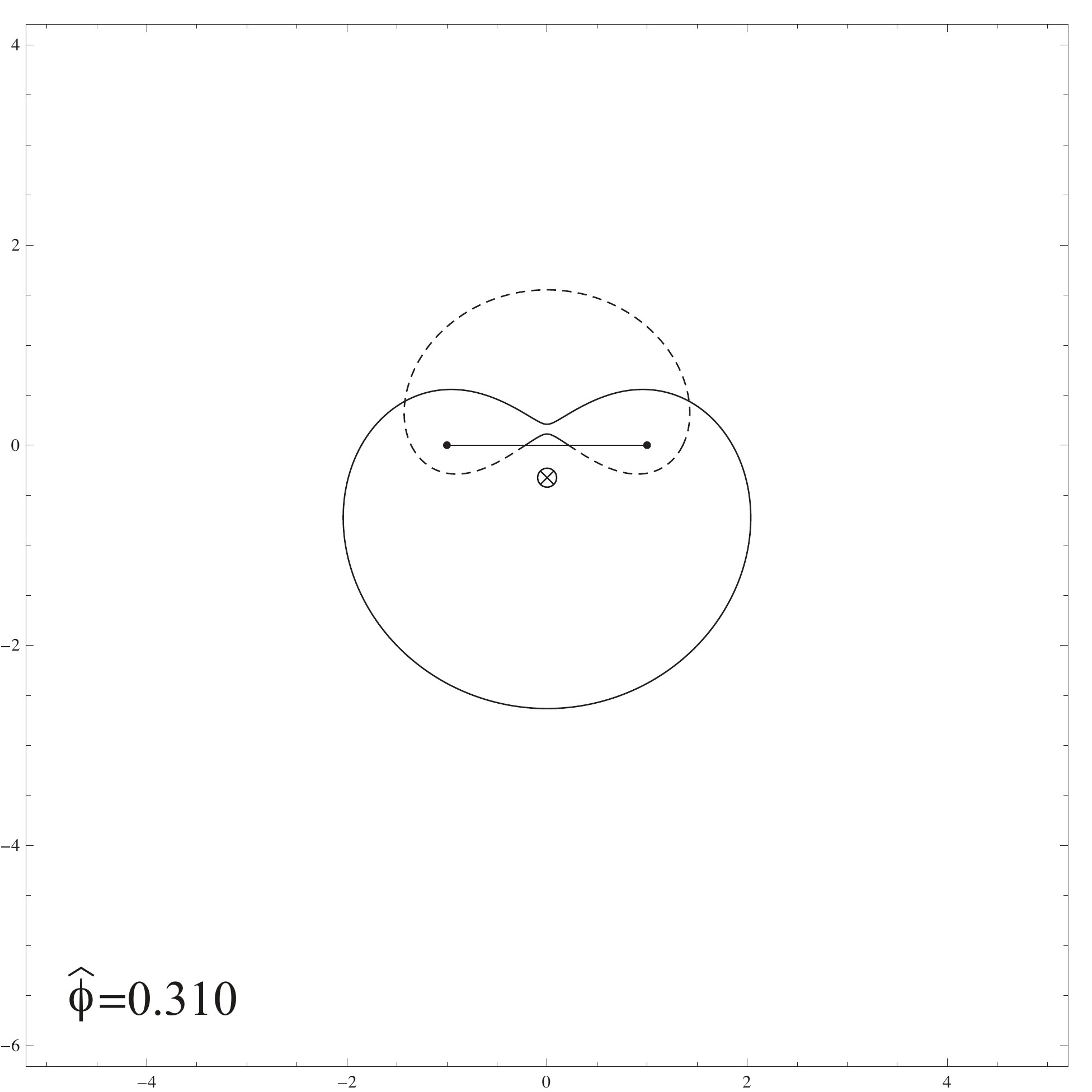} \\
\includegraphics[height=0.312\textheight]{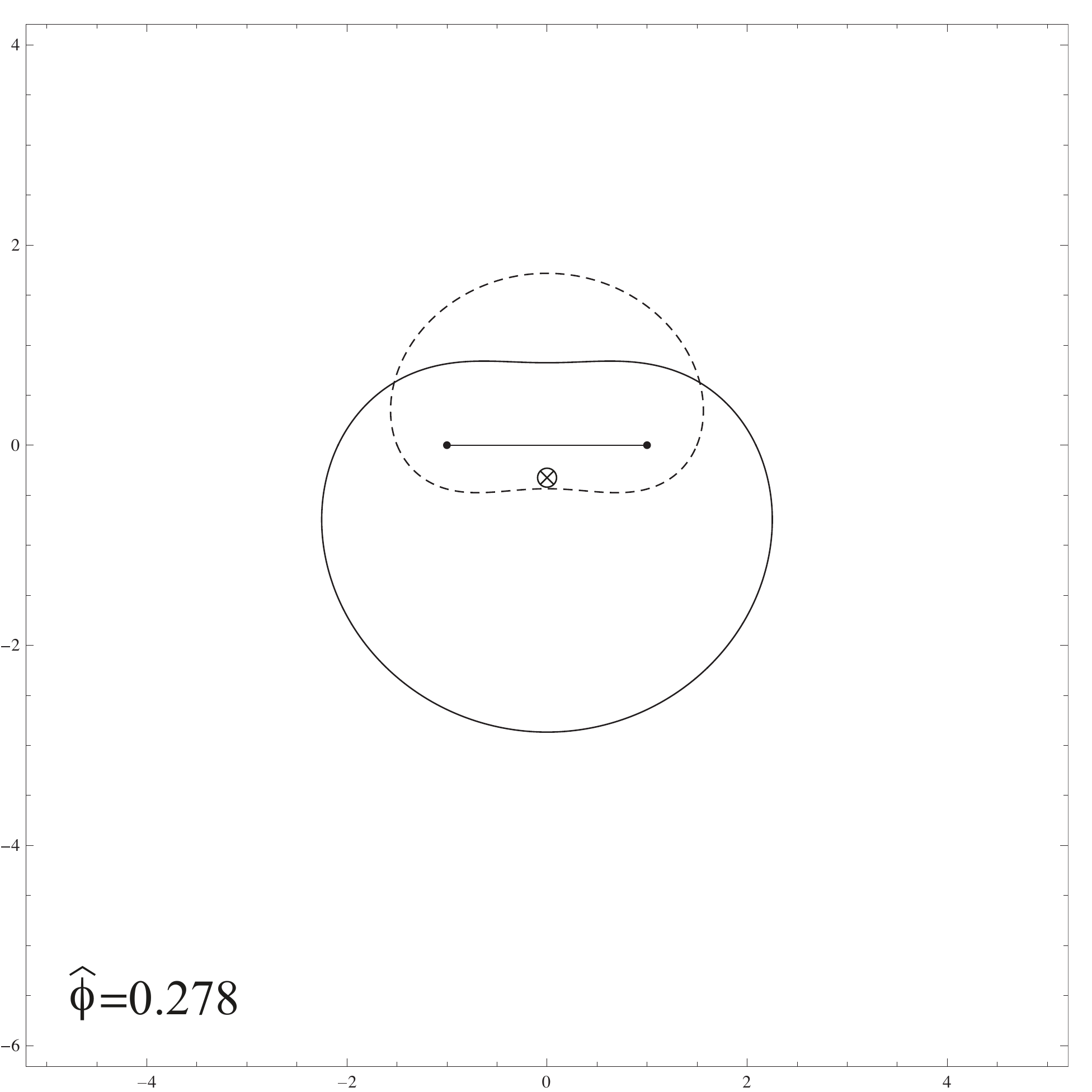} \hspace*{0.2cm}
\includegraphics[height=0.312\textheight]{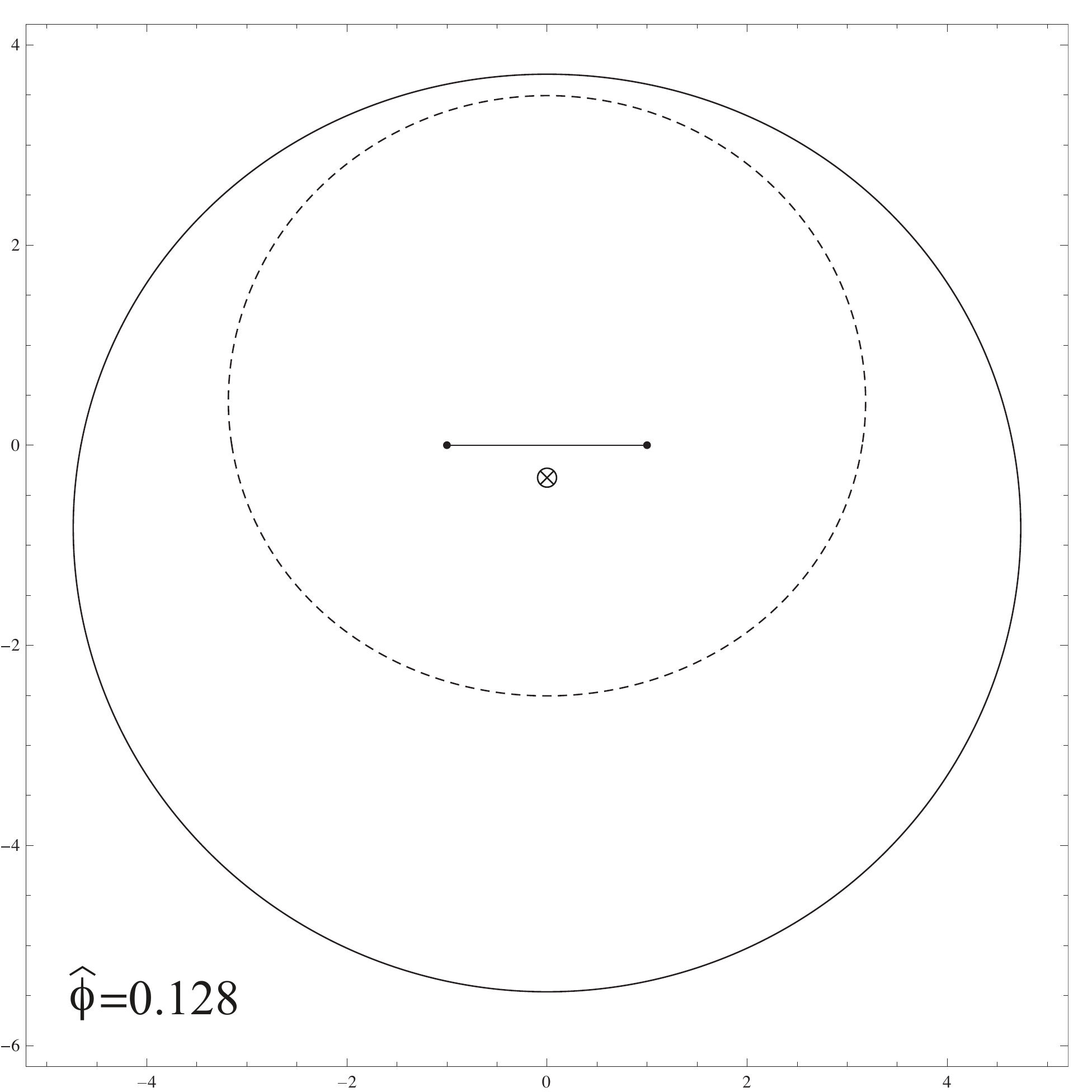} \\
\caption{Evolution of the membrane profiles, showing the different phases of  
the splitting process in Euclidean time in the case $\theta_0=-4\pi/5$}
\label{RPmembranesplittingb}
\end{figure}

Figure~\ref{RPmembranesplittingb} shows the membrane profiles 
corresponding to the choice $\theta_0=-4\pi/5$ 
in~(\ref{RFHobsonspotential}). In this case the point charge is quite close to 
the branch disk and correspondingly the flux/angular momentum gets split 
almost evenly between the two spaces. The sequence of plots shows the 
formation of cusps, which in these examples with axial symmetry occur 
simultaneously at all points on the branch loop. The membrane is then seen 
self-intersecting before it splits into two. As a result, with this choice of 
boundary conditions when the splitting occurs the second membrane has 
already the correct orientation, but the two membranes are still intersecting. 
Notice that, as anticipated in the qualitative description of the previous 
subsection, all the intersections between equipotential surfaces in the figure
(in the third, fourth and fifth panel) involve a dashed and a continuous line, 
\ie they always occur between (portions of) equipotential surfaces belonging 
to different sheets of the Riemann space. The same feature can be observed 
in all the subsequent figures in this and the next subsection.

\begin{figure}[htbp]
\centering
\includegraphics[height=0.312\textheight]{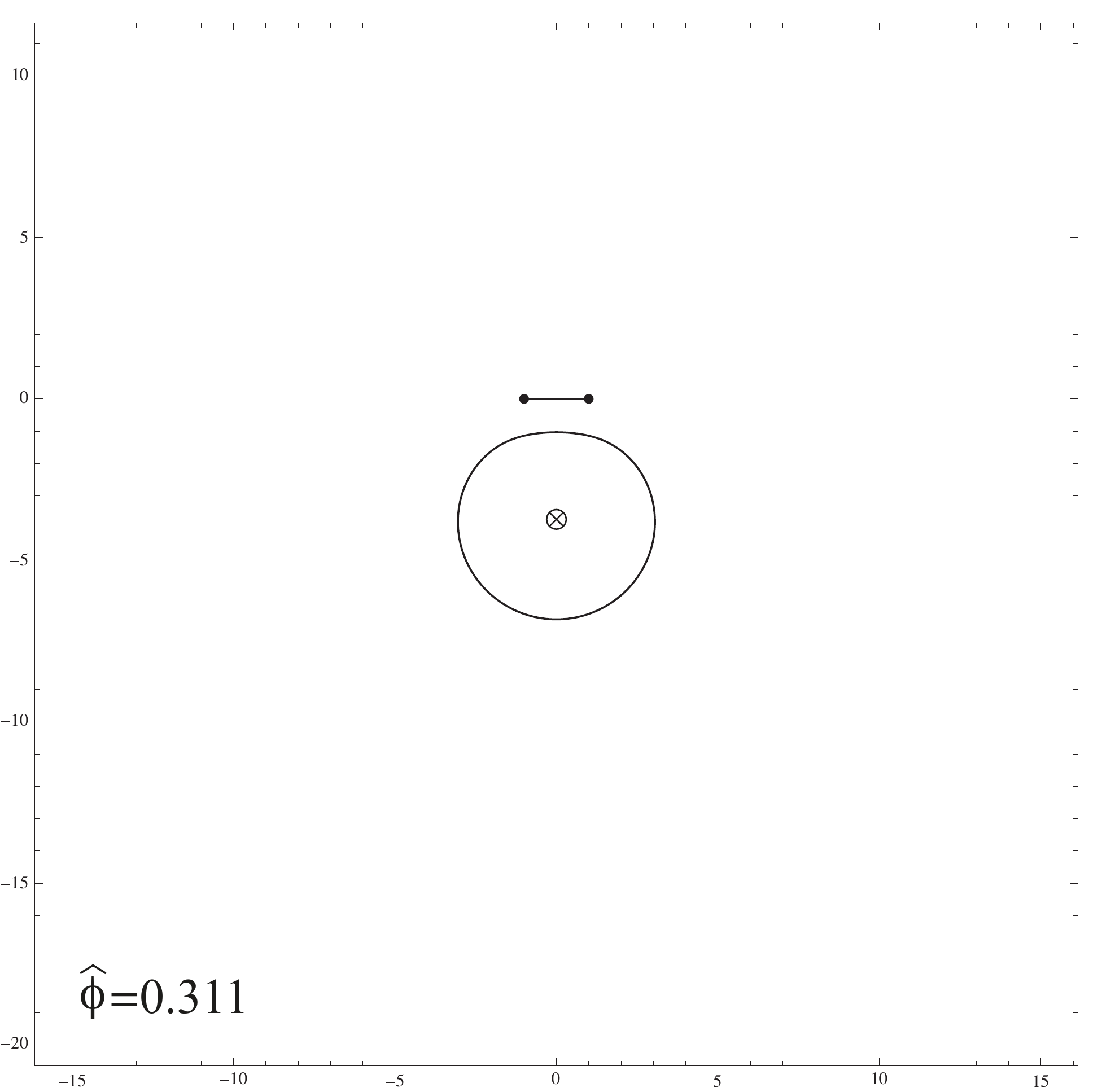} \hspace*{0.2cm}
\includegraphics[height=0.312\textheight]{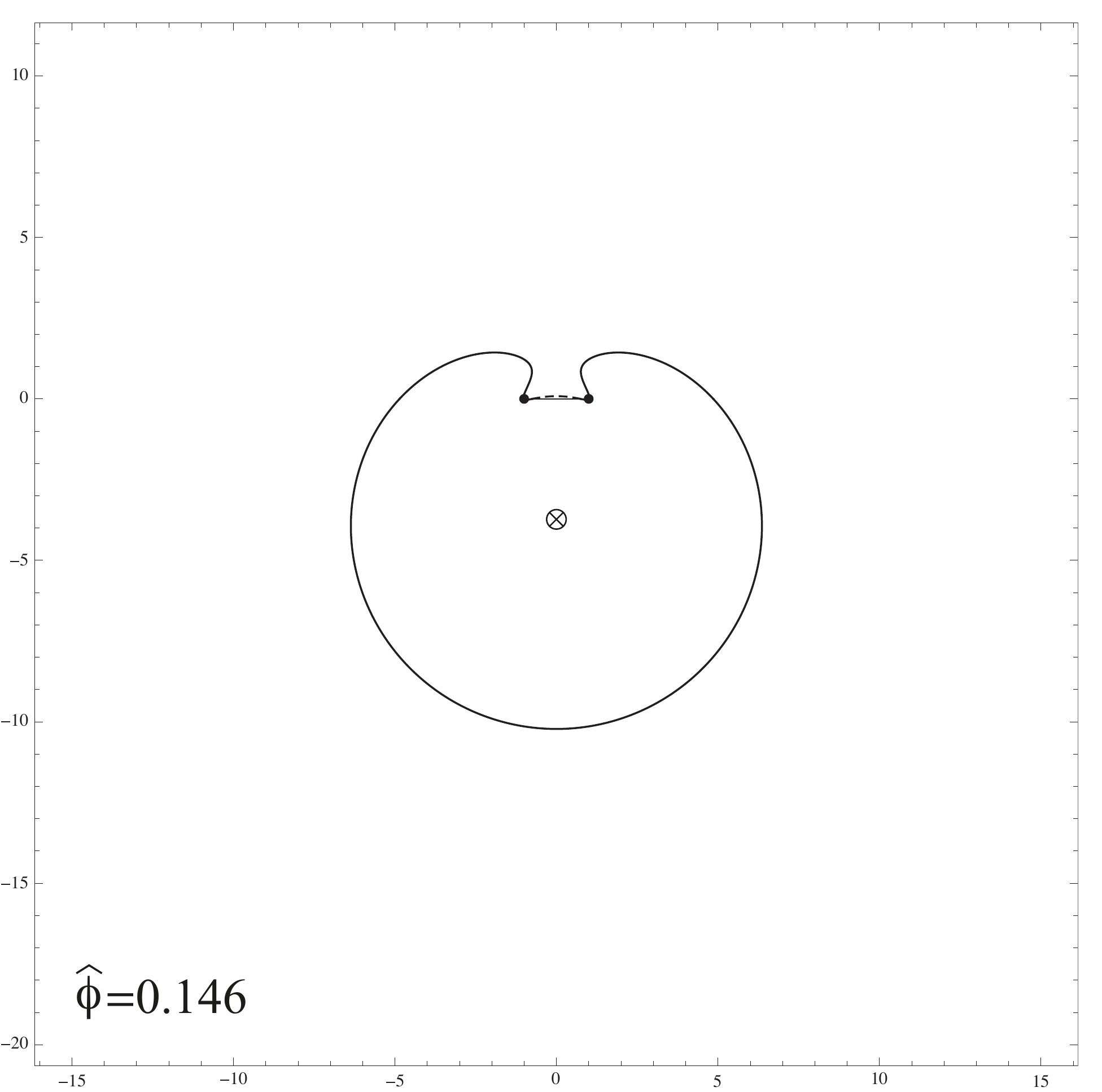} \\
\includegraphics[height=0.312\textheight]{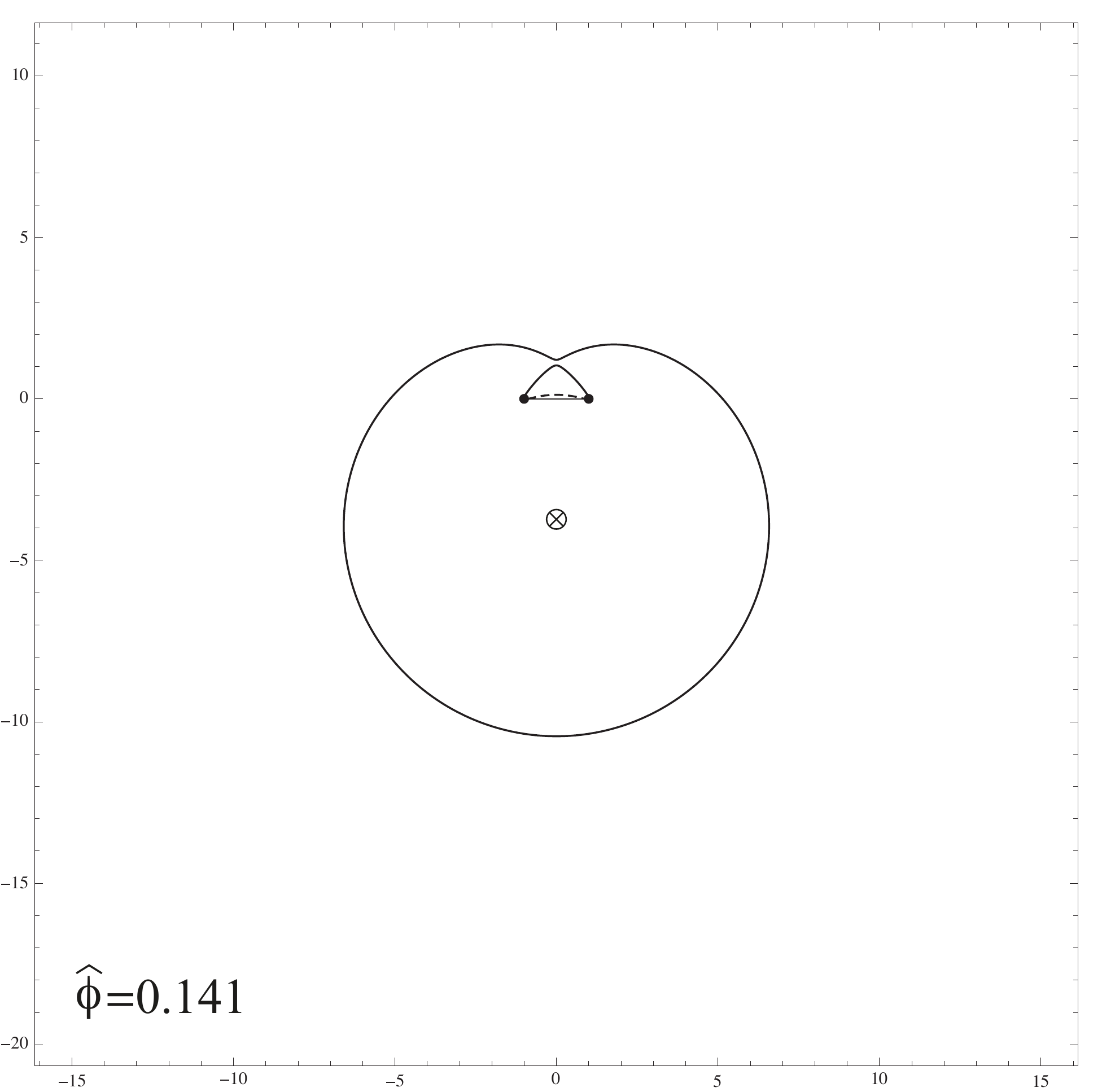} \hspace*{0.2cm}
\includegraphics[height=0.312\textheight]{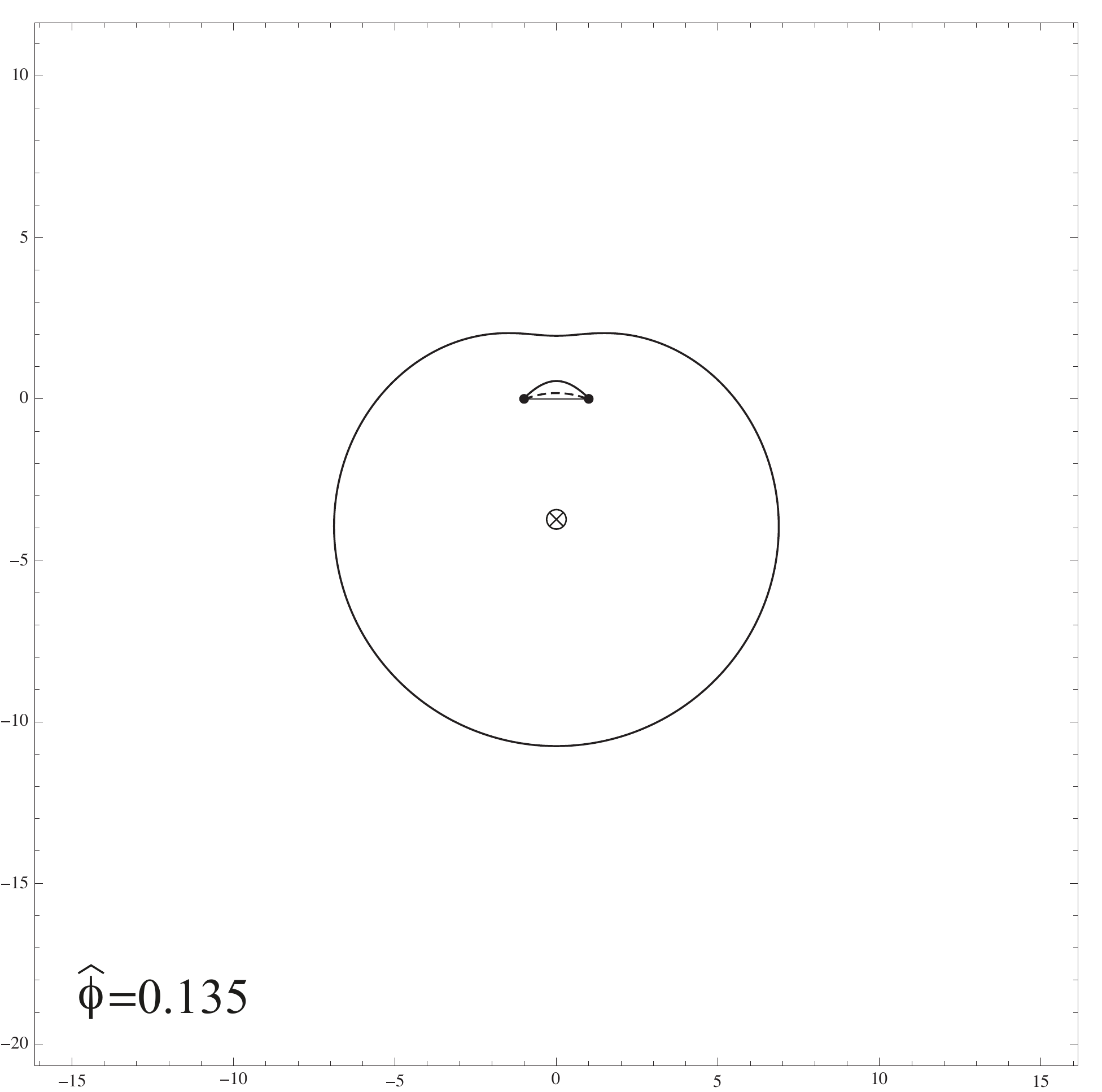} \\
\includegraphics[height=0.312\textheight]{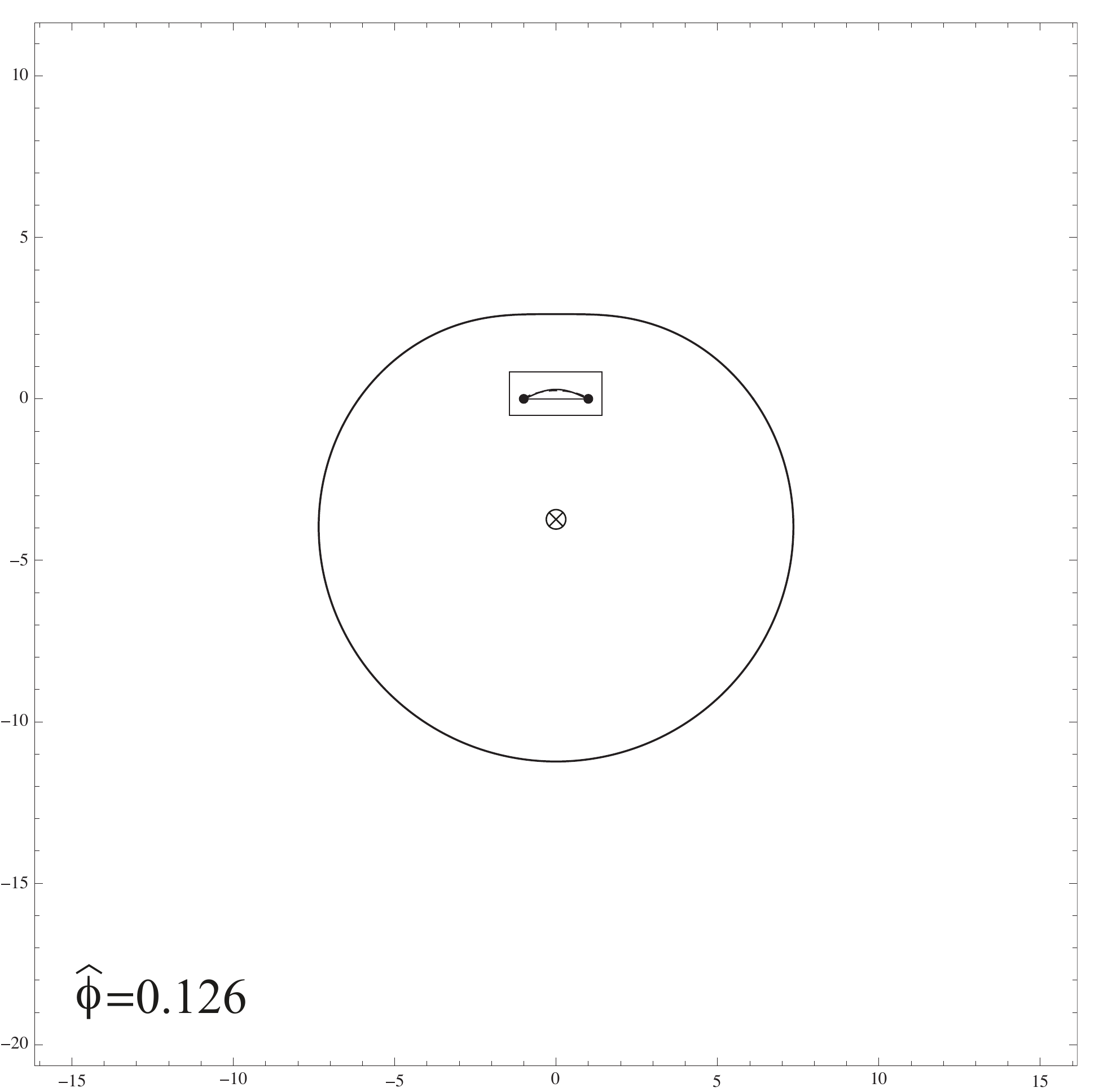} \hspace*{0.2cm}
\includegraphics[height=0.312\textheight]{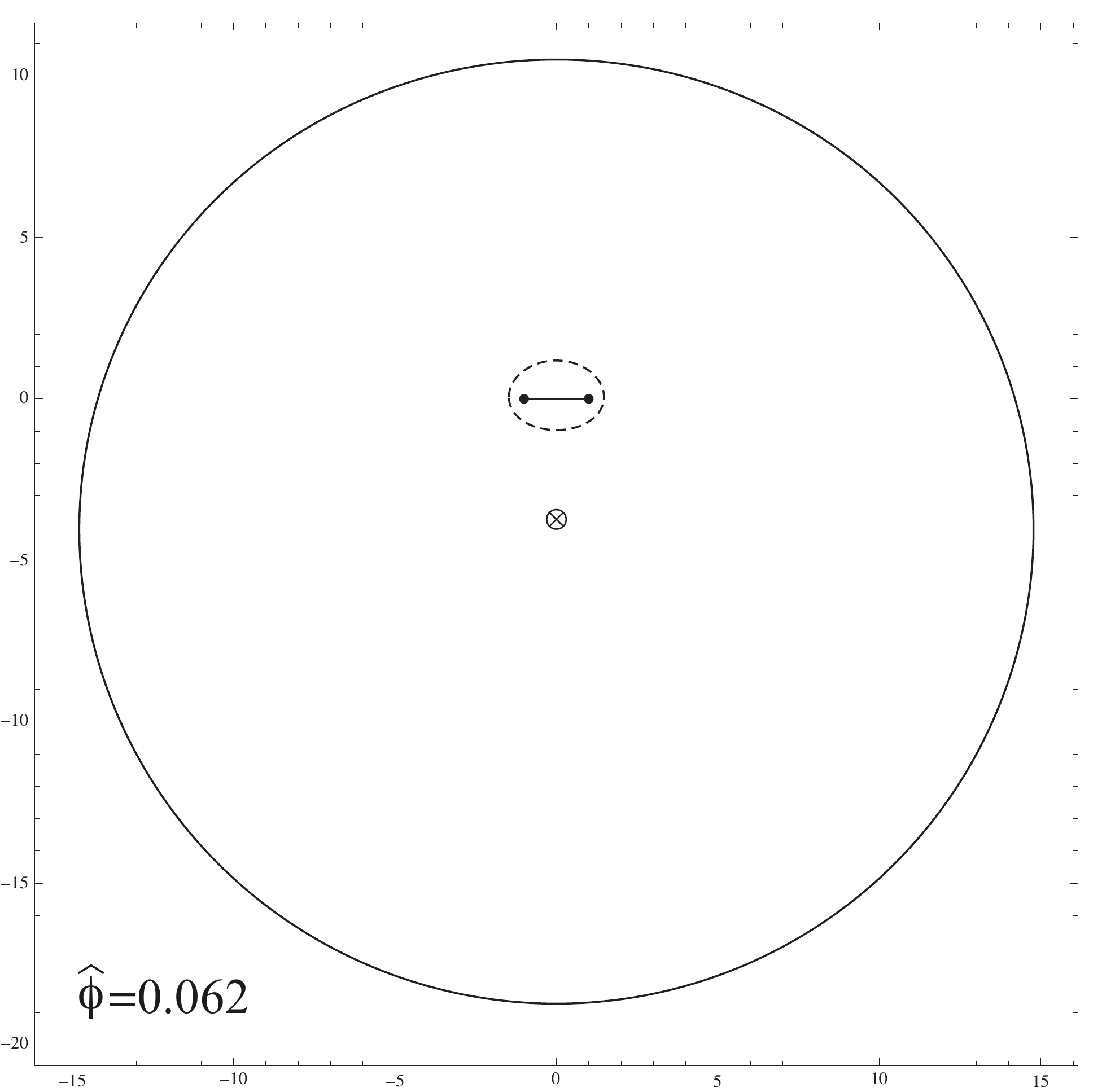} \\
\caption{Evolution of the membrane profiles, showing the different phases of  
the splitting process in Euclidean time in the case $\theta_0=-\pi/6$}
\label{RPmembranesplittingd}
\end{figure}

\begin{figure}[htb]
\centering
\includegraphics[width=0.6\textwidth]{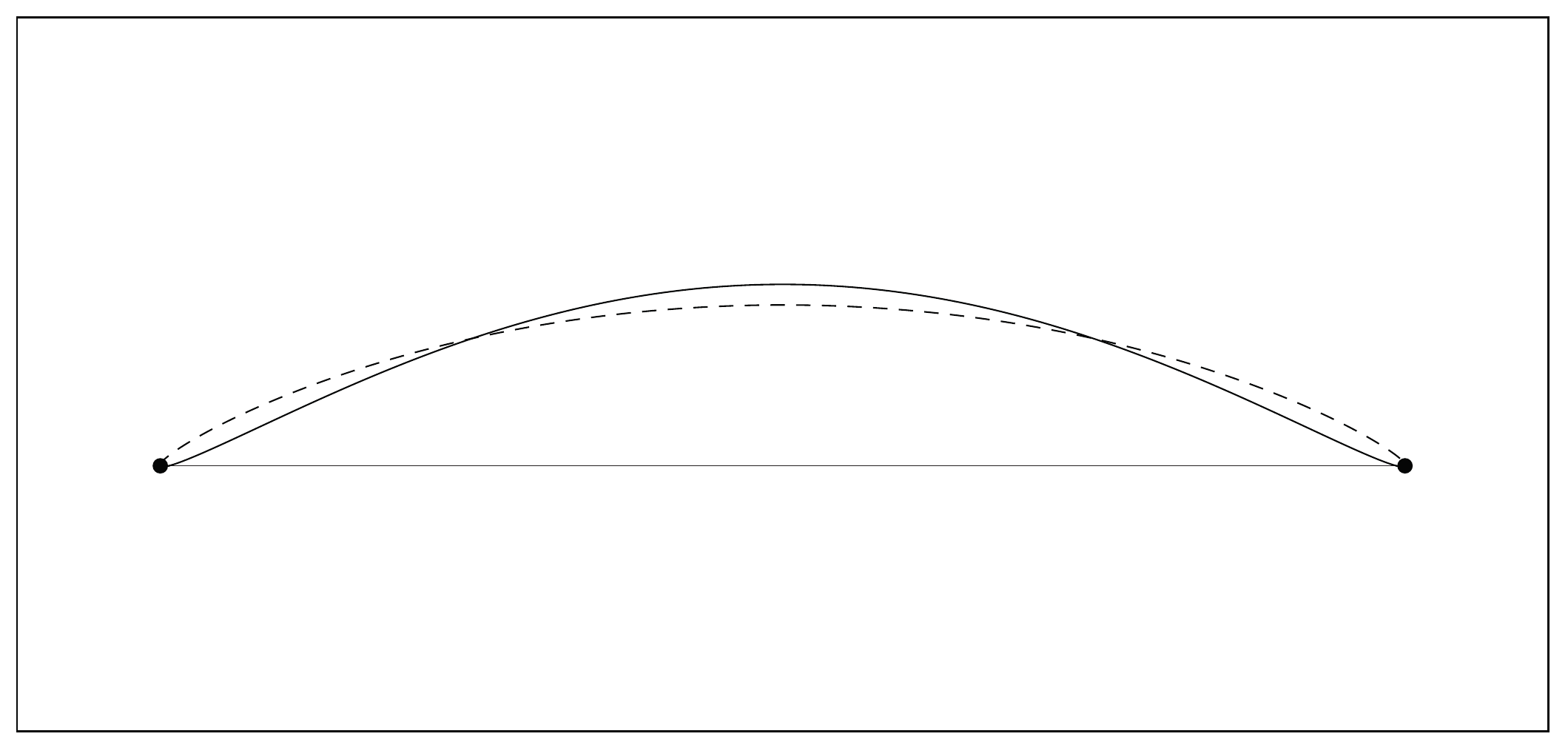}
\caption{Enlargement of the area marked by a rectangle in the fifth panel 
of figure~\ref{RPmembranesplittingd}.}
\label{RPmembranesplittingddetail}
\end{figure}

Figure~\ref{RPmembranesplittingd} shows the evolution of the membrane
profiles for $\theta_0=-\pi/6$, which corresponds to a point charge much 
further from the branch disk. In this case the splitting process follows 
a sequence of steps similar to those depicted qualitatively in 
figures~\ref{RPQualitativeSplit} and~\ref{RPQualitativeSplitRiemann}. 
The initial membrane splits into a membrane and an anti-membrane, which 
subsequently flips its orientation. The cusps are formed after the splitting 
and the subsequent flipping of the orientation of the internal membrane 
requires the latter to self-intersect.   
This takes place in the fifth panel in the figure and is shown more
clearly in figure~\ref{RPmembranesplittingddetail}, which shows an 
enlargement of the rectangular area marked in 
figure~\ref{RPmembranesplittingd}. This example shows a concrete
realisation, in an exact analytic solution, of the crucial step -- illustrated
at stage ($v$) in figures~\ref{RPQualitativeSplit} 
and~\ref{RPQualitativeSplitRiemann} -- which led us to the concept of 
Riemann space. Notice the relative size of the two 
membranes in the final state; in this case (with the point charge further 
away from the disk) the flux is split less evenly than in the previous case 
and correspondingly the second membrane is much smaller.

\begin{figure}[htbp]
\centering
\includegraphics[height=0.312\textheight]{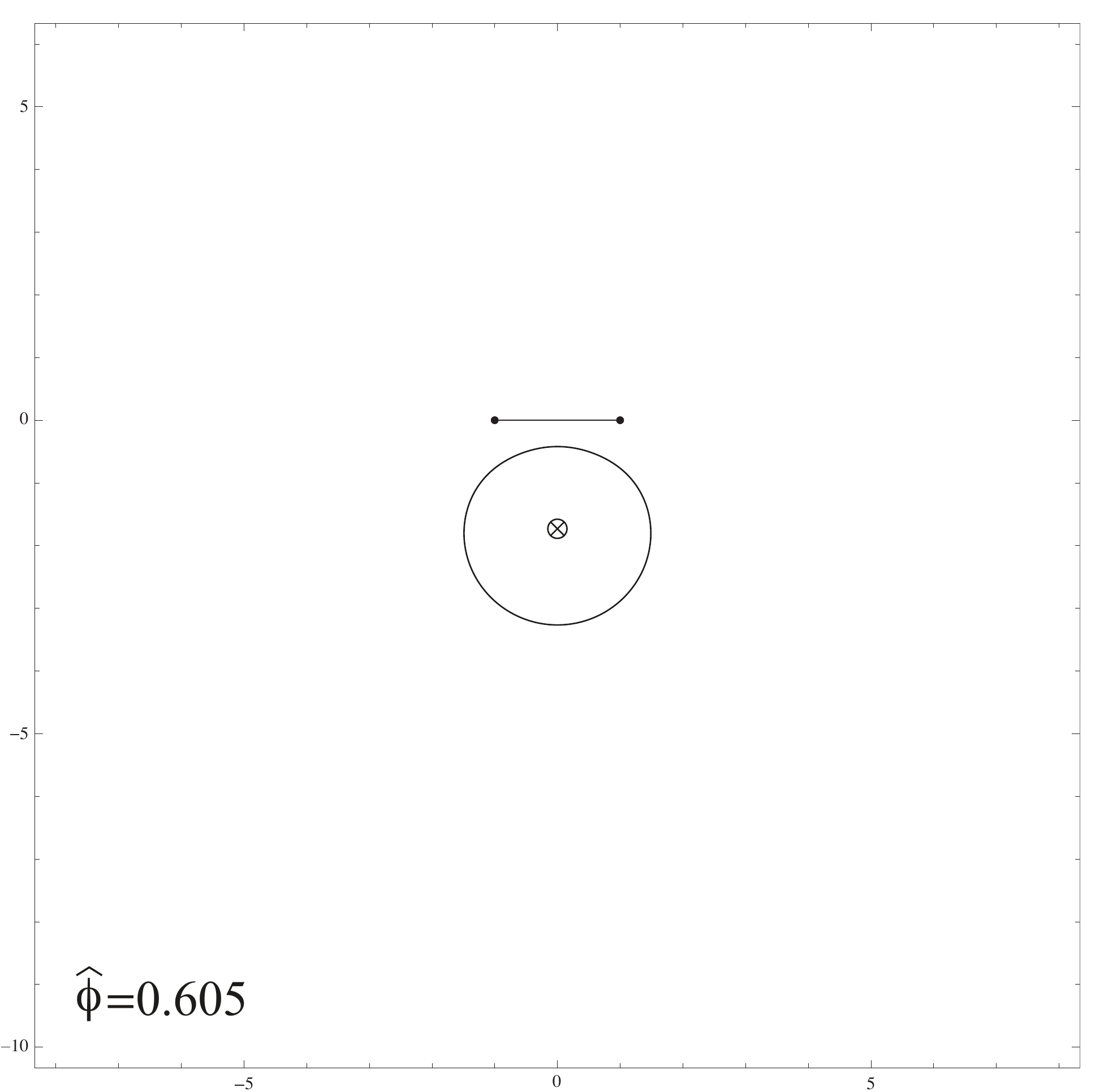} \hspace*{0.2cm}
\includegraphics[height=0.312\textheight]{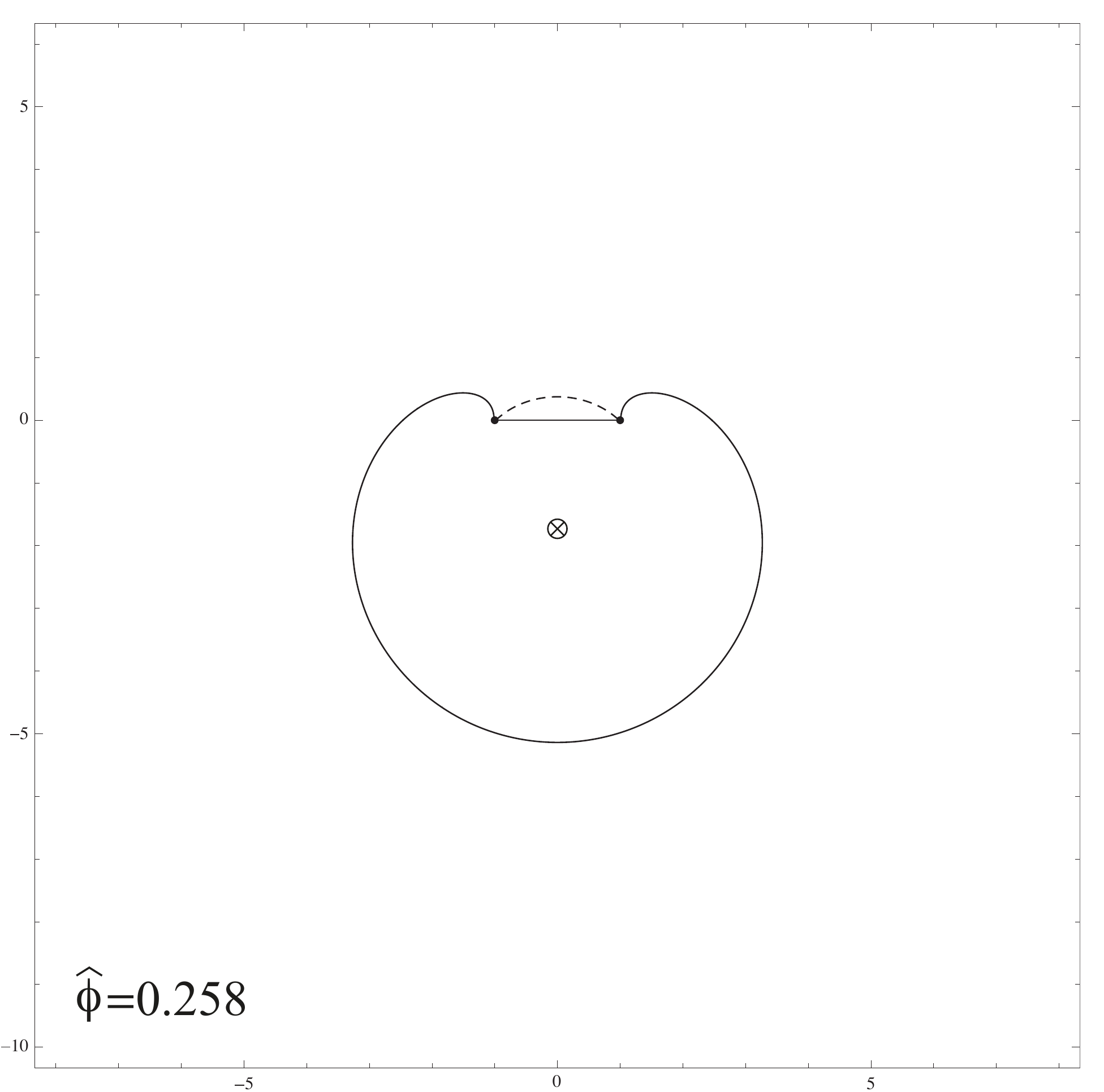} \\
\includegraphics[height=0.312\textheight]{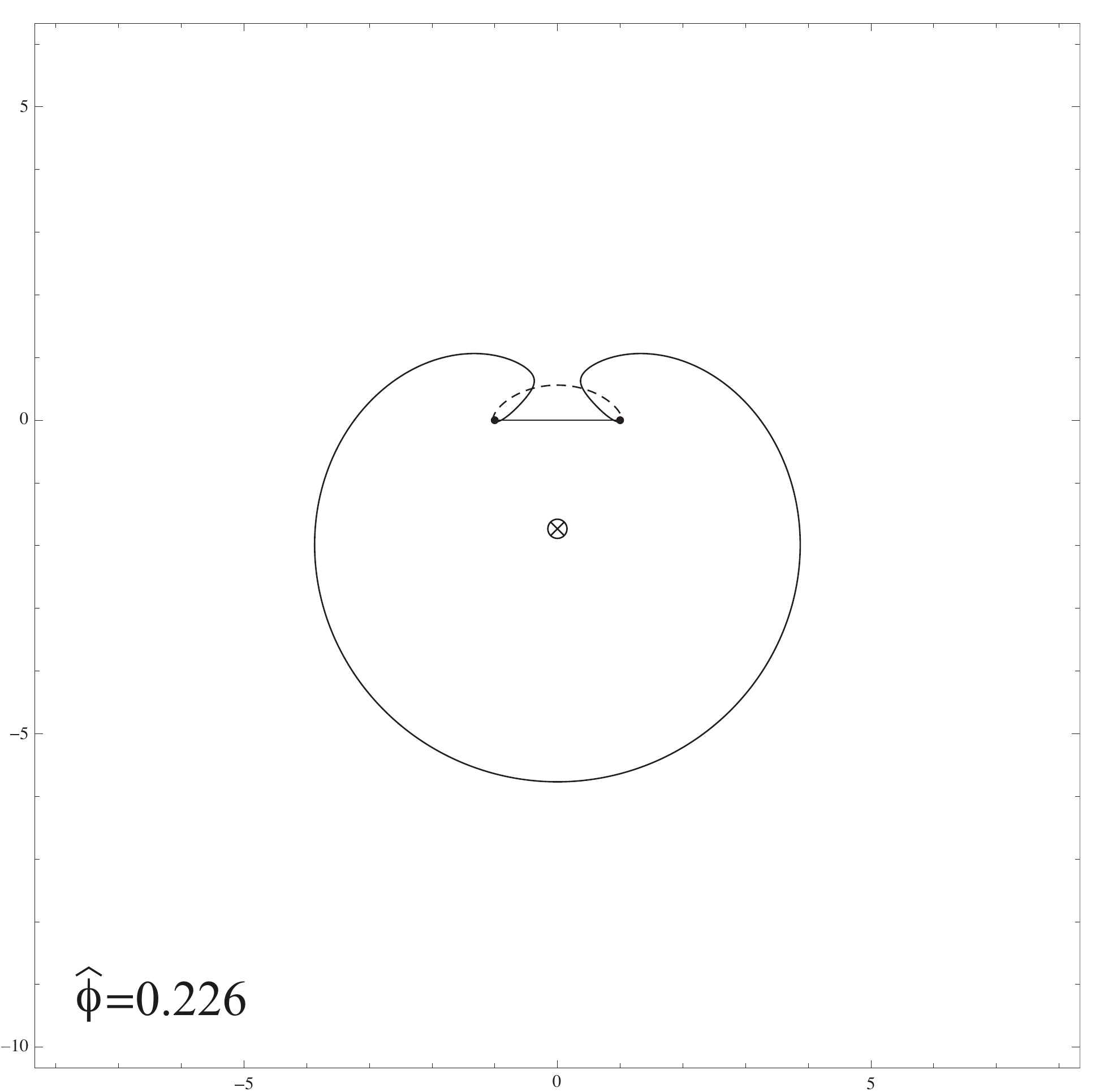} \hspace*{0.2cm}
\includegraphics[height=0.312\textheight]{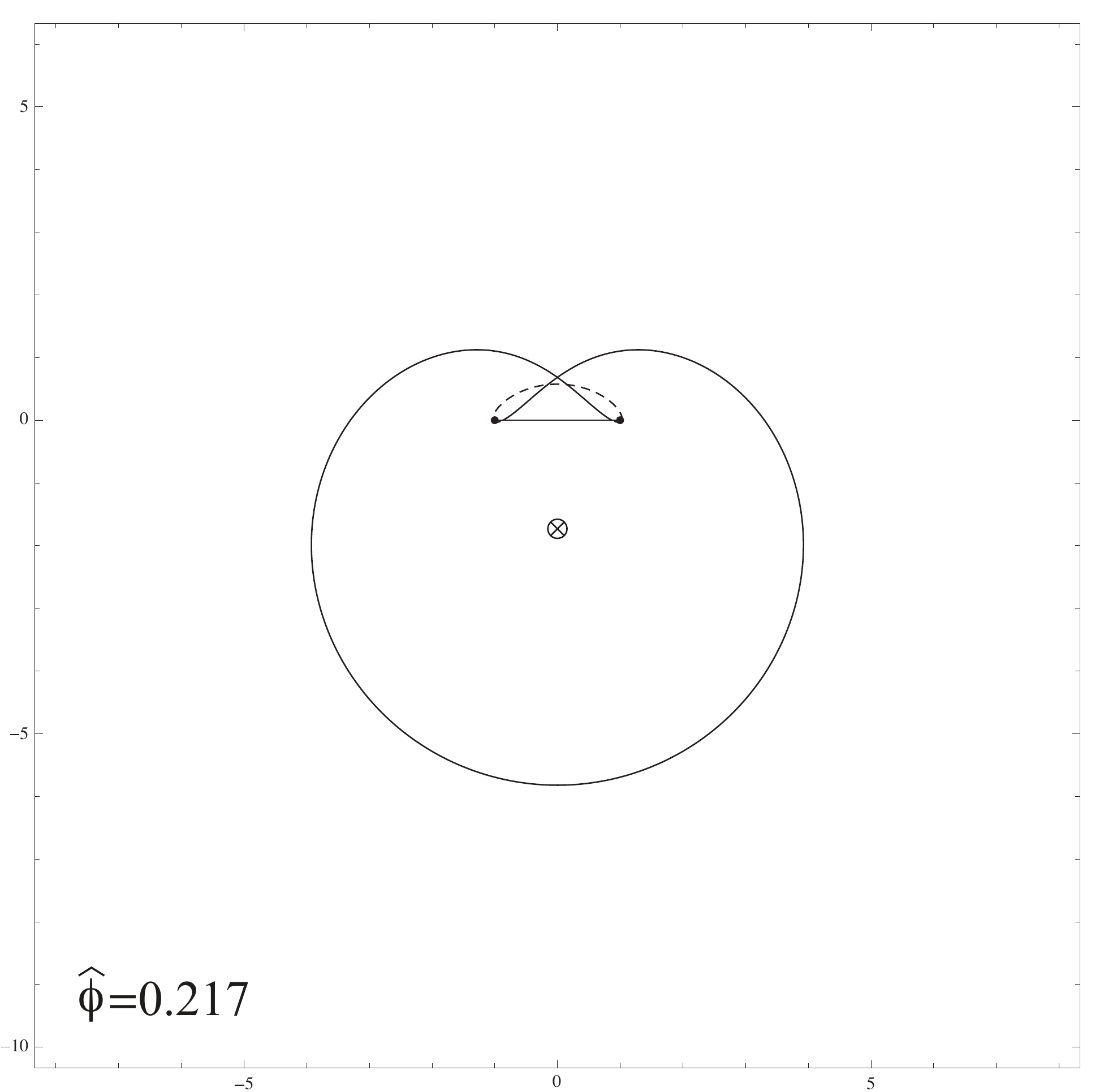} \\
\includegraphics[height=0.312\textheight]{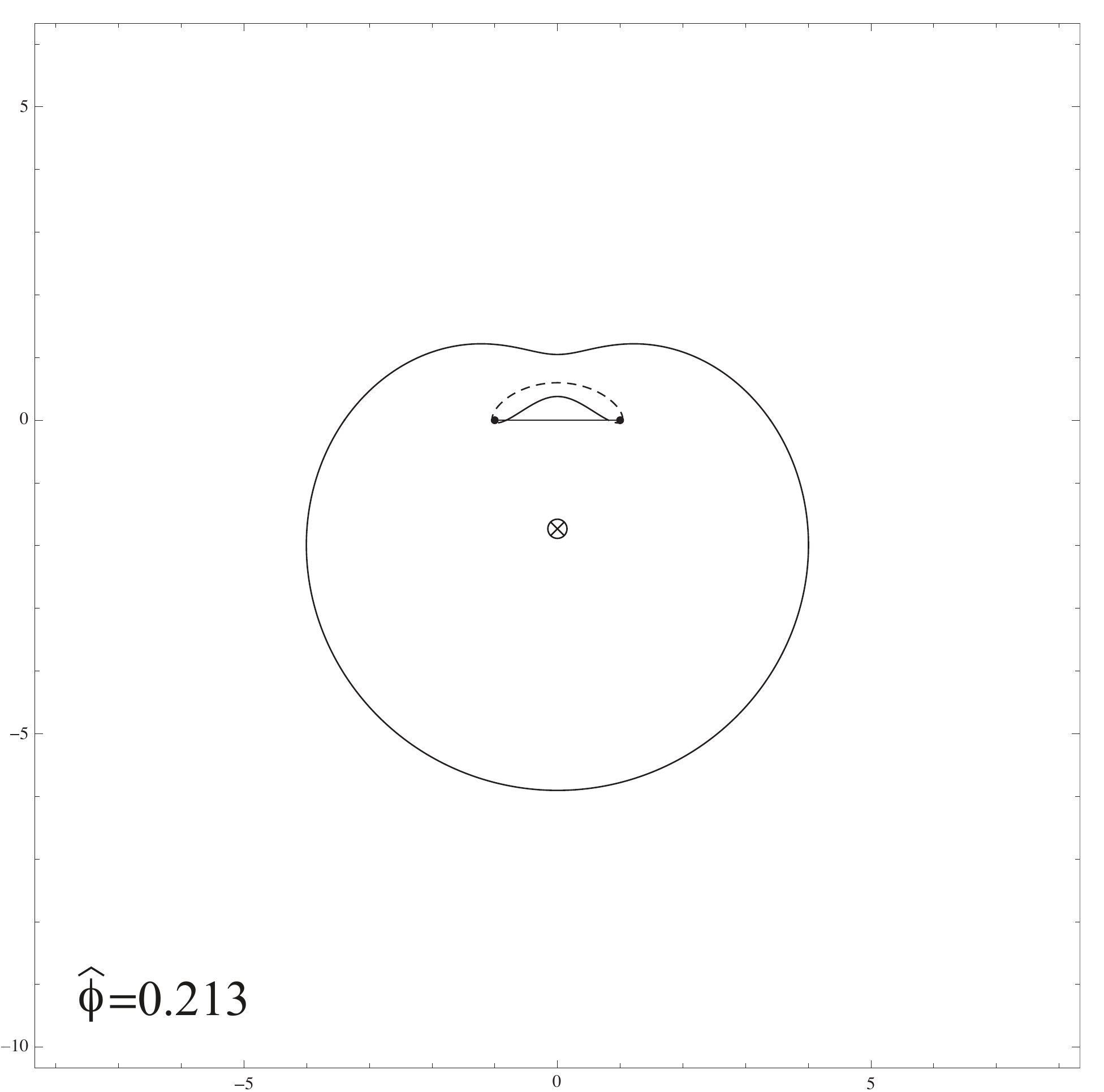} \hspace*{0.2cm}
\includegraphics[height=0.312\textheight]{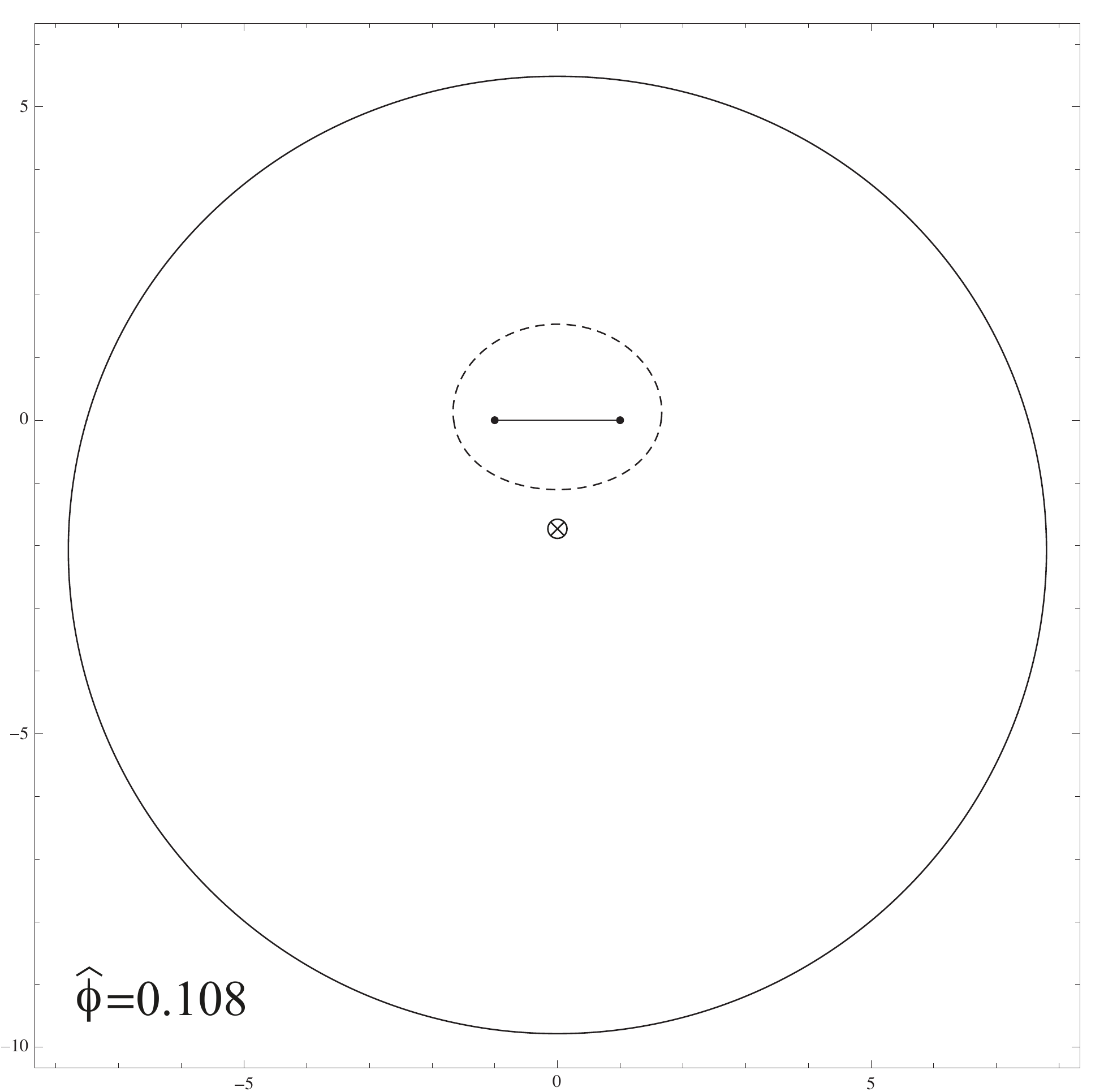} \\
\caption{Evolution of the membrane profiles, showing the different phases of  
the splitting process in Euclidean time in the case $\theta_0=-\pi/3$}
\label{RPmembranesplittingc}
\end{figure}

Figure~\ref{RPmembranesplittingc} shows the splitting process for 
$\theta_0=-\pi/3$, \ie a value intermediate between the cases shown in the 
previous two figures. In this case the splitting and flipping phases occur
simultaneously. At the splitting point the second membrane is still 
self-intersecting and its orientation not completely flipped. The relative 
size of the two membranes in the final state reflects the fact that in this case 
the value of $\theta_0$ (which controls the distance between point charge 
and branch disk) is intermediate between those in the two previous 
examples. It corresponds to a splitting of the flux/angular momentum $J$ 
into $J_1$ and $J_2$ such that the difference $|J_1-J_2|$ is 
larger than in the first case and smaller than in the second case.

If the point charge is not located on the axis of the disk the qualitative
behaviour of the solutions and the associated equipotential surfaces 
remains the same. The processes display similar sequences of steps 
involving splitting and flipping. In this case, however, the cusp singularities 
where $\bm{\nabla}\phi$ diverges do not form simultaneously on the whole 
branch loop, but rather at pairs of points. 

The equipotential surfaces for the potential~(\ref{RFHobsonspotential}),
once expressed in Cartesian coordinates, represent implicit equations of 
the membrane profiles as described by the $z^i(s)$ variables. The evolution
of the shape of the membranes during a splitting process in terms
of the original coordinates, $y^i(t)$, can be reconstructed using the change
of variables~(\ref{RFContiBPSToContiNahm}) in the potential 
$\phi(z^1,z^2,z^3)$~\footnote{In addition one also needs to redefine the 
$y^i$ variables by constant shifts to bring the location of point charge to the 
origin and avoid the run-away behaviour mentioned at the end of 
section~\ref{RSSSolutionCoulomb}.}. The qualitative features of the 
equipotential surfaces parameterised by the $y^i$ coordinates are similar. 
The only significant difference is that the size of the membranes does not 
grow steadily as seen in the previous figures. This is the same difference that 
was observed in section~\ref{RSSSolutionCoulomb} in the case of the 
description of the static spherical solution in terms of the $y^i$ and $z^i$ 
variables. Using the $y^i$ variables one can instead verify 
immediately that the solutions describe a process in which the final 
membranes have radii which add up to the radius of the single membrane in 
the initial state. In the next subsection we will present plots showing the 
evolution of the membrane profiles, using the parametrisation in terms of the 
$y^i$ variables, in the case of solutions interpolating between configurations 
with two membranes in the initial state and two in the final state.

\subsection{General Riemann spaces}
\label{RSSSolutionGeneral}

In the previous subsections we presented explicit solutions to the Laplace
equation describing the most elementary splitting process with a single 
membrane in the initial state and two membranes in the final state. It is 
natural to ask whether a similar approach can be used to construct other 
solutions corresponding to (anti-)instantons of the pp-wave matrix model 
interpolating between states with different numbers of membranes. 
A large class of solutions connecting multi-membrane states was shown to
exist in~\cite{RBBHP}. For our reformulation in terms of the Laplace equation
to be equivalent to the original (continuum) instanton equations, solutions
should exist for all the allowed instanton configurations. Therefore it should 
be possible to identify suitable boundary conditions corresponding to all the
pairs of initial and final states for the original equations, which are permitted 
according to the criteria in~\cite{RBBHP}. Indeed, we will show that the conditions for the existence of BPS instanton solutions obtained 
in~\cite{RBBHP} can be reproduced using our formulation in terms of 
the Laplace equation in a Riemann space.

An analysis of the general features of the solutions we discussed above 
provides hints for the generalisation to more complicated processes. In the 
cases discussed in section~\ref{RSSSolutionHobson} the two membranes in 
the final state correspond to equipotential surfaces in the asymptotic regions 
at infinity, one in each of the two copies of $\R^3$ constituting the Riemann 
space. On the other hand the single membrane in the infinite past 
corresponds to small spheres centred at the location of the point charge. 
These considerations lead us to propose the following general prescription. 
In order to describe a process with $n$ membranes in the final state one 
should consider a Riemann space made of $n$ copies of $\R^3$. Moreover, if 
the initial state contains $m$ membranes one should consider $m$ (positive) 
point charges at the origin of $m$ distinct $\R^3$'s.
The values 
of the point charges, $J_i$, $i=1,2,\ldots,m$, correspond to the angular 
momenta of the $m$ membranes in the initial state. The values of the flux 
of $\bm{\nabla}\phi$ at infinity in each copy of $\R^3$, $J_i'$, 
$i=1,2,\ldots,n$, correspond to the angular momenta of the $n$ membranes 
in the final state. We point out that 
an immediate consequence of this prescription is that by construction the
number, $n$, of membranes in the final state is greater than or equal to the
number, $m$, of membranes in the initial state. Therefore the necessary
condition, $n\ge m$, for the existence of instanton solutions is automatically 
satisfied. 

In the case of a generic Riemann space the relation between the outgoing 
fluxes at infinity, $J'_i$, and the values of the point charges at the origin, 
$J_i$, is non-trivial. 
This is because the way in which the flux of 
$\bm{\nabla}\phi$ gets split among the different spaces depends in a 
complicated way on the number, shape and size of the branch disks and on 
their positions relative to the point charges. In the most elementary case of 
the solution~(\ref{RFHobsonspotential}) we gave explicit formulae for the 
asymptotic fluxes associated with the two membranes in the final state
in~(\ref{RFphi1aympt})-(\ref{RFphi2aymptaxial}).

General multi-sheeted Riemann spaces can involve different combinations of 
branch surfaces. This is the case even in the most elementary example 
discussed in the previous subsection, in which one membrane with angular 
momentum $J$ splits into two membranes with angular momenta $J_1$ and 
$J_2$. We described this process using a Riemann space made of two 
copies of $\R^3$ with a single point charge, $J$, and we presented an exact 
solution involving a circular branch disk. However, the same process receives 
contributions from more complicated Riemann spaces in which the two 
$\R^3$'s are connected by multiple branch disks, so long as the flux at infinity 
in the two spaces is split into the same fractions, $J_1$ and $J_2$.
Of course in order to compute a specific physical transition amplitude it
is necessary to sum all the contributions with the given initial and final
states. The issue of how the different copies of $\R^3$ are connected is 
related to the parameterisation of the moduli spaces of the associated 
instanton solutions and we will briefly comment on this point in 
section~\ref{RSSModuliSpaces}. 

At the present stage we do not have a proof of existence of solutions to the 
three-dimensional Laplace equation in arbitrary Riemann spaces, but we 
propose as a conjecture that solutions should exist for all the boundary 
conditions which according to our prescription correspond to allowed 
(anti-)instantons. The results of~\cite{RBSommerfeld} and~\cite{RBHobson} 
provide a starting point for the analysis of the issue of existence of solutions 
in general Riemann spaces. 
One may view the explicit solutions in~\cite{RBSommerfeld} 
and~\cite{RBHobson} as playing a role similar to that played by
the explicit solution for a spherical conductor obtained by the 
method of images in connection with the theory of the electric potential 
for arbitrarily shaped conductors.
Moreover the Laplace equation in the Riemann 
space can be recast into the form of an integral equation as discussed in  
appendix~\ref{RSAConnectingConditionIntegralEq}. This reformulation may also provide
a useful approach to obtain a proof of existence for the solutions.

\subsubsection{More solutions}
\label{RSSMoreSolutions}

Before discussing how some general properties of the instanton moduli space 
arise within our formulation using the Laplace equation, we now present 
other exact solutions. We use these specific examples to illustrate some 
interesting features arising more generally in the description of instanton 
configurations in terms of equipotential surfaces.

The linearity of the Laplace equation of course implies that linear 
combinations of solutions are also solutions. We can take advantage of
this simple observation to obtain new interesting examples. In particular,
a solution describing a process with two membranes with angular momenta 
$J_1$ and $J_2$ in the initial state and two in the final state -- which based 
on our prescription requires a two-sheeted Riemann space with a point 
charge in each copy of $\R^3$ -- can be obtained as a linear 
combination of the potential~(\ref{RFHobsonspotential}) with charge $J_1$ 
and the analogous potential with point charge $J_2$ at $\bm{z}=\bm{z_0}$ in 
the second space. To obtain the potential due to a point 
charge at the point corresponding to $\bm{z}_0$ in the second space, we 
recall that in peripolar coordinates the first sheet is parameterised by
$\theta\in[-\pi,\pi]$ and the second sheet by $\theta\in[\pi,3\pi]$. Therefore 
the potential induced by a charge located at the 
point in the second $\R^3$ corresponding to $\bm{z}_0$ is obtained 
substituting $\theta_0\to\theta_0+2\pi$ in~(\ref{RFHobsonspotential}). 
The resulting solution in the two sheeted Riemann space
with point charges in
both spaces is
\begin{align}
\phi(\bm{z},\bm{z}_0) 
&= \frac{J_1}{4\pi|\bm{z}-\bm{z}_0|} 
\left[ \frac{1}{2} + \frac{1}{\pi} \arcsin\!\left( 
\cos \left(\frac{\theta - \theta_0}{2}\right) 
\sqrt{\frac{2}{\cosh \alpha +1}}  \right) \right] \notag \\
&\, + \frac{J_2}{4\pi|\bm{z}-\bm{z}_0|} 
\left[ \frac{1}{2} - \frac{1}{\pi} \arcsin\!\left( 
\cos \left(\frac{\theta - \theta_0}{2}\right) 
\sqrt{\frac{2}{\cosh \alpha +1}}  \right) \right] 
\label{RFscatteringpotential} \\
&= \frac{1}{4\pi |\bm{z}-\bm{z}_0|}  
\left[ \frac{(J_1+J_2)}{2} + \frac{(J_1-J_2)}{\pi} \arcsin\!\left( 
\cos \left(\frac{\theta - \theta_0}{2}\right) 
\sqrt{\frac{2}{\cosh \alpha +1}}  \right) \right] , \nn
\end{align}
where $|\bm{z}-\bm{z}_0|$ and $\cosh\a$ are given 
in~(\ref{RFHobsonsdistance}) and~(\ref{RFHobsonscoshalpha}) 
respectively and we have chosen the radius of the branch loop to be 
$a=1$. The sign difference between the first two lines 
in~(\ref{RFscatteringpotential}) comes from the shift in $\theta_0$. 
Recalling that $\cosh\a$ is independent of $\theta_0$, we get
\begin{equation}
\arcsin\!\left[\cos \left(\frac{\theta - (\theta_0+2\pi)}{2}\right) 
\sqrt{\frac{2}{\cosh \alpha +1}}  \right] = 
- \arcsin\!\left[\cos \left(\frac{\theta - \theta_0}{2}\right) 
\sqrt{\frac{2}{\cosh \alpha +1}}  \right], 
\label{RFExplanationOf2PiShift}
\end{equation}
because the shift by $\pi$ in the argument flips the sign of the cosine
and the inverse sine function is odd. Notice that if the two charges are equal, 
$J_1=J_2$, (\ref{RFscatteringpotential}) reduces to a simple Coulomb 
potential in both sheets of the Riemann space. This is consistent with our 
general prescription. With equal charges the fluxes of $\bm{\nabla}\phi$ from 
the first space into the second and from the second space into the first are 
equal, irrespective of the position, $\bm{z}_0$, of the charges. So the flux at 
infinity in the two spaces is the same and hence the final state is the same as 
the initial state. Therefore this case corresponds simply to a stable vacuum 
configuration with two membranes of the same size. Note that, up to an 
overall scale, the potential~(\ref{RFscatteringpotential}) can also be obtained 
from the superposition of the solution~(\ref{RFHobsonspotential}) and a 
simple Coulomb potential.

We now focus on the special case of axially symmetric solutions 
(corresponding to $\rho_0=\v_0=0$) with $J_1=2J$, $J_2=J$ for 
which~(\ref{RFscatteringpotential}) becomes
\begin{equation}
\phi(\bm{z},\bm{z}_0) = \frac{J}{4\pi |\bm{z}-\bm{z}_0|}  
\left[ \frac{3}{2} + \frac{1}{\pi} \arcsin\!\left( 
\cos \left(\frac{\theta - \theta_0}{2}\right) 
\sqrt{\frac{2}{\cosh \rho +1}}  \right) \right] \, ,
\label{RFscatteringpotential32}
\end{equation}
where we used the fact that $\cosh\alpha=\cosh\rho$ when $\bm{z}_0$ is
on the axis of the disk.

The final states in the process described by the 
potential~(\ref{RFscatteringpotential32}) depend on the way in which the total 
flux at infinity is divided between the two spaces, which in turn is controlled by 
the value of $\theta_0$. Different choices for $\theta_0$ give rise to solutions 
displaying various interesting features and below we present a few selected
examples. 

The following figures~\ref{RPmembranescatteringdb},
\ref{RPmembranescatteringe} and~\ref{RPmembranescatteringd}  
show the evolution of the two membranes described by the 
solution~(\ref{RFscatteringpotential32}) for $\theta_0=-2\pi/3$, 
$\theta_0=-\pi/10$ and $\theta_0=-\pi/6$ respectively. 
As in the previous subsection, the plots show equipotential surfaces
for the rescaled potential $\hat\phi=(4\pi/J)\phi$.
The figures depict the equipotential surfaces for  $\phi(y^1,y^2,y^3)$ 
expressed in terms of the coordinates, $y^i(t)$, in the 
original instanton equations~(\ref{RFContinuumBPSEq}). The conventions 
used are the same as in previous figures. The 
(portions of) membranes in the 
first copy of $\R^3$ are shown as continuous contours and those in the 
second $\R^3$ as dashed contours. Since here we are using the $y^i$ 
coordinates, the contours represent the evolution of the profiles of the 
membranes from $t=-\infty$ to $t=+\infty$. Notice that with this 
parameterisation of the solution the membranes do not expand steadily.
However, their relative radii change between the initial and final state because 
of the transfer of angular momentum. The use of these variables makes 
manifest the conservation of angular momentum ($J_1'+J_2'=J_1+J_2$), 
which in the figures is reflected in the fact that the radii of the spheres in the 
initial and final states add up to the same total. In the figures we have not 
displayed the branch loop as in these coordinates its position is not constant. 

In the case shown in figure~\ref{RPmembranescatteringdb}
the value of 
$\theta_0=-2\pi/3$ and thus the point charges are relatively close to the 
branch disk. As a result there is a large fraction of flux passing from the first 
$\R^3$ into the second. This corresponds to a large transfer of angular 
momentum and thus a significant change in the relative radii of the two 
membranes. In the sequence shown in the figure the second and third steps 
show both membranes extending across both spaces. In the third panel the 
outer membrane develops a self-intersection. The small region marked by a 
square is shown enlarged in figure~\ref{RPmembranescatteringbdetail}, 
where the self-intersection is more clearly visible. 
The membranes then
split and reconnect in the way shown in the fourth panel of 
figure~\ref{RPmembranescatteringdb} resulting in two equipotential 
surfaces each contained entirely in one copy of $\R^3$.

\begin{figure}[htbp]
\centering
\includegraphics[height=0.312\textheight]{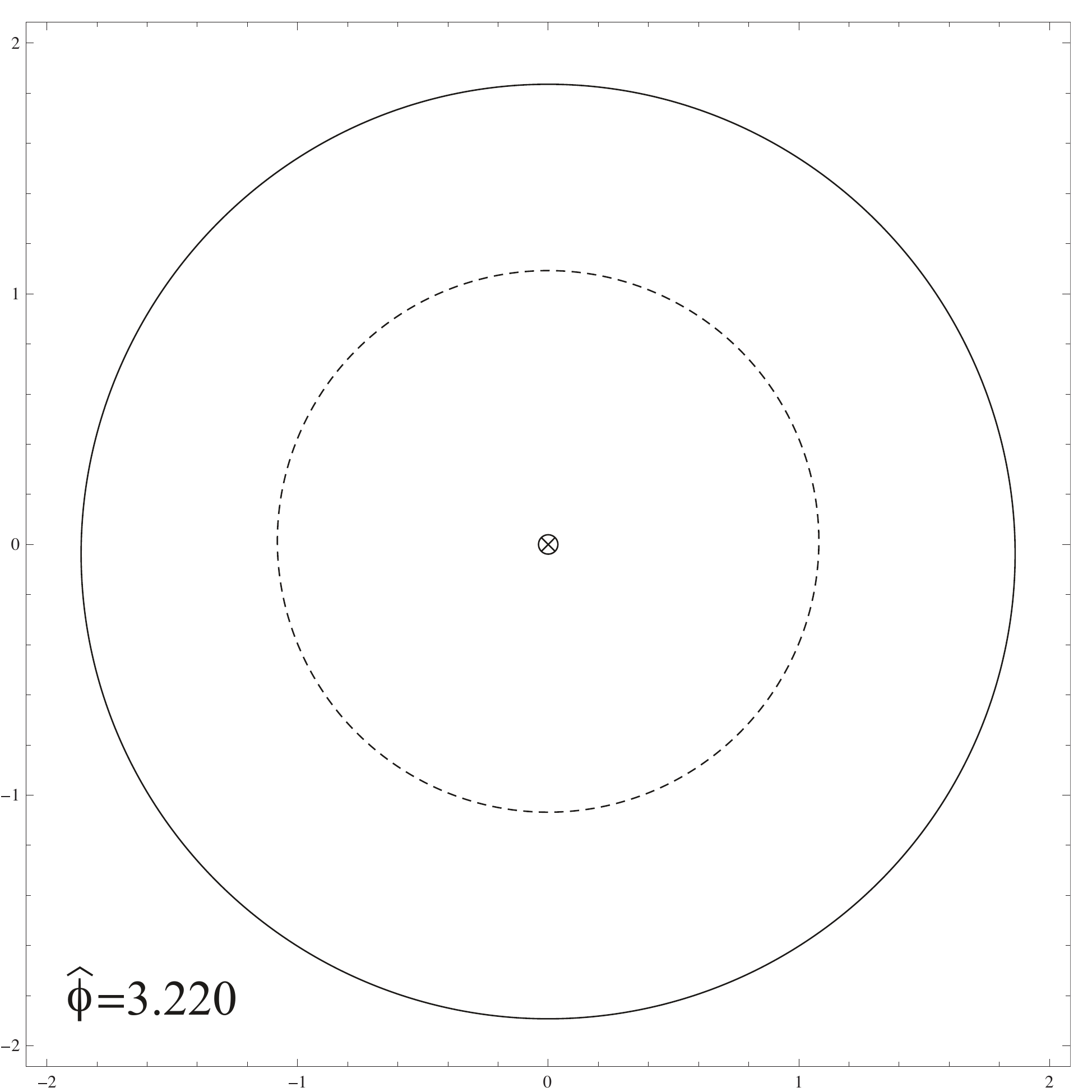} \hspace*{0.2cm}
\includegraphics[height=0.312\textheight]{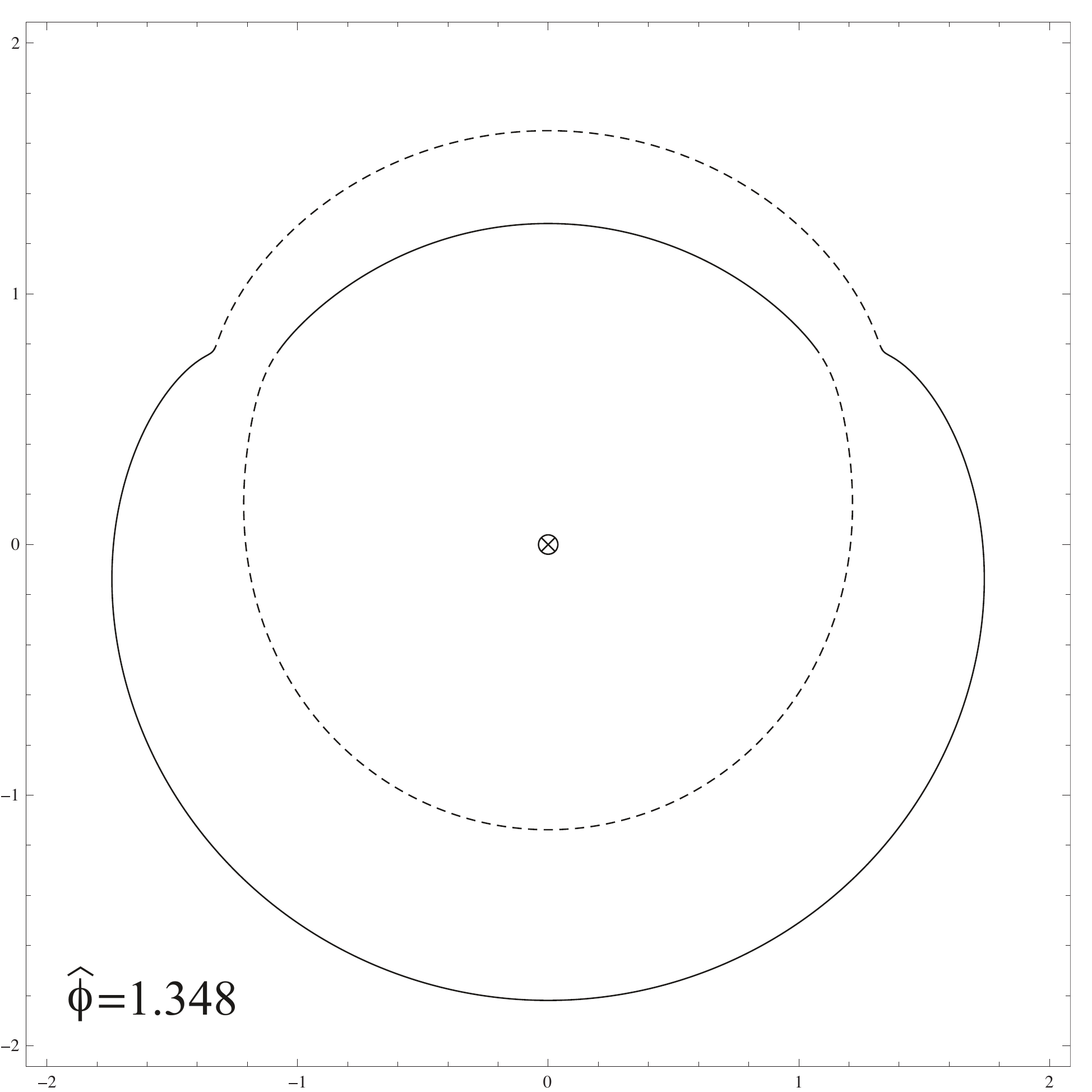} \\
\includegraphics[height=0.312\textheight]{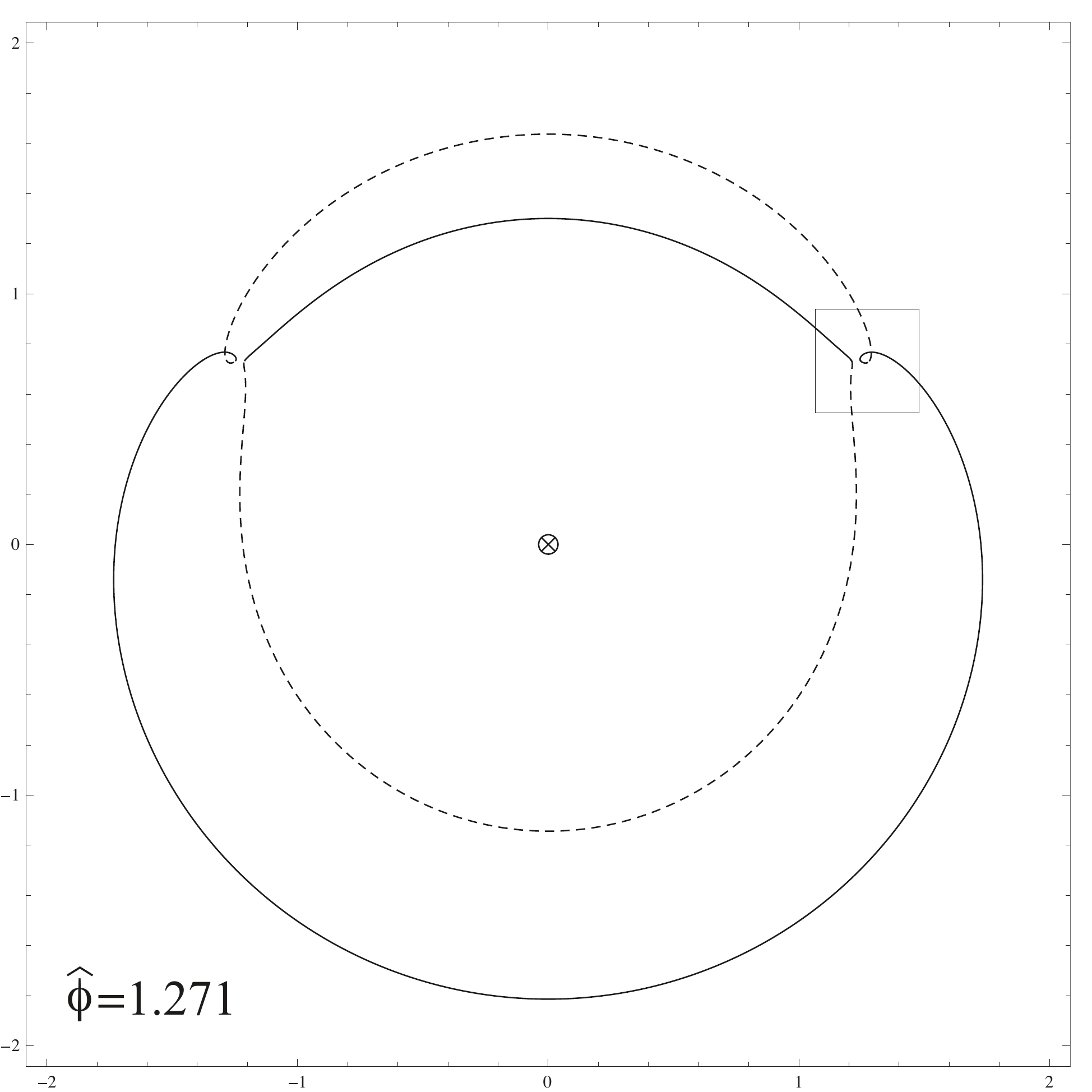} \hspace*{0.2cm}
\includegraphics[height=0.312\textheight]{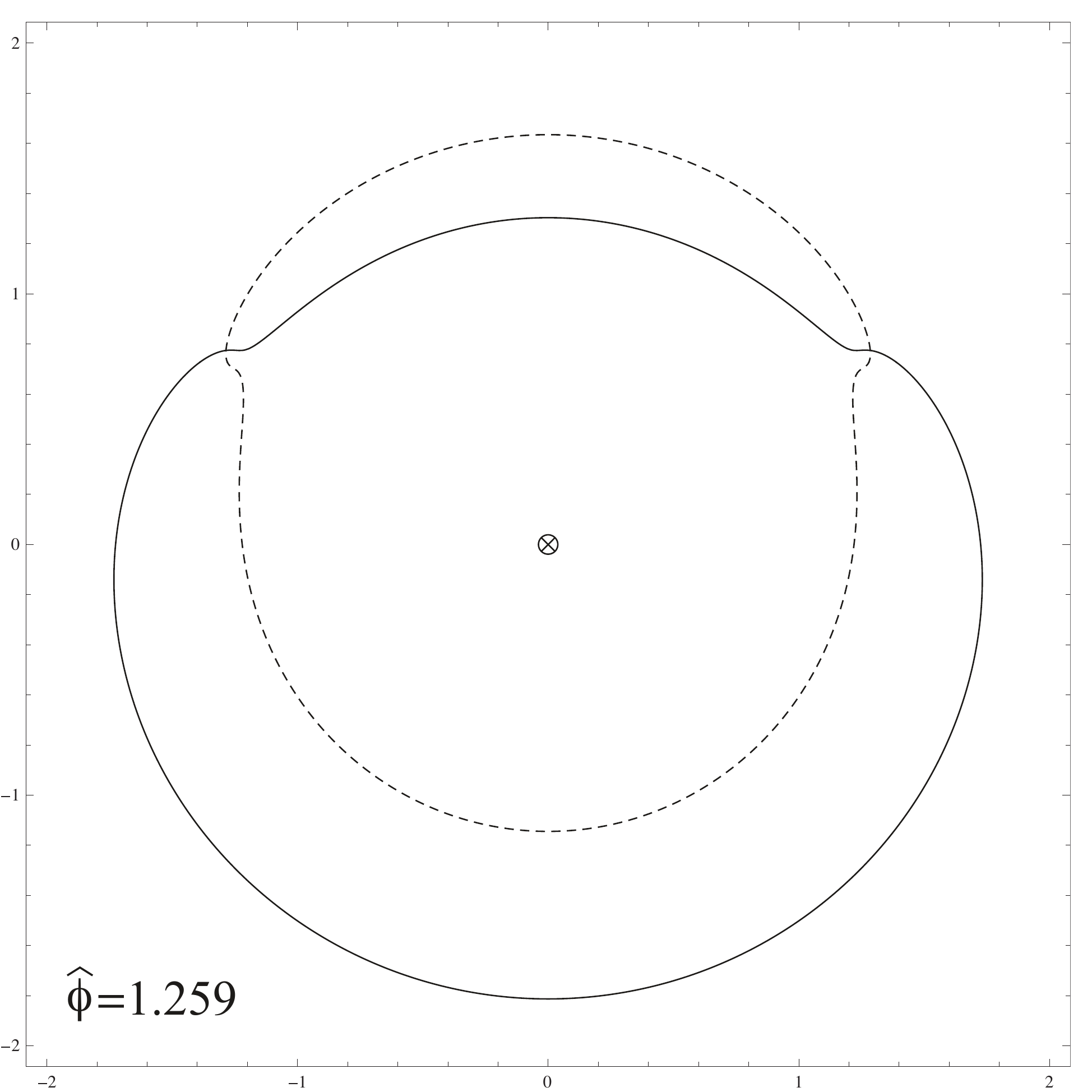} \\
\includegraphics[height=0.312\textheight]{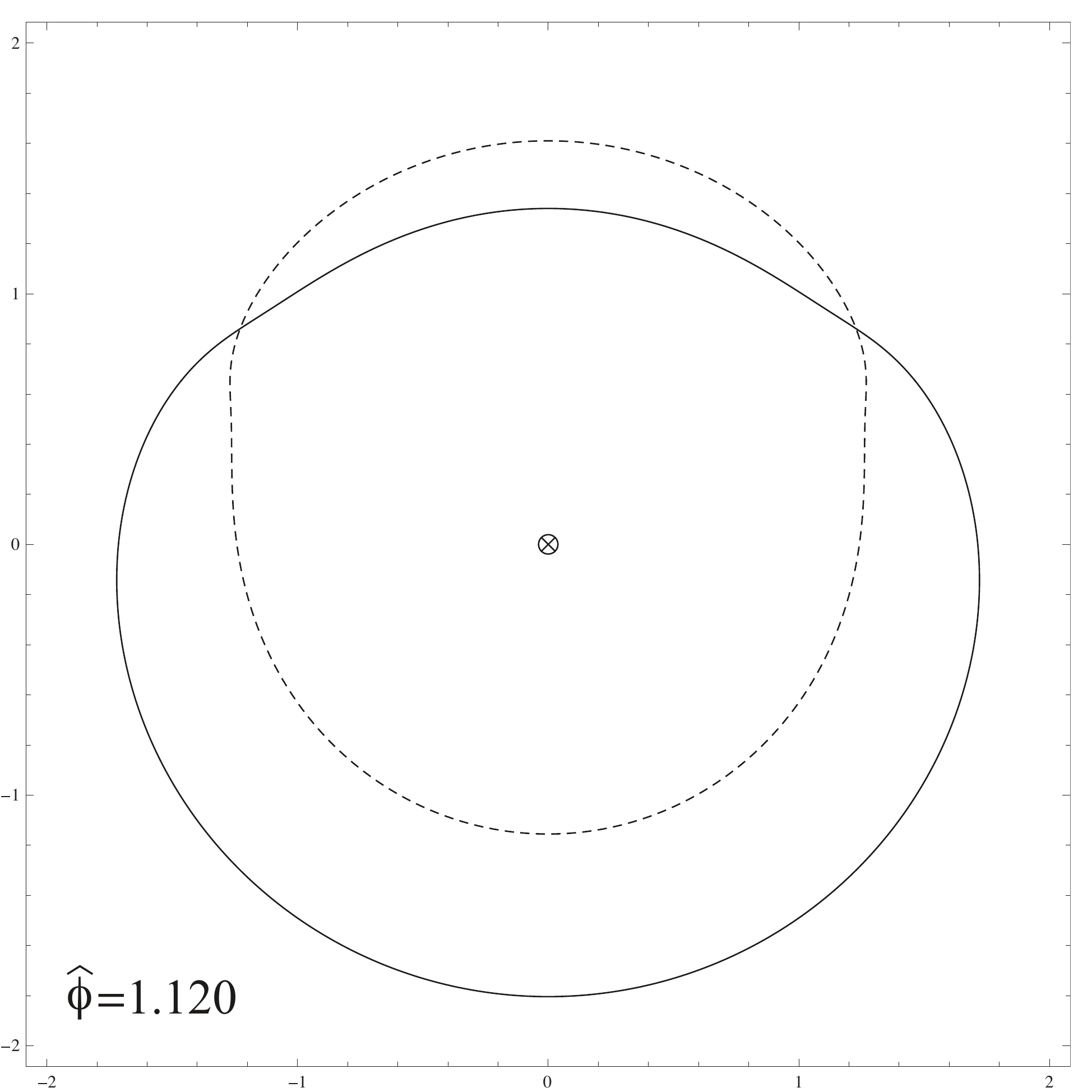} \hspace*{0.2cm}
\includegraphics[height=0.312\textheight]{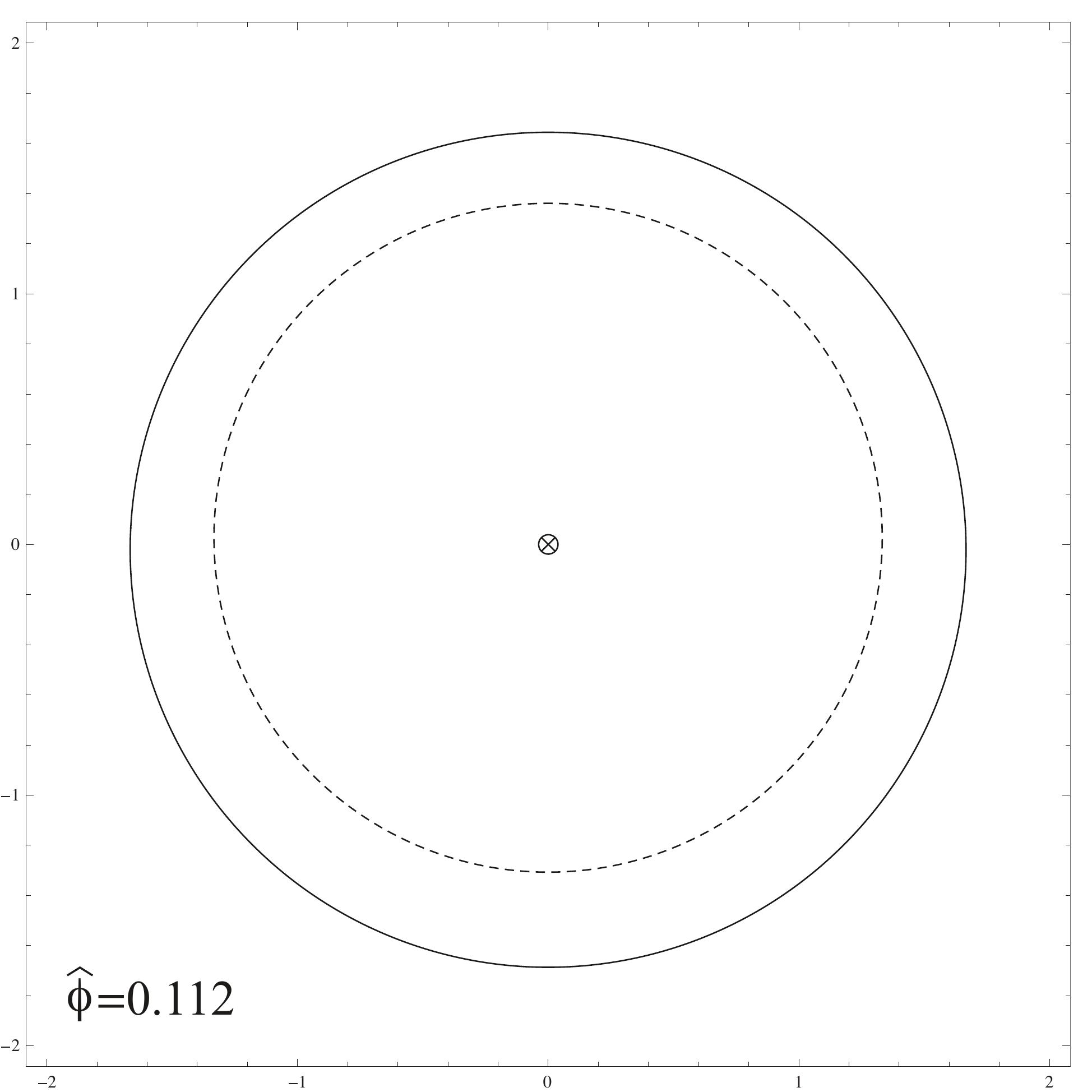} \\
\caption{Evolution of the membrane profiles for a process with two
membranes in the initial state and two in the final state  
($\theta_0=-2\pi/3$).}
\label{RPmembranescatteringdb}
\end{figure}

\begin{figure}[htb]
\centering
\includegraphics[width=0.3\textwidth]{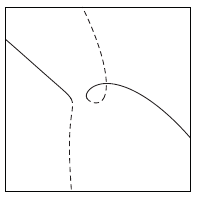}
\caption{Enlargement of the area marked by a square in the third panel 
of figure~\ref{RPmembranescatteringdb}.}
\label{RPmembranescatteringbdetail}
\end{figure}

In the case shown in  figure~\ref{RPmembranescatteringe} the point charges 
are at $\theta_0=-\pi/10$, \ie further from the branch disk and this means that  
there is a smaller transfer of angular momentum. With this choice of boundary
conditions we observe another interesting feature: for some intermediate
values of Euclidean time there are three membranes involved in the process. 
A small membrane detaches from the larger membrane in the first $\R^3$ and 
subsequently gets absorbed by the other membrane. In the intermediate 
steps this third  membrane crosses the branch disk from the first $\R^3$ into 
the second, flipping its orientation. The self-intersection involved in the 
flipping of the membrane orientation is shown in 
figure~\ref{RPmembranescatteringedetail}, which is an enlargement of the 
small rectangle marked in the fourth panel of 
figure~\ref{RPmembranescatteringe}. 

\begin{figure}[htbp]
\centering
\includegraphics[height=0.312\textheight]{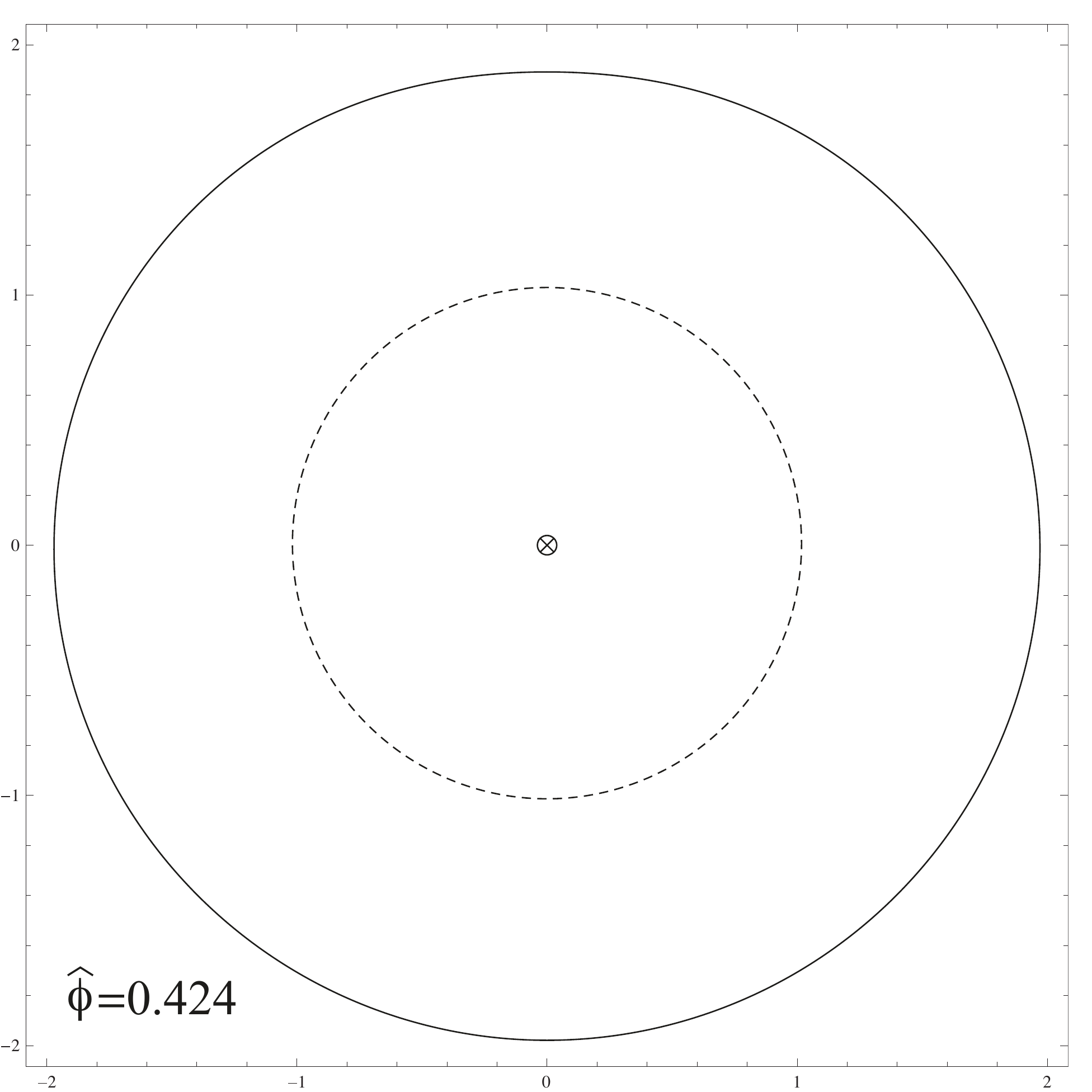} \hspace*{0.2cm}
\includegraphics[height=0.312\textheight]{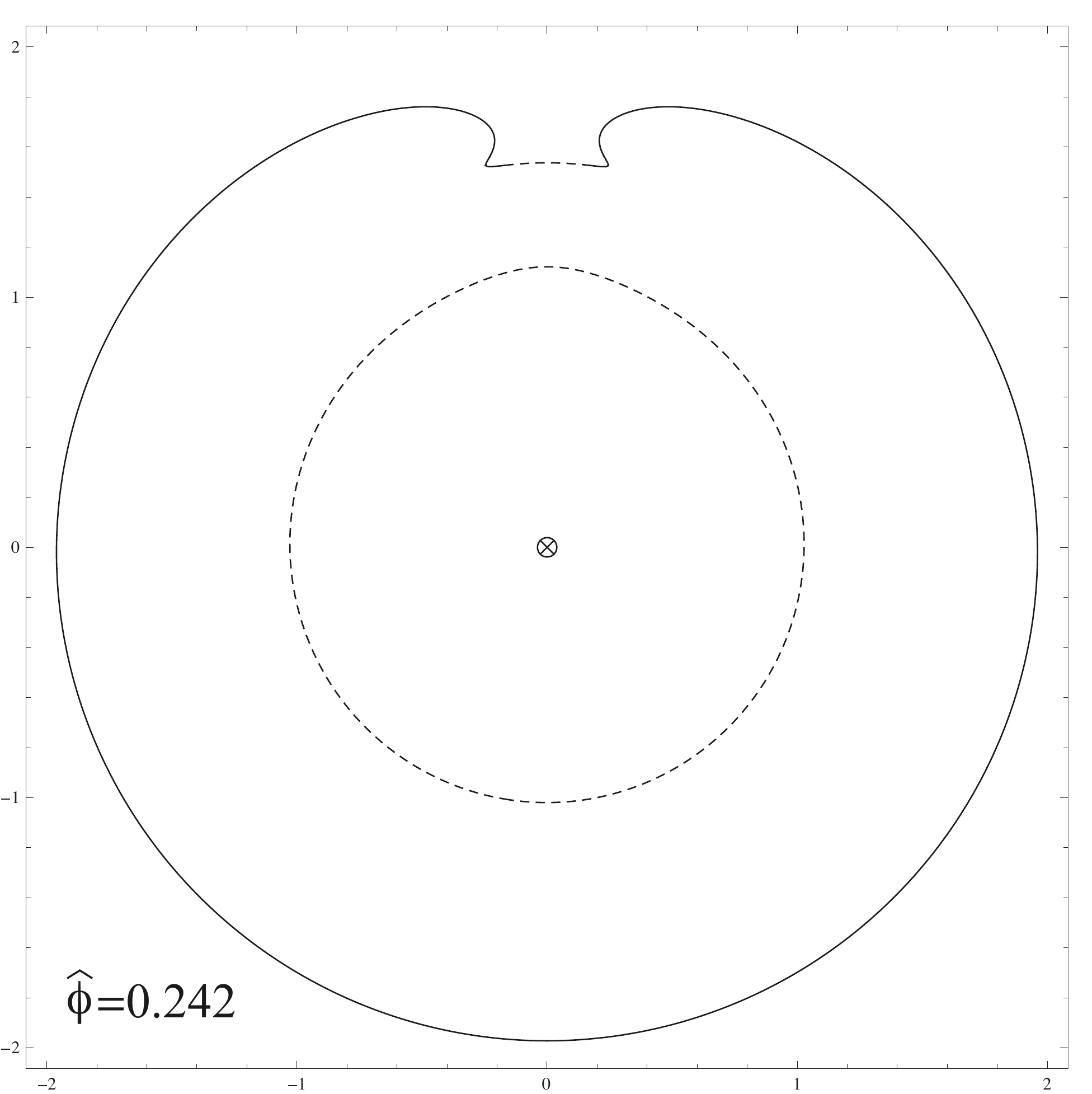} \\
\includegraphics[height=0.312\textheight]{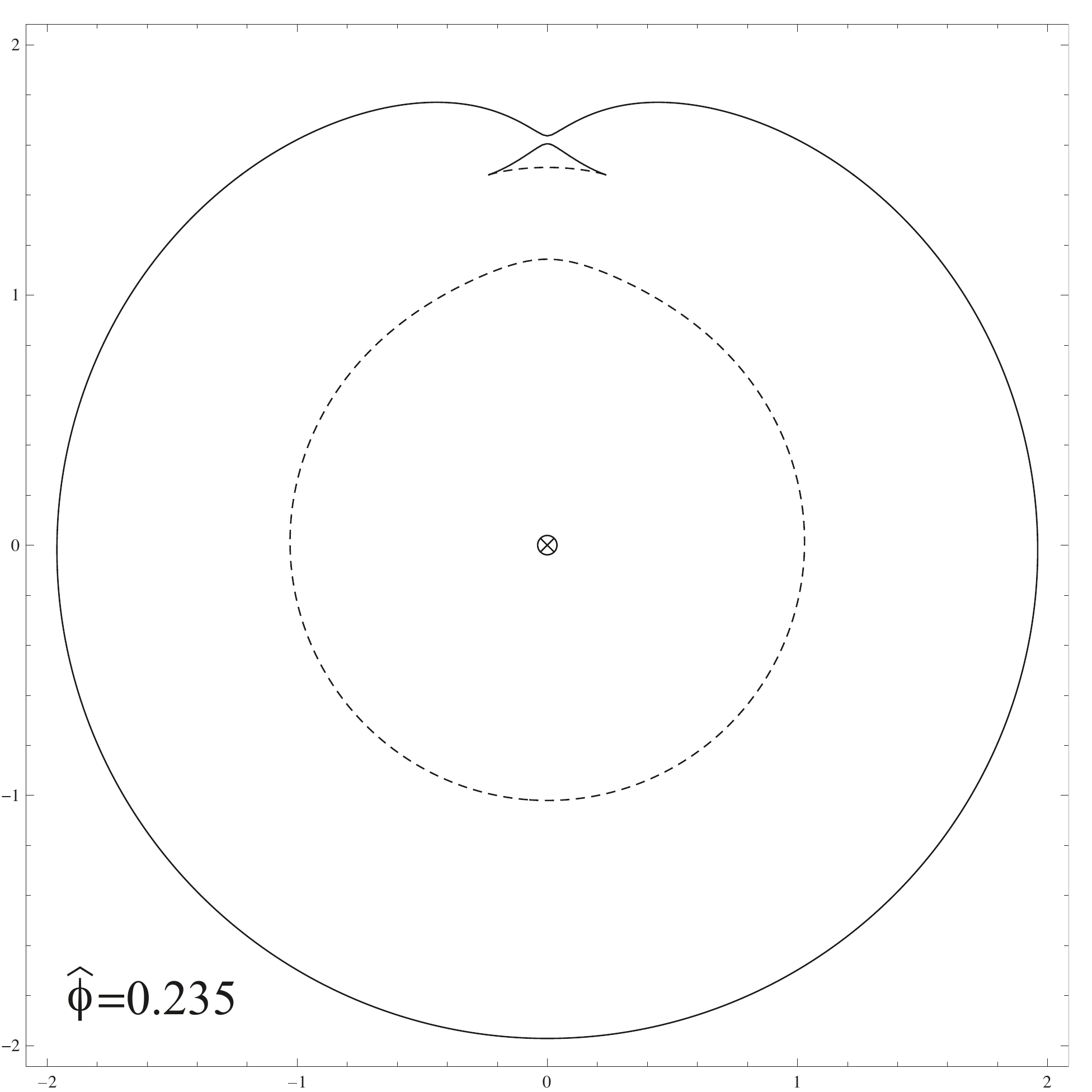} \hspace*{0.2cm}
\includegraphics[height=0.312\textheight]{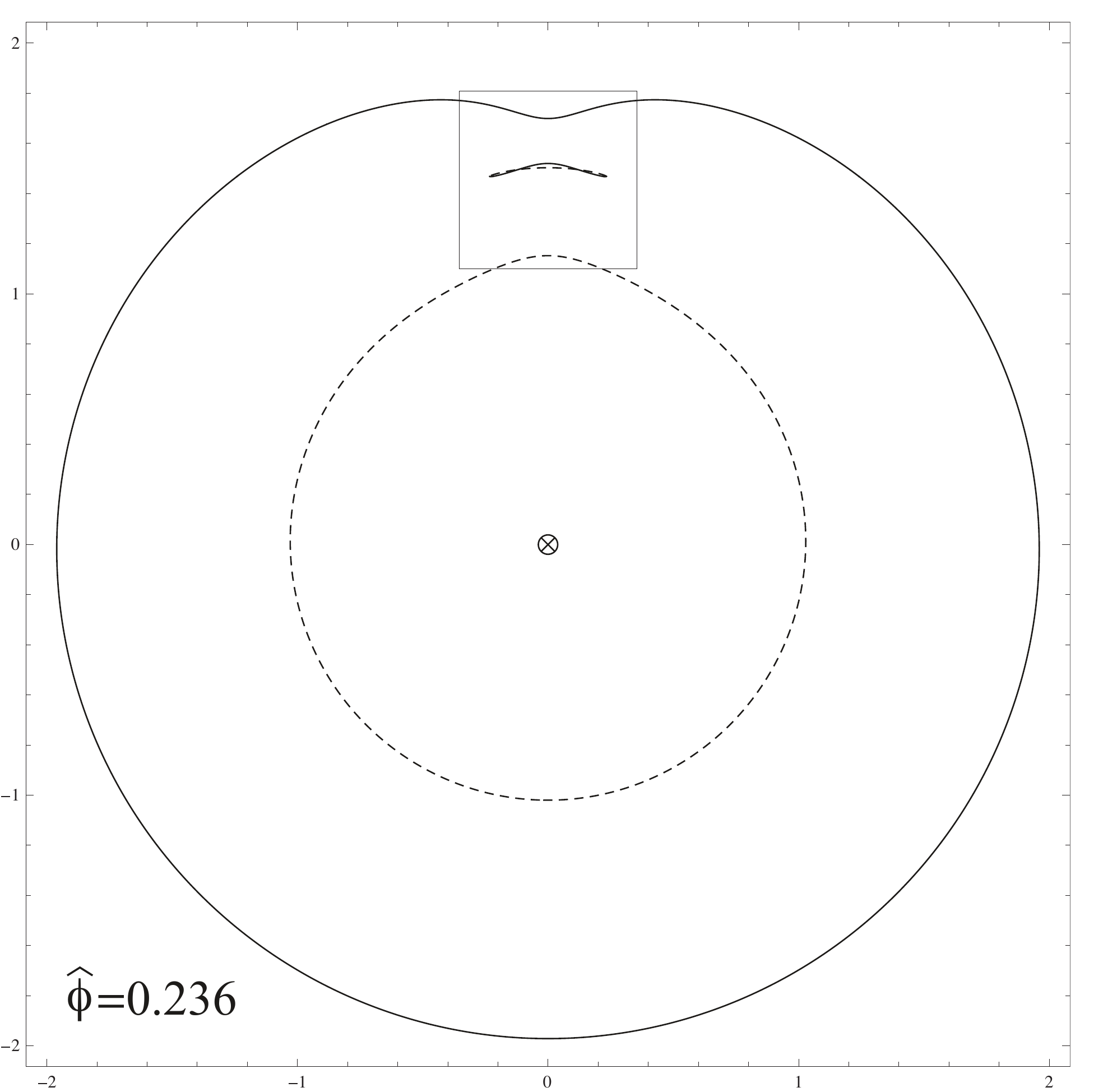} \\
\includegraphics[height=0.312\textheight]{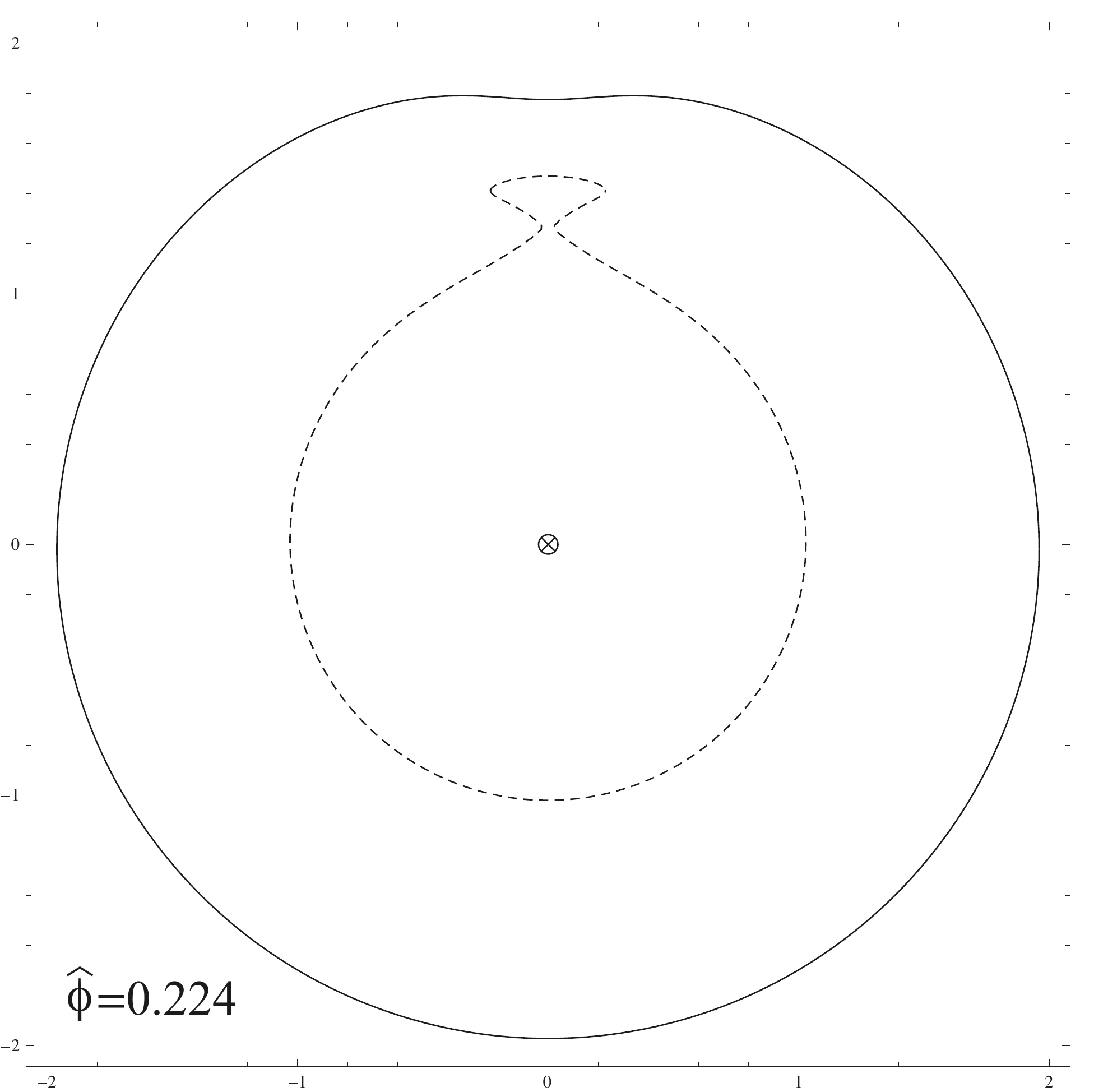} \hspace*{0.2cm}
\includegraphics[height=0.312\textheight]{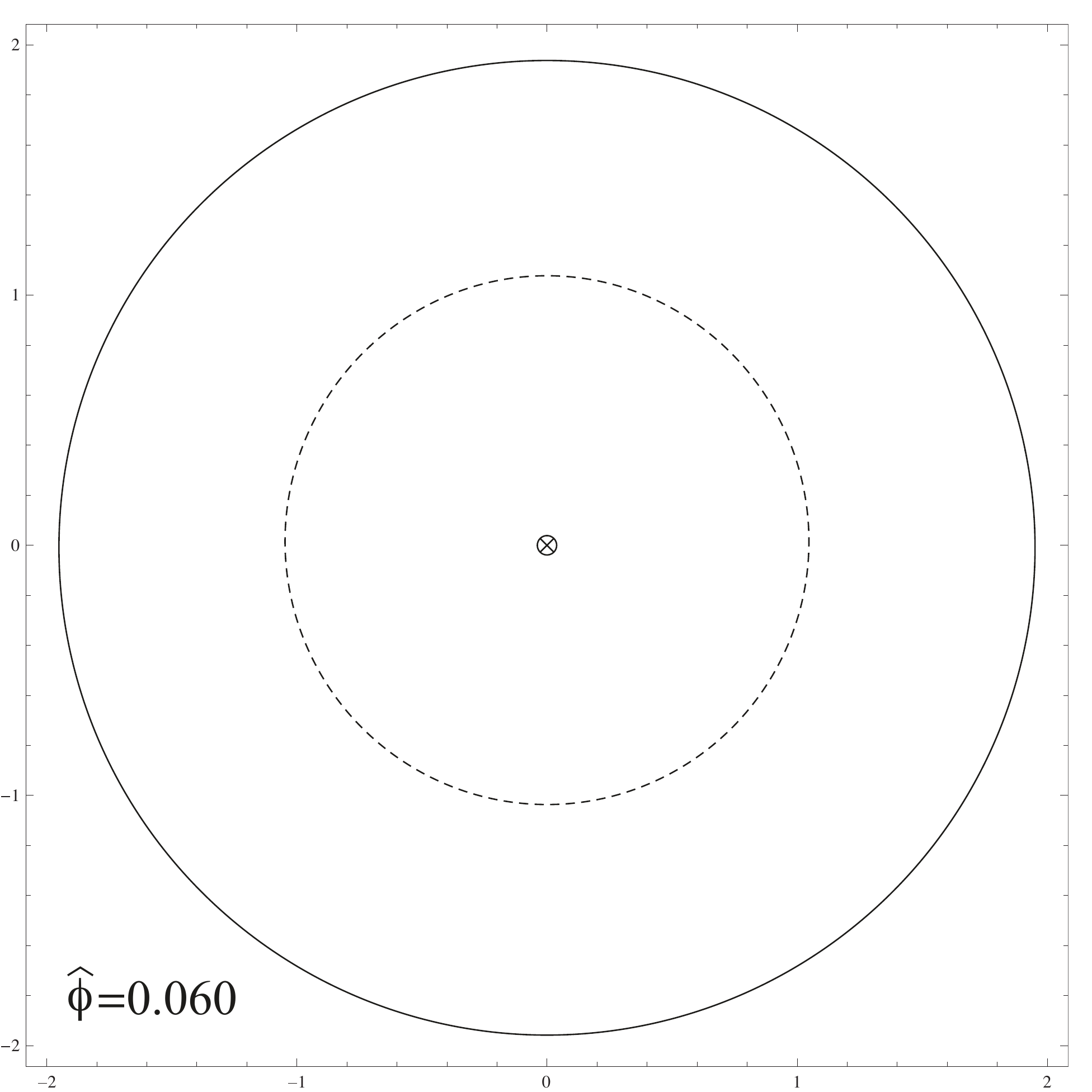} \\
\caption{Evolution of the membrane profiles for a process with two
membranes in the initial state and two in the final state  
($\theta_0=-\pi/10$).}
\label{RPmembranescatteringe}
\end{figure}

\begin{figure}[htb]
\centering
\includegraphics[width=0.3\textwidth]{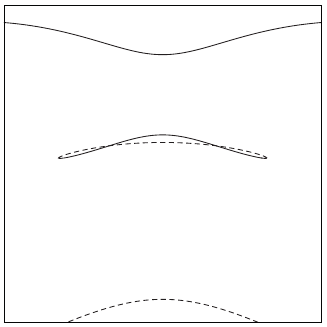}
\caption{Enlargement of the area marked by a rectangle in the fourth panel 
of figure~\ref{RPmembranescatteringe}.}
\label{RPmembranescatteringedetail}
\end{figure}

Figure~\ref{RPmembranescatteringd} depicts a case intermediate between 
the previous two examples, corresponding to $\theta_0=-\pi/6$. In this 
case the transfer of angular momentum does not involve the exchange of
a third membrane, but the interplay between the equipotential surfaces
is rather interesting. In figure~\ref{RPmembranescatteringddetail} we 
show a detailed view of the area within the rectangles marked in the
third and fourth panels in figure~\ref{RPmembranescatteringd}, presenting
a sequence of contour plots for values of the potential between 
$\hat\phi=0.380$ and $\hat\phi=0.378$. The fourth panel in 
figure~\ref{RPmembranescatteringddetail}  shows clearly that in this case 
the splitting takes place not at a point, but simultaneously at points along a 
ring. This is of course not a generic feature. It is a degenerate case which 
only occurs in axially symmetric solutions for certain values of the 
parameters. A similar splitting ring is also present in the previously 
discussed case of figure~\ref{RPmembranescatteringdb}. 

\begin{figure}[htbp]
\centering
\includegraphics[height=0.312\textheight]{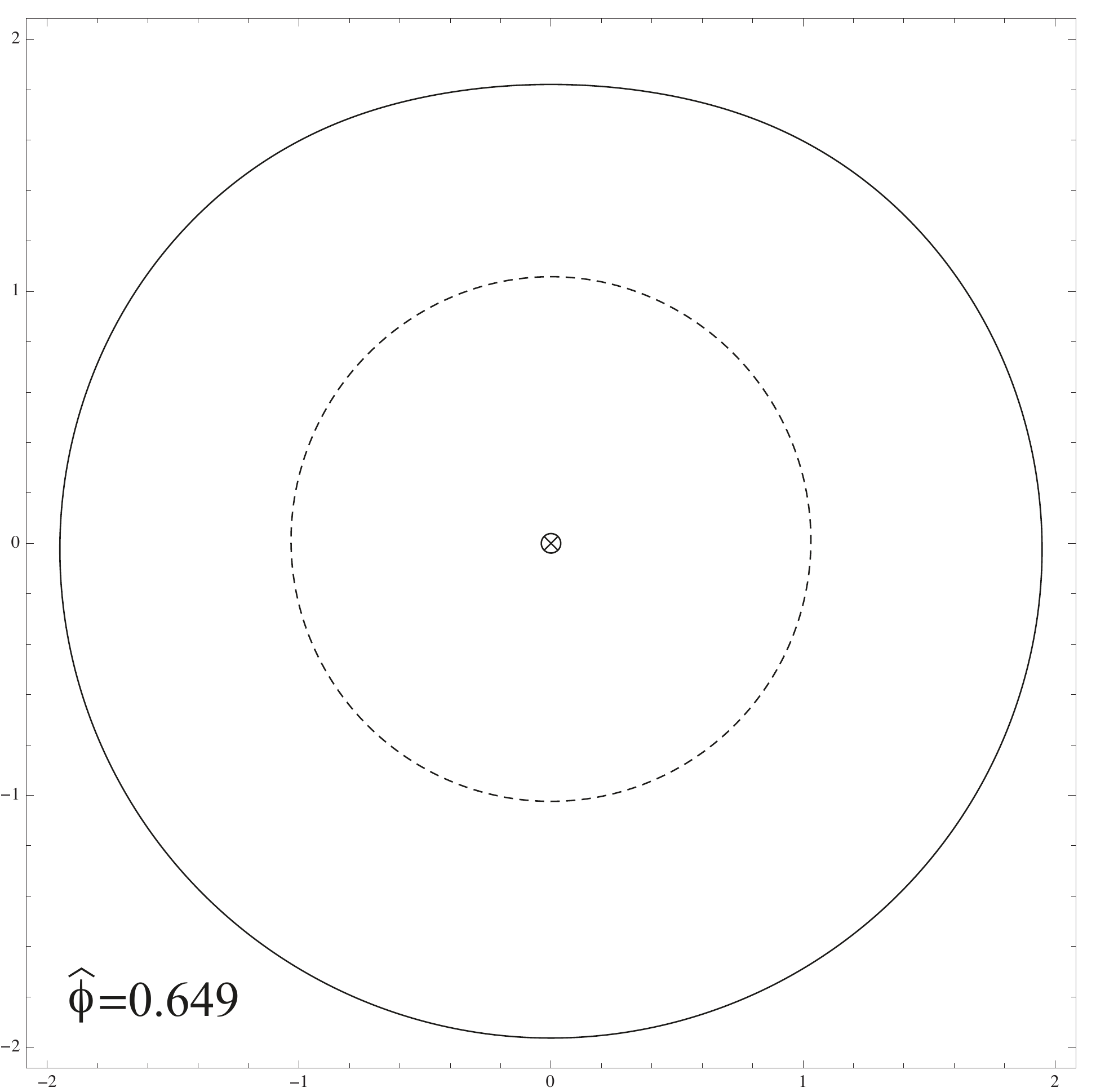} \hspace*{0.2cm}
\includegraphics[height=0.312\textheight]{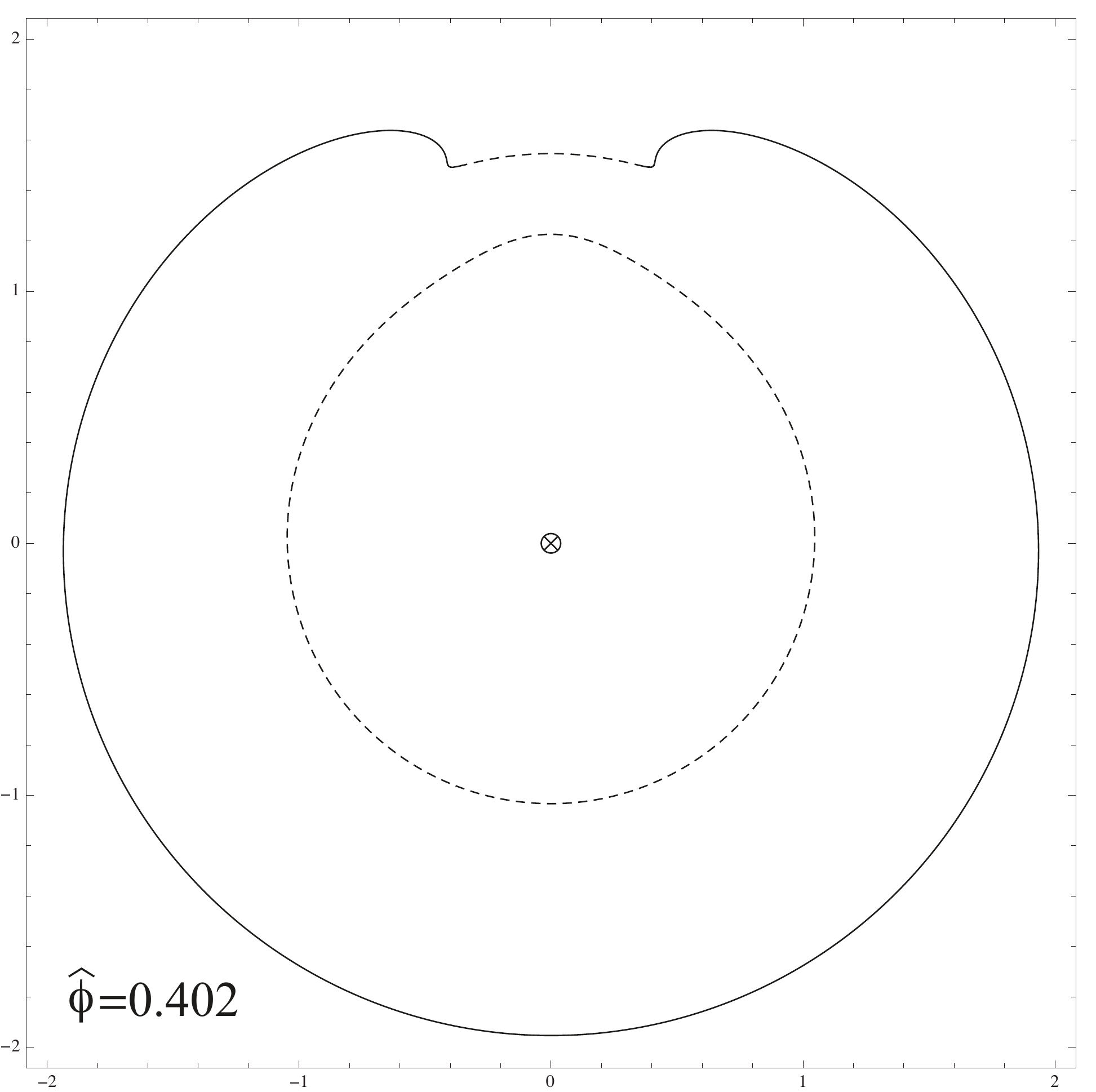} \\
\includegraphics[height=0.312\textheight]{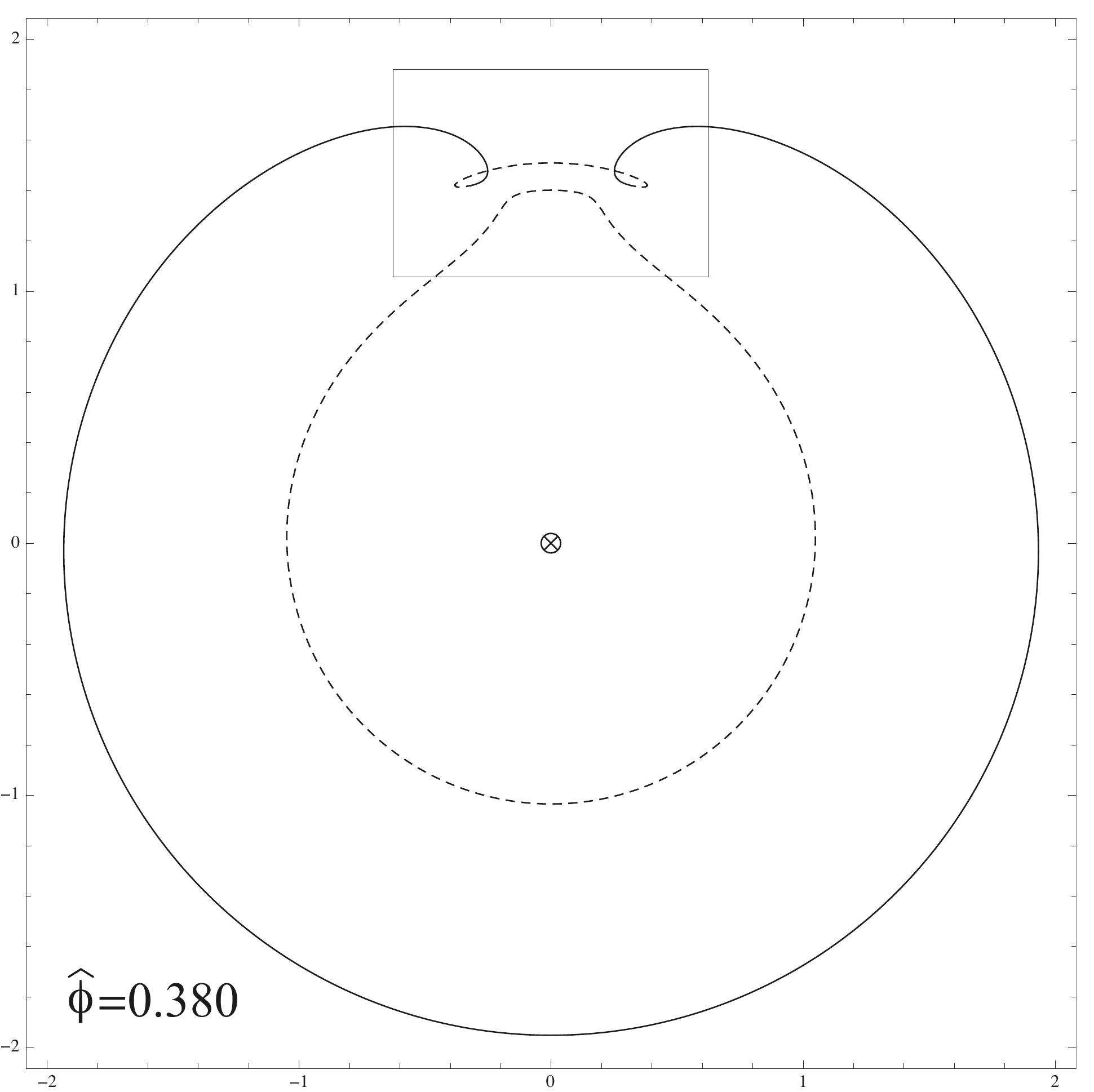} \hspace*{0.2cm}
\includegraphics[height=0.312\textheight]{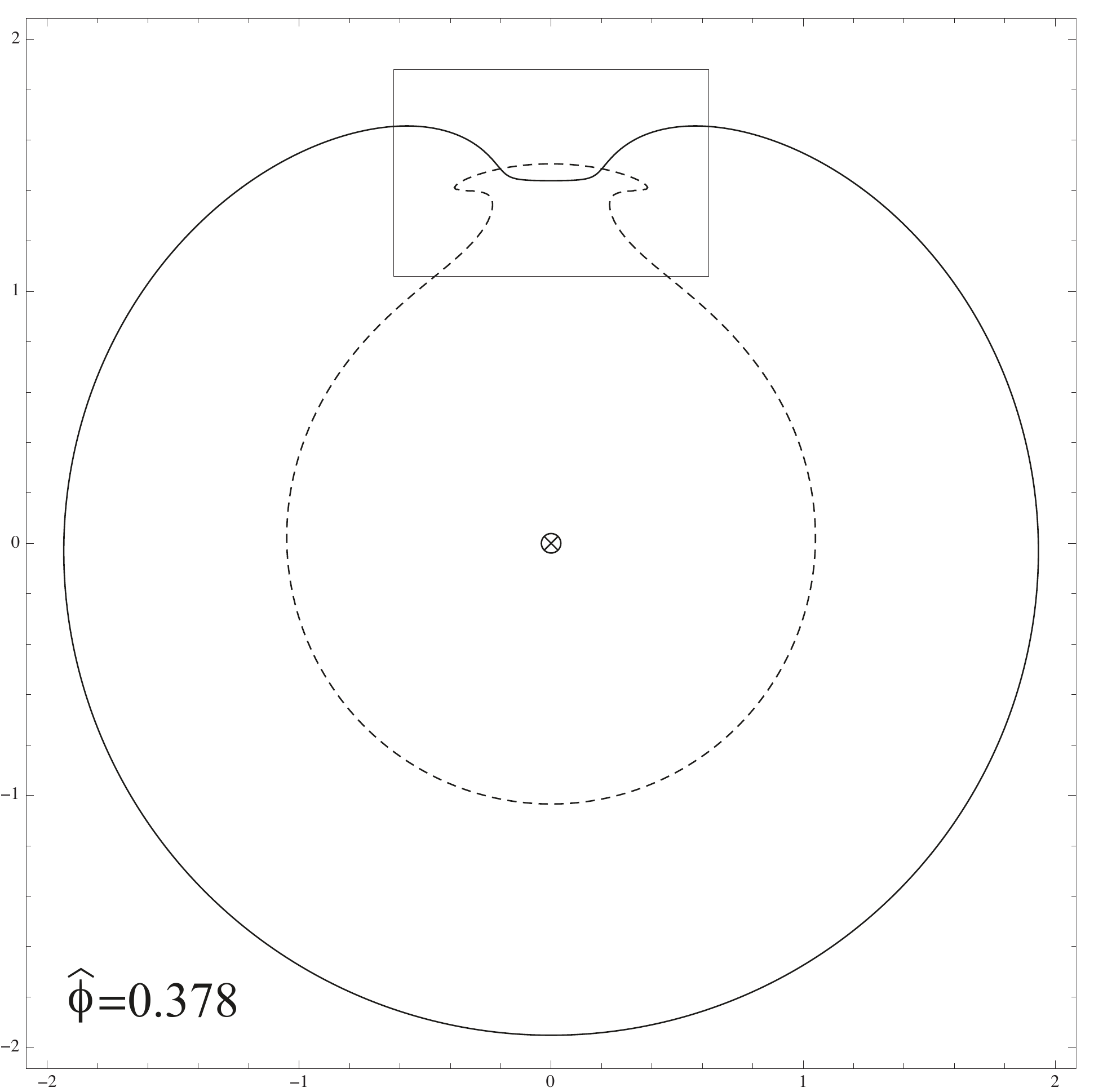} \\
\includegraphics[height=0.312\textheight]{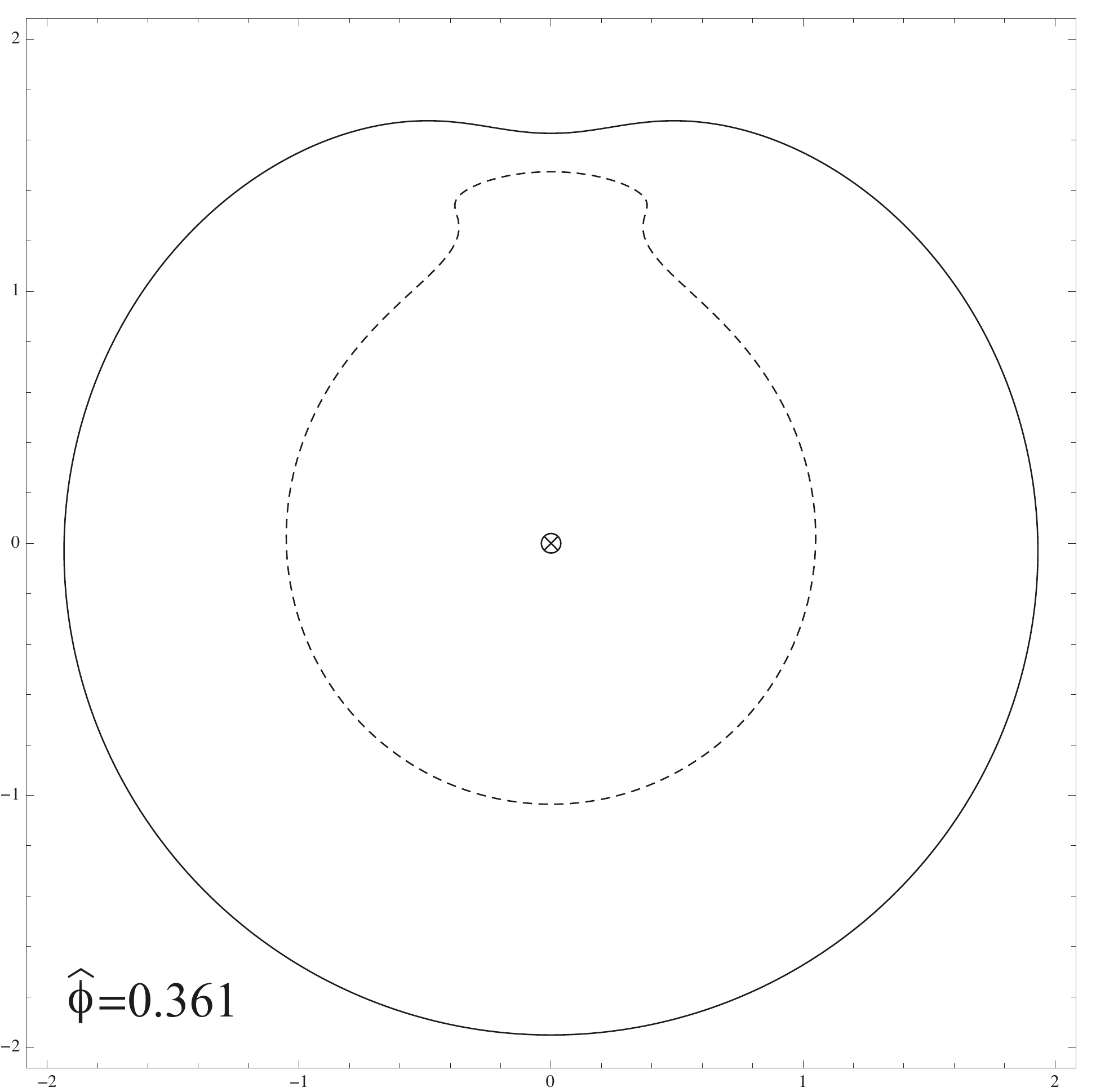} \hspace*{0.2cm}
\includegraphics[height=0.312\textheight]{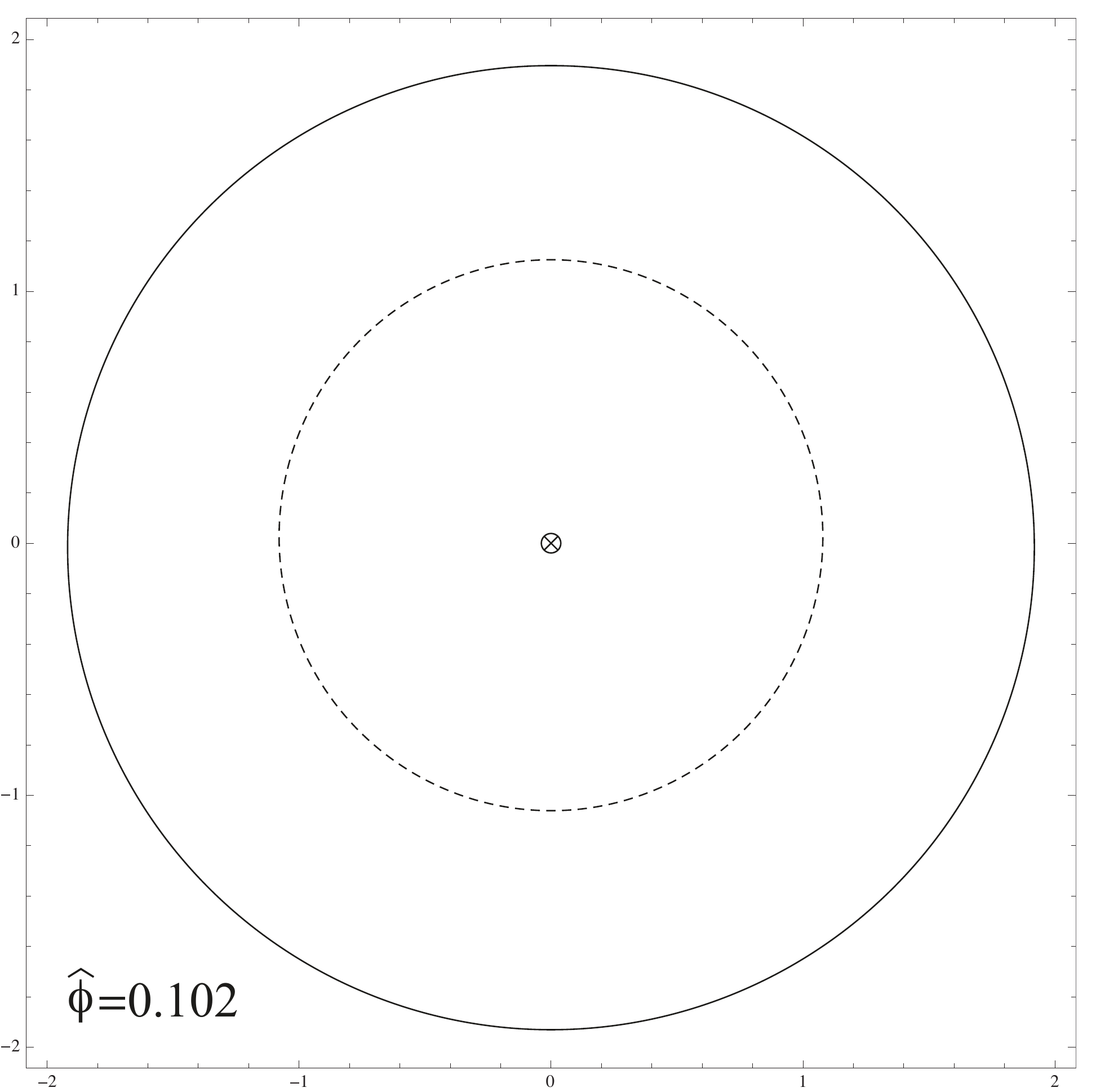} \\
\caption{Evolution of the membrane profiles for a process with two
membranes in the initial state and two in the final state  
($\theta_0=-\pi/6$).}
\label{RPmembranescatteringd}
\end{figure}

\begin{figure}[htbp]
\centering
\includegraphics[width=0.47\textwidth]{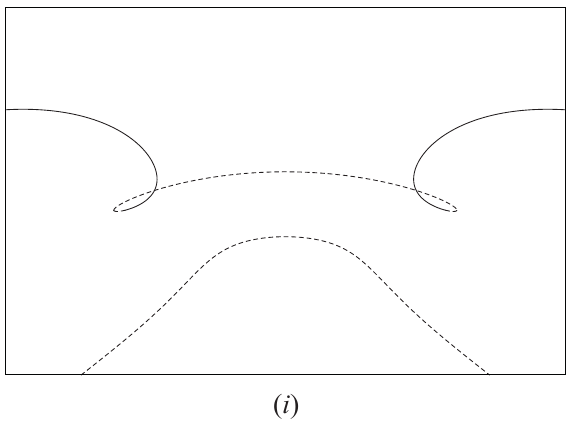} 
\hspace*{0.2cm}
\includegraphics[width=0.47\textwidth]{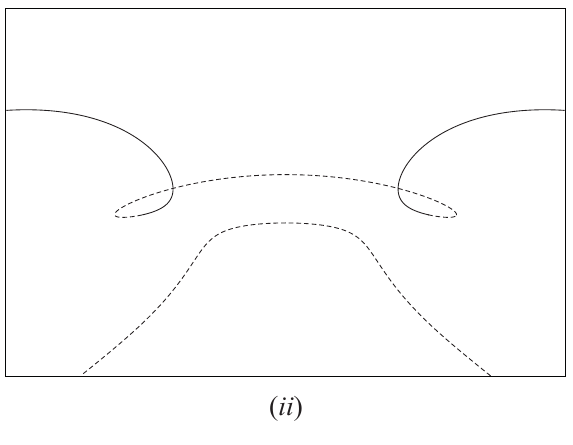} \\[1mm]
\includegraphics[width=0.47\textwidth]{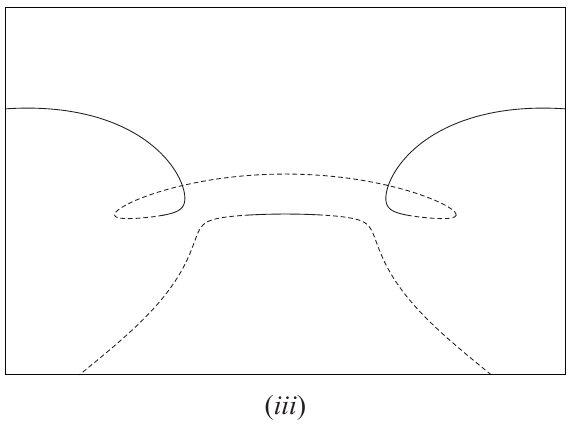} 
\hspace*{0.2cm}
\includegraphics[width=0.47\textwidth]{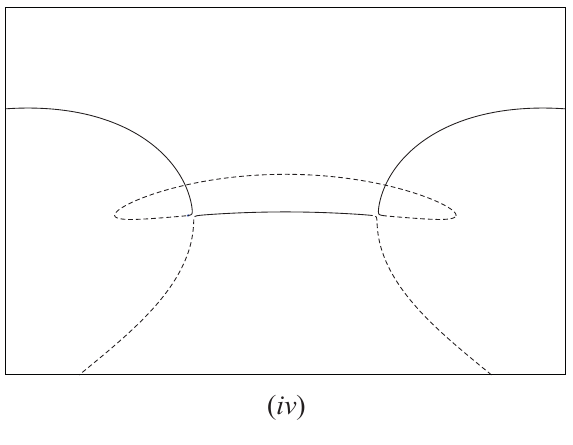} \\[1mm]
\includegraphics[width=0.47\textwidth]{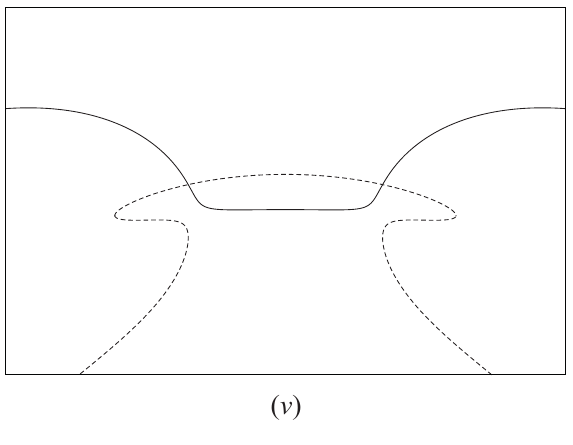} 
\hspace*{0.2cm}
\includegraphics[width=0.47\textwidth]{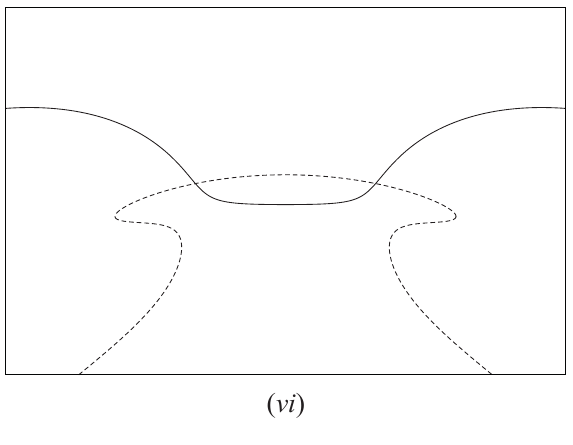} \\
\caption{Enlargement of the area marked by rectangles in the third and fourth 
panels of figure~\ref{RPmembranescatteringd}, showing intermediate steps
between the two.}
\label{RPmembranescatteringddetail}
\end{figure}

It is quite remarkable that the rather striking behaviour of membranes 
illustrated by the above examples may be reproduced by solutions of 
such a simple and well studied equation as the Laplace equation.

\subsubsection{Comments on the moduli space of solutions}
\label{RSSModuliSpaces}

In this subsection we discuss some features of the moduli space 
of instantons associated with general Riemman spaces. 
As we previously noted, the number of copies
of $\R^3$, $n$, corresponds to the number of membranes 
in the final state. 
Positive charges, $J_i \ge 0$ $(i=1, \ldots, n)$, can only be placed at the
origin of these spaces.
The charge $J_i$ corresponds to the angular momentum  
of the $i$-th membrane in the initial state.
Some of the $J_i$'s may be zero and 
the number of non-zero $J_i$'s is the
number of membranes in the initial state.

The sheets of the Riemann space are connected by  branch disks, 
bounded by  branch loops.
By definition when a branch disk connects two copies of $\R^3$ 
the disk is located at the same place, with the same shape,
in the two copies~\footnote{If this condition were not met,
we would get unphysical discontinuities in the
instanton solutions in the $\bm{y}$-space,
reconstructed using (\ref{RFContiBPSToContiNahm}).}. 
Except for this condition, the number of branch loops connecting the 
copies of $\R^3$, their positions and their shapes are arbitrary.
In the following we collectively refer to the number of copies of $\R^3$ and the number, the positions and the shapes of the branch loops as the 
geometry of the Riemann space.
Our conjecture is that for any geometry of the Riemann space
(and the distribution of positive charges at the origins) we have a unique
solution to the Laplace equation satisfying the boundary conditions.

A point on the instanton moduli space 
is characterised by 
specifying the geometry of the Riemann space.
This will give an interpretation of the 
moduli space of solutions of the BPS instanton equation (\ref{RFBPSEq})
discussed in \cite{RBBHP}
when $J$ is large (but finite).
Stated differently, the moduli space 
associated with (\ref{RFBPSEq}) can be considered as a regularisation
of the moduli space of branch loops.

The outgoing flux at infinity in the $i$-th copy of $\R^3$,
which we call $J_i'$,
corresponds to the angular momentum of the $i$-th 
membrane in the final state.
 
A necessary and sufficient condition 
for the existence of solutions to the instanton equations 
interpolating between the initial state
characterised by $J_i$, $i=1,\ldots,n$, and the 
final state characterised by $J_i'$, $i=1,\ldots,n$, 
was established in \cite{RBBHP}.
Below we will show that the same condition can be 
derived from our approach taking advantage, in particular, of the linearity
of the Laplace equation.

The linearity 
implies that for a fixed geometry of the Riemann space,
one has a linear relationship between $J_i$ and $J_i'$,
\begin{equation}
J_i' = \sum_j K_{ij} J_j \, .
\label{RFDefMatrixK}
\end{equation}
The matrix $K$ characterises the flow of the flux lines
for a given Riemann space.
(It is reminiscent of the coefficients of capacity in
the  electrostatic theory of conductors, which are determined by the shape 
and positions of the conductors.)
In general the matrix $K$ is not symmetric.
We will also use a matrix-vector notation
\begin{equation}
\bm{J}'= K \bm{J} \, ,
\end{equation}
where $\bm{J}=(J_1, \ldots, J_n)$,
$\bm{J'}=(J'_1, \ldots, J'_n)$.

Some simple but important properties follow from the definition of $K_{ij}$.
The first property is the positivity of the elements of the matrix $K$
\begin{equation}
K_{ij} \ge 0. \label{RFPositivityK}
\end{equation}
In order to see this, we consider the case in which there is only one unit 
charge, located at the origin of the $j$-th copy of $\R^3$, in the entire 
Riemann space. The flux flowing to the $i$-th space from the $j$-th space 
equals $K_{ij}$. Since there is a positive charge in the $j$-th space and no 
charge in the $i$-th space, it is clear that the flux should always flow from 
the $j$-th to the $i$-th space, not the other way around. This implies 
$K_{ij} \ge 0$. 

Other important properties of $K$ are the sum rules,
\begin{equation}
\sum_i K_{ij} = 1 \,  \quad \forall\,j \, , 
\label{RFSumRule1}
\end{equation}
\begin{equation}
\sum_j K_{ij} = 1\, \quad \forall\,i \, . 
\label{RFSumRule2}
\end{equation}
The property (\ref{RFSumRule1}) simply follows from the conservation 
of the number of flux lines (\ie the Laplace equation and Gauss' theorem).
The property (\ref{RFSumRule2}) can be deduced as follows.
We consider a special potential function in the Riemann space defined 
by the requirement that in each copy of $\R^3$ it is identical to the Coulomb 
potential associated with a unit charge at the origin. Such a potential solves 
the Laplace equation and satisfies all the boundary 
conditions~\footnote{
We note that this is a generalisation of the observation given below
(\ref{RFExplanationOf2PiShift}).
}. 
The existence of this solution implies that if $J_i=(1,1,1,\ldots,1)$,
then $J_i'=(1,1,1,\ldots,1)$ in (\ref{RFDefMatrixK}),
which is equivalent to (\ref{RFSumRule2}). 

The positivity (\ref{RFPositivityK})  and the sum rules 
(\ref{RFSumRule1})-(\ref{RFSumRule2}) imply rather strong constraints
on the allowed combinations of $J_i$'s and $J_i'$'s, including 
conditions equivalent to the criteria given in \cite{RBBHP} for the 
existence of solutions.

First, one can show that 
\begin{equation}
\sum_i {J_i'}^p \le \sum_i J_i^p  
\label{RFConditionPnorm}
\end{equation}
for any $p\ge 1$.
By using the vector norm $||\bm{v}||_p=(\sum_i |v_i|^p)^{1/p}$,
the matrix norm is defined by~\cite{RBGolubVanLoan}
\begin{equation}
||A||_p=\underset{\bm{v}\neq 0}{\mathrm{max}} \frac{||A\bm{v}||_p}{||\bm{v}||_p}.
\end{equation}
Hence in order to establish  (\ref{RFConditionPnorm}) 
it is sufficient to show
\begin{equation}
||K||_p \le 1. \label{RFMatrixNormK}
\end{equation}
Using an inequality for matrix norms (cf. formula (1.11) in \cite{RBHigham}),
\begin{equation}
||A||_p \le ||A||_{1}^{1/p} ||A||_{\infty}^{1-1/p}
\end{equation}
and the well-known formulae~\cite{RBGolubVanLoan}
\begin{equation}
||A||_1
=
\underset{1\le j\le n}{\mathrm{max}} 
\sum_{i=1}^{n}|A_{ij}|,\quad
||A||_\infty 
=
\underset{1\le i\le n}{\mathrm{max}} 
\sum_{j=1}^{n}|A_{ij}|,
\end{equation}
we see that $||K||_1=1$ and $||K||_\infty=1$ and hence 
(\ref{RFMatrixNormK}) and (\ref{RFConditionPnorm}) are established.
For $p=3$, we have
\begin{equation}
\sum_i {J_i'}^3 \le \sum_i J_i^3  \, ,
\label{RFJCubedDecreasing}
\end{equation}
which gives a continuum counterpart of the necessary condition 
(\ref{RFinstaction}) for the existence of instantons 
(which can also be derived directly by defining the
continuum version of (\ref{RFsuperpotential})).

\begin{figure}[htb]
\centering
\includegraphics[width=\textwidth]{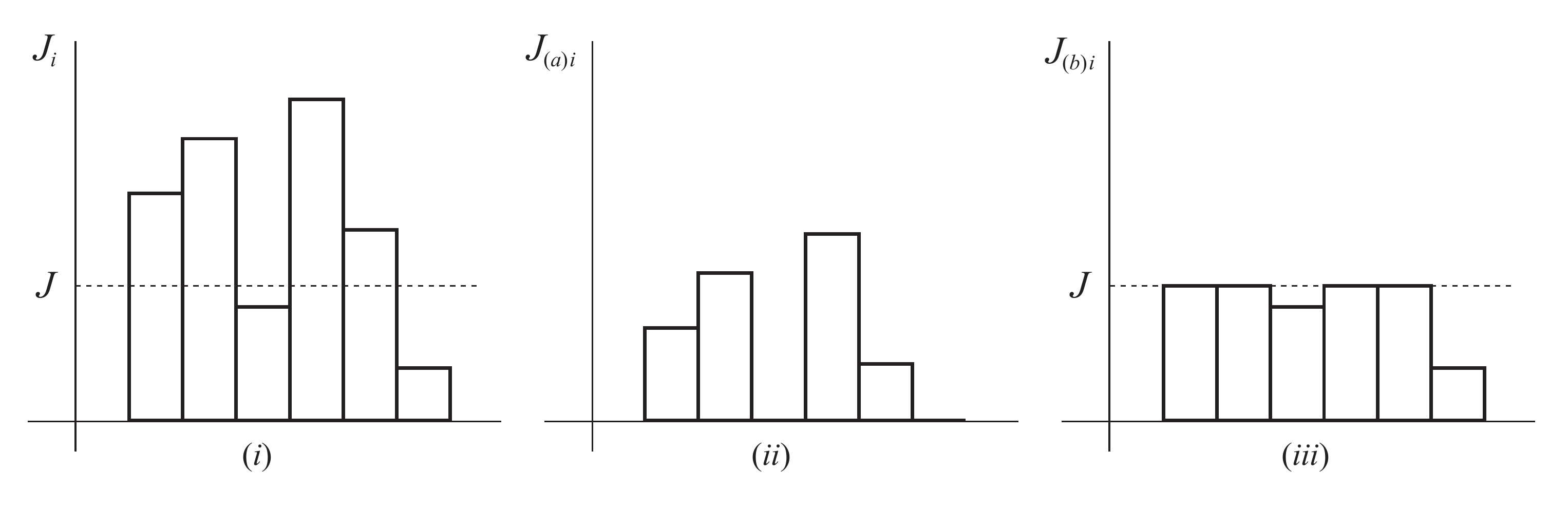}
\caption{
The angular momenta  of membranes $\bm{J}=(J_1,\ldots, J_n)$ can 
be represented as
a histogram as shown in ($i$). The area of the histogram $\bm{J}$ below 
a given height $J$ is denoted by $\mathcal{A}(J;\bm{J})$.
We define histograms $\bm{J}_{(a)}$ and $\bm{J}_{(b)}$ by cutting the
original histogram $\bm{J}$ along the horizontal line and taking 
the part above and below it, as shown in ($ii$) and ($iii$).
}
\label{RPHistogram}
\end{figure}

One can introduce an ``entropy'' associated with the distribution of the 
total angular momentum among the individual membranes defined as
\begin{equation}
\sum_i  \left(-J_i\log J_i\right)  \qquad {\rm and} \qquad 
\sum_i  \left(-J'_i\log J'_i \right)
\label{RFEntropydef}
\end{equation}
for the initial and final states respectively. Then by taking 
$p=1+\epsilon$ in (\ref{RFConditionPnorm}) and using the conservation 
law $\sum J_i=\sum J'_i$, 
one can show that the entropy is always non-decreasing,
\begin{equation}
\sum_i \left( -J'_i \log{J'_i} \right)
\ge
\sum_i \left( -J_i \log{J_i} \right) .
\label{RFEntropyIncrease}
\end{equation}
Let us now deduce the necessary and sufficient
condition for the existence of instantons given 
in \cite{RBBHP}.  The condition can be phrased as follows 
(cf. figure 3 in \cite{RBBHP}).
Since the $J_i$'s are non negative, they can be represented by a
histogram as shown in figure \ref{RPHistogram} ($i$). A similar histogram 
can be constructed with the $J'_i$'s in the final state.
We draw a horizontal line at a height $J\ge 0$ on the histogram 
and denote the area of the histogram 
below the value $J$ by
\begin{equation}
\mathcal{A}(J;\bm{J}).
\end{equation}
The condition in \cite{RBBHP} can then be expressed as
\begin{equation}
\mathcal{A}(J; \bm{J}') 
\ge \mathcal{A}(J; \bm{J}) \quad \forall \, J.
\label{RFBHPCondition}
\end{equation}
(We note that we do not assume $J_1\ge J_2\ge\ldots\ge J_n$
or $J'_1\ge J'_2\ge\ldots\ge J_n$
here.)
Thus, the area of the support of the histogram $J_i'$ is 
always larger than or equal to that of the histogram $J_i$.
An interpretation of this condition is that the histogram $J_i'$ 
should be more smeared compared to the histogram 
$J_i$, which is natural in view of (\ref{RFEntropyIncrease})~\footnote{
We note that the condition (\ref{RFBHPCondition})
is stronger than the condition (\ref{RFJCubedDecreasing}).
For example, 
the case with
$\bm{J}=(5,5,0)$ and $\bm{J'}=(6, 2, 2)$ 
is allowed by (\ref{RFJCubedDecreasing}),
but it is prohibited by (\ref{RFBHPCondition}).
}.

In order to prove (\ref{RFBHPCondition}) we introduce two 
auxiliary histograms, $J_{(a)i }$ and $J_{(b)i}$,
generated cutting the histogram $J_i$ at a fixed height $J$.
As shown in figure \ref{RPHistogram}, 
we define $J_{(a)i}$ and $J_{(b)i}$
as the histograms above and below the horizontal line
at the height $J$, respectively.
By definition we have
\begin{equation}
\bm{J}=\bm{J}_{(a)}+\bm{J}_{(b)}.
\end{equation}
Moreover
\begin{equation}
\bm{J'} = K \bm{J} = K \bm{J}_{(a)} + K \bm{J}_{(b)}.
\end{equation}
We first focus on
the histogram defined by $K \bm{J}_{(b)}$.
All columns in this histogram have height smaller than or equal to $J$, \ie
$(K \bm{J}_{(b)})_i \le J$. To see this we recall that
(\ref{RFSumRule2}) implies that
\begin{equation}
K 
\left[
 \begin{array}{c}
    J\\
    \vdots \\
    J 
 \end{array}
\right]
=
\left[
 \begin{array}{c}
    J\\
    \vdots \\
    J 
 \end{array}
\right].
\end{equation}
Since all elements of $\bm{J}_{(b)}$ are smaller than or equal to $J$ 
by definition and all elements of $K$ are positive (\ref{RFPositivityK}),
the elements of $K \bm{J}_{(b)}$ cannot be larger than $J$.

The sum rule  (\ref{RFSumRule1}) implies that area of the histogram 
$K \bm{J}_{(b)}$ is the same as the area of the histogram $\bm{J}_{(b)}$,
which by definition equals $\mathcal{A}(J;\bm{J})$.
Now recall that $\bm{J}'= K \bm{J}_{(a)} + K \bm{J}_{(b)}$.
Since the height of the histogram
$K \bm{J}_{(b)}$ is less than or equal to $J$,
all of the histogram $K \bm{J}_{(b)}$ contributes to 
$\mathcal{A}(J;\bm{J'})$.
The area $\mathcal{A}(J;\bm{J'})$ has a further contribution
from $K \bm{J}_{(a)}$, 
which is always non-negative,
because of the positivity of $K$ and $J_{(a)}$.
Thus, we have
\begin{equation}
\mathcal{A}(J;\bm{J'}) \ge 
\mathcal{A}(J;K \bm{J}_{(b)}) 
=
\mathcal{A}(J;\bm{J}),
\end{equation}
and
(\ref{RFBHPCondition}) is proven.

Another interesting property of the moduli space
of solutions to the BPS instanton equation (\ref{RFBPSEq}),
which was established in \cite{RBBHP}, 
is the additivity rule (\ref{RFAdditivity}) for the dimension of the moduli 
spaces. A possible interpretation
of this result, which is instructive if perhaps not mathematically rigorous, 
is the following. 
We assume that we have two instanton solutions,
interpolating between vacua A and B,
and between vacua B and C, such that
both of them are reasonably localised in time.
It is natural to expect that one can construct 
an instanton solution interpolating
between vacuum A and vacuum C, 
by stitching together the two original solutions
taken to be well separated in time.
The number of parameters of the composite instanton 
thus constructed should be given by the
sum of the numbers of parameters of the two original instantons.

This construction of a composite instanton 
from two instantons well separated in time
has a natural counterpart in our description of the instantons using
Riemann spaces. 
The following argument relies on viewing a solution in a multi-sheeted 
Riemann space as a composite of solutions built using pairs of connected 
copies of $\R^3$. Within this picture the basic idea underlying the 
construction is that, at least in the asymptotic region at infinity, a branch disk 
with incoming flux can be viewed as a point charge. 
Here we wish to point out the essential features of this construction 
without going into the details.
In the following it is crucial that 
the time variable  in the BPS instanton
corresponds to the logarithm of
the distance from the origin in our description.
We start with two instanton solutions,
each associated with the branch disks $D_1$ and $D_2$.
Because of the 
scale invariance of the Laplace equation, 
one can move one of the disks, \eg $D_2$, further away from the origin
by rescaling simultaneously its size and its distance from the origin.
The physics associated with the second disk, such as  
the number of flux lines going into the disk and the profile of the potential 
function 
do not change under the rescaling.
We refer to the rescaled disk as $D_2'$.
Now we consider a new Riemann space equipped with two branch disks, 
$D_1$ and $D_2'$.
Since the disk $D_2'$ is sufficiently far away, its existence does not affect
the potential produced by the point charge at the origin and the first branch 
disk $D_1$.
Furthermore, from the viewpoint of the second disk $D_2'$, 
the disk $D_1$ and the point charge appear as a new point charge 
at the origin whose value can be reconstructed using the original charge and 
the number of flux lines going through the disk $D_1$.
In this manner, the
potential on the Riemann space with two disks ($D_1$, $D_2'$) can 
be constructed by using the potentials corresponding to the two 
separate solutions associated with the disks $D_1$ and $D_2$.

As previously mentioned, the calculation of physical transition amplitudes
using a semi-classical approximation involves the integration over
the collective coordinates parameterising the instanton moduli space.
From the point of view of the description of the moduli space in terms
of the geometry of Riemann spaces, this means that one should sum the
contributions of all possible combinations of branch surfaces giving rise
to the same initial and final states. Such a sum may include not only
branch disks of more complicated shapes than the circular ones we
discussed, but also in principle branch surfaces with more complicated
topology, \eg consisting of higher genus surfaces with boundaries.
It would be interesting to study the characteristics of the resulting Riemann spaces.

\section{Conclusions and discussion}
\label{RSConclusion}

In this paper we have initiated a study of the interactions associated with 
membrane splitting/joining processes in M-theory. We considered a large
angular momentum sector of M-theory in AdS$_4\times S^7$, which can be
described using the matrix model approach in the pp-wave approximation.
The vacua of the pp-wave matrix model consist of collections of concentric
membranes and contributions to splitting or joining processes arise from 
tunnelling amplitudes between states built on vacua with different numbers
of membranes. Classical tunnelling configurations are instantons, \ie finite 
action solutions to the Euclidean equations of motion of the pp-wave matrix 
model. Relying on the same ideas underlying the definition of the matrix
model as regularisation of the supermembrane theory, we have presented a 
method for the construction of approximate solutions to the instanton 
equations. Using a continuum formulation in which large matrices are 
approximated by functions, we have shown that the problem of constructing 
solutions to the instanton equations of the pp-wave matrix model can be 
reduced to that of solving the three-dimensional Laplace equation.  A 
remarkable outcome of our analysis is that in order to describe membrane 
splitting/joining processes it is necessary to study the Laplace equation in a 
Riemann space, which is a three-dimensional generalisation of the familiar 
concept of Riemann surface. The interpolating configurations at different 
points in Euclidean time arise as equipotential surfaces for the solutions to the 
Laplace equations in a Riemann space with suitable boundary conditions. 
The use of Riemann spaces provides a very compelling answer to various 
issues arising in the description of membrane splitting processes in terms of 
the Laplace equation. 

The approach we developed in this paper presents
some similarities to the description of splitting/joining interactions
in the matrix string formulation of type IIA string 
theory~\cite{RBDVV} in terms of instantons in two-dimensional SYM 
theory~\cite{RBMatrixStringInstantons1,RBMatrixStringInstantons2,
RBMatrixStringInstantons3}. 
There are also some analogies between our work 
and~\cite{RBSasakuraSugimoto}, in which 1/4 BPS  
solitons in $\mathcal{N}=4$ SYM
are described using multi-pronged $(p,q)$ string states. The 
latter are constructed using the two-dimensional Laplace equation
which arises from configurations of M2-branes stretched between 
M5-branes.

In section~\ref{RSSolutions} we presented exact solutions to the Laplace
equation, which represent approximate solutions to the original BPS instanton
equations valid for large $J$. The instanton configurations in the matrix 
model are saddle points for the Euclidean path integral of the theory 
and can be used to  compute contributions to physical transition amplitudes 
using a semi-classical approximation. As a next step it will be important to 
deduce the exact form of the solutions in the matrix model, associated with 
the approximate solutions we obtained using the continuum approximation. 
The connection between the instanton equations and the Laplace equation in 
the continuum approximation is established mapping the former to the Nahm 
equations. This means that in the process of reconstructing the matrix model 
solutions we should be able to take advantage of the integrability 
properties of the Nahm equations. 

Semi-classical contributions to physical transition amplitudes in the 
Lorentzian signature pp-wave matrix model should be obtained from
a saddle point approximation about the instanton configurations followed
by an analytic continuation. The semi-classical calculation involves an 
integration over the moduli space of the BPS instantons, whose properties
were studied in~\cite{RBBHP,RBYeeYi}. Completing the semi-classical
analysis will allow us to extract effective interaction vertices for
membranes in the pp-wave background. A first interesting issue to address
is the determination of the associated coupling constants. Naively one 
would expect instanton induced transition amplitudes to be exponentially
suppressed, due to the contribution of the classical instanton action. For
instance the most elementary amplitude, corresponding to the splitting of
a single membrane with angular momentum $J$ into two membranes with
angular momenta $J_1$ and $J_2$, carries a weight
\begin{equation}
\er^{-S_E} = \er^{-\frac{JJ_1J_2}{8N}} \, ,
\label{RFinstantonweight}
\end{equation}
where $J_1 \sim J_2 \sim J$ and $N^{1/3} \ll J \ll N^{1/2}$ for the applicability 
of the pp-wave approximation in a weak coupling regime. 
However, it is probably premature to conclude that the coupling constant 
measuring the strength of membrane interactions is exponentially small.
The dependence on the parameters can be modified by factors arising from 
the integration over the instanton moduli space, which has a dimension that 
grows with $J$ -- for example it is $4J_2$ in the case of the single membrane 
splitting, see~(\ref{RFDimM12}). Volume factors can modify the exponential 
behaviour~(\ref{RFinstantonweight}) and it is possible, at large $J$ in 
particular, for the complete physical transition amplitudes to have a power-law 
dependence on the parameters $J$ and $N$, which might be more natural 
from the point of view of a description of membrane interactions in target 
space. A similar mechanism, leading to a power-law behaviour, has been 
recently suggested to arise in the calculation of solitonic contributions to 
certain scattering amplitudes~\cite{RBPapagRoyst1} and may be relevant for 
the understanding of the connections between the maximally supersymmetric 
Yang-Mills theory in five dimensions and the $(2,0)$ superconformal theory 
in six dimensions~\cite{RBPapagRoyst2,RBLambPapagSchmidt}. 

The particular solution~(\ref{RFBHPSolution}) constructed in~\cite{RBBHP},
which approaches the configuration corresponding to 
the $J$-dimensional representation 
$\ul{1}\oplus\ul{1}\oplus\cdots\oplus\ul{1}$ at $t=+\infty$, was studied
in the context of the pp-wave matrix model in~\cite{RBYeeYi}. According
to a proposal presented in~\cite{RBMSVR}, the configuration associated
with the representation $\ul{1}\oplus\ul{1}\oplus\cdots\oplus\ul{1}$ 
should be interpreted as a single M5-brane and therefore the 
solution~(\ref{RFBHPSolution}) would contribute to an amplitude coupling
M2- and M5-branes. It would be interesting to determine whether such a 
coupling has a different weight in terms of the $J$ and $N$ parameters, 
compared to those involving only M2-branes.  

The study of instanton effects in another maximally supersymmetric theory,
$\calN=4$ SYM with SU($N$) gauge group, suggests that  the large $J$ 
limit, which plays a central role in our continuum approximation, may also 
lead to important simplifications in the semi-classical calculation of physical 
transition amplitudes. In the case of $\calN=4$ SYM at large $N$ the 
integration over the multi-instanton moduli space is dominated by special 
configurations and a saddle point approximation makes it possible to compute 
the leading contributions at large $N$ for arbitrary instanton 
number~\cite{RBDoreyEtal}. It would be very interesting to understand if a 
similar saddle point based approach can be applied in the present case for 
large $J$. 

The following considerations, albeit somewhat speculative,
may provide insights into possible mechanisms leading to 
the emergence of a saddle point approximation for large $J$.
Among the collective coordinates parameterising the instanton moduli space
there are variables  associated with the SU($J$) gauge orientations, which
are the remnant in the matrix model of the invariance of the membrane theory 
under area preserving diffeomorphisms (APD's). When computing 
contributions to gauge invariant observables the integrations over these 
colour-related variables can be carried out and in general produce a 
non-trivial measure for the integrals over the remaining collective coordinates, 
that can be referred to as the `physical' ones. At large $J$ this measure can 
be very peaked and give a large weight to special regions of the physical 
moduli space. In such a situation it is possible to expect the emergence of a 
good saddle point approximation. Given the fact that the SU($J$) gauge 
symmetry descends from the APD invariance in the continuum, it may be 
natural to expect that the configurations acquiring the largest weight in the 
physical moduli space should be the smoothest ones, since they are the ones 
for which the finite SU($J$) symmetry is closer to the infinite dimensional 
group of APD transformations. In our description of instanton 
configurations in terms of the Laplace equation on a Riemann space the
collective coordinates on the instanton moduli space are associated with
the geometry of the Riemann space. Then the above reasoning suggests 
that contributions arising from the smoothest branch loops may be the 
dominant ones. Hence for the specific transition amplitude corresponding to 
the splitting of one membrane into two, the Riemann space with circular 
branch disk used in the Hobson solution that we employed in 
section~\ref{RSSolutions} should presumably provide the dominant 
contribution.

Supersymmetry of the pp-wave matrix model is important in various 
aspects of the calculation of instanton contributions to the transition 
amplitudes that we are discussing. It is expected to be responsible for 
the cancellation of the determinants arising from the integration
over non-zero mode fluctuations in the instanton background. Moreover the
instanton moduli space includes fermionic collective coordinates associated
with supersymmetries broken by the instanton configurations. Although the 
particular features of the supersymmetry algebra in the pp-wave background 
make the analysis rather subtle, we expect the presence of fermion zero 
modes to restrict the set of  allowed instanton amplitudes. In addition to the 
conditions for the existence of interpolating instanton configurations 
discussed in~\cite{RBBHP}, further selection rules should arise from the 
integration over the fermionic collective coordinates. This was shown 
in~\cite{RBYeeYi} to be the case for the special 
solution~(\ref{RFBHPSolution}) obtained in~\cite{RBBHP}. In general the 
results of~\cite{RBDSVR2}, showing that the vacua of the pp-wave matrix 
model are non-perturbatively protected, indicate that instanton induced 
transition amplitudes can be non-zero only when excited states are involved.
It should be possible to gain further insights into the general selection rules 
which constrain these processes by studying the fermion zero modes in the 
background of the general matrix  model solutions corresponding to those 
discussed in this paper in the continuum approximation. Similarly to what 
happens in the case of $\calN=4$ SYM~\cite{RBinstadims} it should be 
possible to carry out this analysis even without a complete calculation of the 
transition amplitudes. 

Various properties of the moduli space of solutions of the BPS instanton 
equations were discussed in~\cite{RBBHP,RBYeeYi}. The collective 
coordinates parameterising the matrix model solutions have counterparts in 
the parameters charactering the solutions to the Laplace equation that we 
used our approach. The description in terms of the Laplace equation 
provides a geometric and intuitive interpretation of some of these 
parameters and the linearity of the equation is useful in explaining certain 
features of the solutions. According to our proposal the configurations 
interpolating between different vacua of the pp-wave matrix model have an 
approximate description in terms of equipotential surfaces of potential 
functions solving the Laplace equation in certain Riemann spaces. Any given 
solution is characterised by the states at $t=\pm\infty$ in the original picture, 
\ie by the asymptotic SU(2) representations defined by the dimensions, 
$J_i$, $i=1,\ldots,m$, and $J'_k$, $k=1,\ldots,n$, of the irreducible blocks at 
$t=-\infty$ and $t=+\infty$ respectively. In our description the values of the 
$J_i$ integers for the initial state correspond to point charges at the origin in 
different copies of $\R^3$, while the $J'_k$'s for the final state correspond to 
the fluxes of the gradient of the potential at infinity in each of the $\R^3$'s 
which constitute the Riemann space. 
In such a picture the information on the
parameters characterising the solutions should be encoded in the geometry
of the Riemann space, \eg the shape and relative position of the branch 
disks. In section~\ref{RSSSolutionGeneral} we discussed some properties
of the moduli space of the BPS instantons from the point of view of the 
Riemann space picture. It would be interesting to have a more complete
understanding of the results of~\cite{RBBHP} within the framework 
proposed in this paper. 
In \cite{RBBHP}, it was shown that the BPS equations,
with the same boundary conditions considered in this paper,
describe domain wall solutions in the $\calN=1^*$ SYM
theory and also the D3-D1 system in the context of the AdS/CFT
correspondence. It would be interesting to consider the implications of our 
results in these contexts.

The matrix model, providing a regularisation of the supermembrane
theory, represents the proper framework for a full quantum mechanical 
calculation of physical transition amplitudes of the type we discussed in 
this paper. However, it may be possible to carry out the semi-classical 
calculation using as 
starting point the path integral for the continuum 
supermembrane theory. In the semi-classical approximation the calculation
is reduced to an integration over the instanton moduli space. It may then
be possible to introduce an appropriate regularisation, suitable to make the  
integration over the moduli space of the continuum solutions well defined. 
In this way an approximate result, valid for large $J$, might be attained 
working directly with our continuum solutions without reconstructing the 
explicit form of the corresponding matrix model configurations. Such a 
possibility is made more plausible by the large amount of supersymmetry
in the theory. The use of this continuum approximation can probably be 
justified more straightforwardly if indeed the integration over the moduli space
is dominated by a saddle point for large $J$.  

The study of interactions between M-theory objects using the matrix model 
formulation has so far been mostly limited to processes involving no 
longitudinal momentum transfer. These calculations allow to extract effective
supergravity interactions from perturbative calculations in the matrix model. 
A comprehensive review of these results and a detailed list of references 
can be found in~\cite{RBTaylorReview}. A contribution to the scattering of 
membranes with minimal momentum transfer in the M-theory direction was 
carried out in~\cite{RBPolchinskiPouliot}. The process studied 
in~\cite{RBPolchinskiPouliot} -- described in terms of transfer of a D0-brane 
between two parallel D2-branes -- can be considered as analogous (in a 
special kinematical regime) to the one corresponding to the solution with two 
membranes in the initial state and two in the final state discussed in 
section~\ref{RSSSolutionGeneral} -- more specifically the example shown in
figure~\ref{RPmembranescatteringe}. However, we emphasise that the 
approach proposed in this paper is more general, since it allows us to 
describe processes with an arbitrary number of membranes in the initial and 
final states. Moreover we can describe interactions involving the exchange
of a large amount of longitudinal momentum. In particular, in cases such
as the example shown in figure~\ref{RPmembranescatteringe} we have
an exchange of a genuine M-theory object, which in the D0-brane picture
would require considering a configuration involving a large number of 
D0-branes. In a calculation of the type presented 
in~\cite{RBPolchinskiPouliot} this would mean evaluating contributions
from sectors with large instanton number. 
Further work on interactions involving longitudinal momentum transfer was 
done in~\cite{RBP11transfer}. 

Our results on membrane splitting/joining transitions provide a foundation for 
the systematic study of more general interactions of membranes in M-theory. 
The existence of classical solutions corresponding to configurations that
interpolate between states containing different numbers of membranes,
supports the conclusion that the matrix model is capable of describing 
splitting/joining interactions, reinforcing the early indications offered by the 
results of~\cite{RBPolchinskiPouliot,RBP11transfer}. To provide a truly 
convincing argument it is of course necessary to complete the calculation of 
the physical transition amplitudes in semi-classical approximation. 
Moreover, it is important to be able to check the results by independent 
means and the AdS/CFT duality of~\cite{RBABJM} provides a way of 
achieving this. Based on the dictionary for the large $J$ sector established 
in~\cite{RBKSS1}, the vertices associated with membrane splitting/joining
interactions can be related to correlation functions of monopole operators 
in the dual CFT. As pointed out in~\cite{RBKSS1}, the use of radial 
quantisation is a convenient tool for the study of monopole operators and their 
correlators in the ABJM theory. The relevant monopole operators are 
analogous to disorder operators in two-dimensional 
CFT. To compute CFT quantities that can be
related to membrane interaction vertices one needs to consider amplitudes
involving multiple insertions of such operators. It is an interesting possibility 
that Riemann spaces may be useful for the analysis of correlation 
functions of this type in the radially quantised ABJM theory. Moreover,
as is the case for the study of the spectrum~\cite{RBKSS1}, 
the presence of the large parameter $J$ should make it possible to 
develop an approximation scheme allowing a reliable comparison with the 
matrix model results. A similar approximation, valid in a sector characterised  
by a large global charge $J$, was recently discussed in~\cite{RBHORW}
in the case of a different class of strongly coupled three-dimensional theories. 

An alternative way of comparing the effective membrane interaction vertices
to the dual CFT may be based on the method proposed 
in~\cite{RBGomisMoriyamaPark}
for string theory in a pp-wave background. 
In this approach membrane couplings would be related to matrix elements
of the dilation operator of the ABJM theory between states associated with 
monopole operators. It should be possible to compute such matrix elements
in semi-classical approximation, using tunnelling configurations
interpolating between monopole states characterised by different GNO 
charges. It is natural to expect that the Nahm equations might play a role in 
a gauge theory calculation of this type as well. 
We note that certain BPS equations associated with the description of 
bound states of M2- and M5-branes in the ABJM theory were studied 
in~\cite{RBNosakaTerashima, RBSakaiTerashima}, 
where interesting relations to the Nahm equations were pointed out. 

It would be very interesting to generalise the type of analysis that 
we developed for the pp-wave matrix model to other cases, relevant 
for M-theory in different backgrounds. The cases of the AdS$_4\times S^7$ 
and AdS$_7\times S^4$ spaces are of course of interest. For these 
backgrounds no matrix model formulation is known, however, the 
supermembrane action has been 
constructed~\cite{RBdeWitPeetersPlefkaSevrin}. Therefore at least a study 
of instanton solutions using our approach might be possible,
since the continuum version of the instanton equations
can be also understood directly from the membrane theory without reference
to the matrix model as discussed in section 
\ref{RSSClassicalMembranePerspective}~\footnote{
For the membrane instantons in the AdS space, solutions 
with run-away behaviour (discussed at the end of section \ref{RSSSolutionCoulomb})
may be relevant in view of the interpretation of the change of variables
mentioned in footnote \ref{RFootnoteChangeVariable}.
}.
Another obvious 
case to consider is flat space, which may be approached starting from the 
general pp-wave matrix model described in 
appendix~\ref{RSAGeneralisedFormulae} and using the mass parameter $
\mu$ as an infra-red regulator. The $\mu\to 0$ limit would, however, be very 
subtle and require extreme caution. 

The emergence of the notion of Riemann space in connection with membrane 
interactions is very suggestive. We were led to introduce this concept in 
order to describe the evolution of membrane configurations in terms of 
solutions to the Laplace equation. It is an intriguing possibility that 
our results may be an indication of more general features of the dynamics of 
membranes and that Riemann spaces may turn out to be a central ingredient 
in the description of membrane interactions, in the same way as Riemann 
surfaces are essential in the formulation of string perturbation theory as 
a genus expansion.

\vsp{0.7}
\ndt
{\bf Acknowledgments}

\vsp{0.3}
\ndt
We would like to thank 
Sudarshan~Ananth, 
Yuhma~Asano, 
Nathan Berkovits,
Massimo Bianchi,
Veselin Filev, 
Masafumi~Fukuma,
Masanori~Hanada, 
Shinji~Hirano, 
Jens~Hoppe, 
Nobuyuki~Ishibashi,
Goro~Ishiki, Satoshi~Iso, 
Hiroshi~Itoyama,
Hikaru~Kawai, Yoichi~Kazama,
Shota~Komatsu,
Tsunehide~Kuroki,
Marianne Leitner,
Andrey Mikhaylov,
Tristan McLoughlin,
Takeshi~Morita,
Sanefumi~Moriyama,
Werner Nahm, 
Horatiu Nastase, 
Naoki~Sasakura, 
Niels~Obers, 
Jo\~ao~P.~Rodrigues,
Yuji~Sato, Fumihiko~Sugino, Shigeki~Sugimoto, 
Shotaro~Sugishita,
Asato~Tsuchiya,
Tamiaki~Yoneya for encouragement, discussions and 
comments.
HS would like to thank Jens~Hoppe 
for discussions on interactions of membranes and the matrix regularisation,
in particular for letting him know about references
\cite{RBBHP, RBWard, RBHoppeLaplaceEq}.
SK visited the ICTP-SAIFR, S\~ao Paulo, Brazil, during the completion of
this work. He would like to thank all members of the institute and Nathan
Berkovits in particular for the warm hospitality.
YS would like to thank the members in the Radboud University, Nijmegen, 
where part of this work was done, for the kind hospitality.

\appendix

\section{
Geometric proof of the equivalence between Laplace and 
continuum Nahm equations 
}
\label{RSAGeometricalDerivationCNahmtoLaplace}
In this appendix we provide a proof of the
equivalence between the three-dimensional Laplace equation 
and the continuum Nahm equations~(\ref{RFContinuumnahm}). 
The latter are sometimes also referred to as SU($\infty$) Nahm equations. 
The argument that we present is more geometric than those in 
\cite{RBWard, RBHoppeLaplaceEq}. 
We start with the continuum Nahm equations,
\begin{equation}
\frac{\partial z^i}{\partial s} 
= - \alpha \frac{1}{2}\e^{ijk} \{ z^j, z^k \},
\label{RFContinuumNahm}
\end{equation}
where $\alpha$ is a positive constant 
and the Lie bracket is defined as 
$\{f,g \}=\frac{\partial f_1}{\partial \s^1}\frac{\partial g}{\partial \s^2} 
- \frac{\partial g}{\partial \s^1}\frac{\partial f}{\partial \s^2}$ 
for arbitrary functions, $f$ and $g$.
This equation describes the evolution in time $s$ of a two-dimensional 
surface parametrised by $(\s^1,\s^2)$. 
Locally this evolution defines a function $\phi(z^i)=s(z^i)$ on a 
region of the three-dimensional
$z$-space swept by the evolving surface.
We prove that the function $\phi=s$ satisfies the Laplace equation.

\begin{figure}[htb]
\centering
\includegraphics[width=0.9\textwidth]{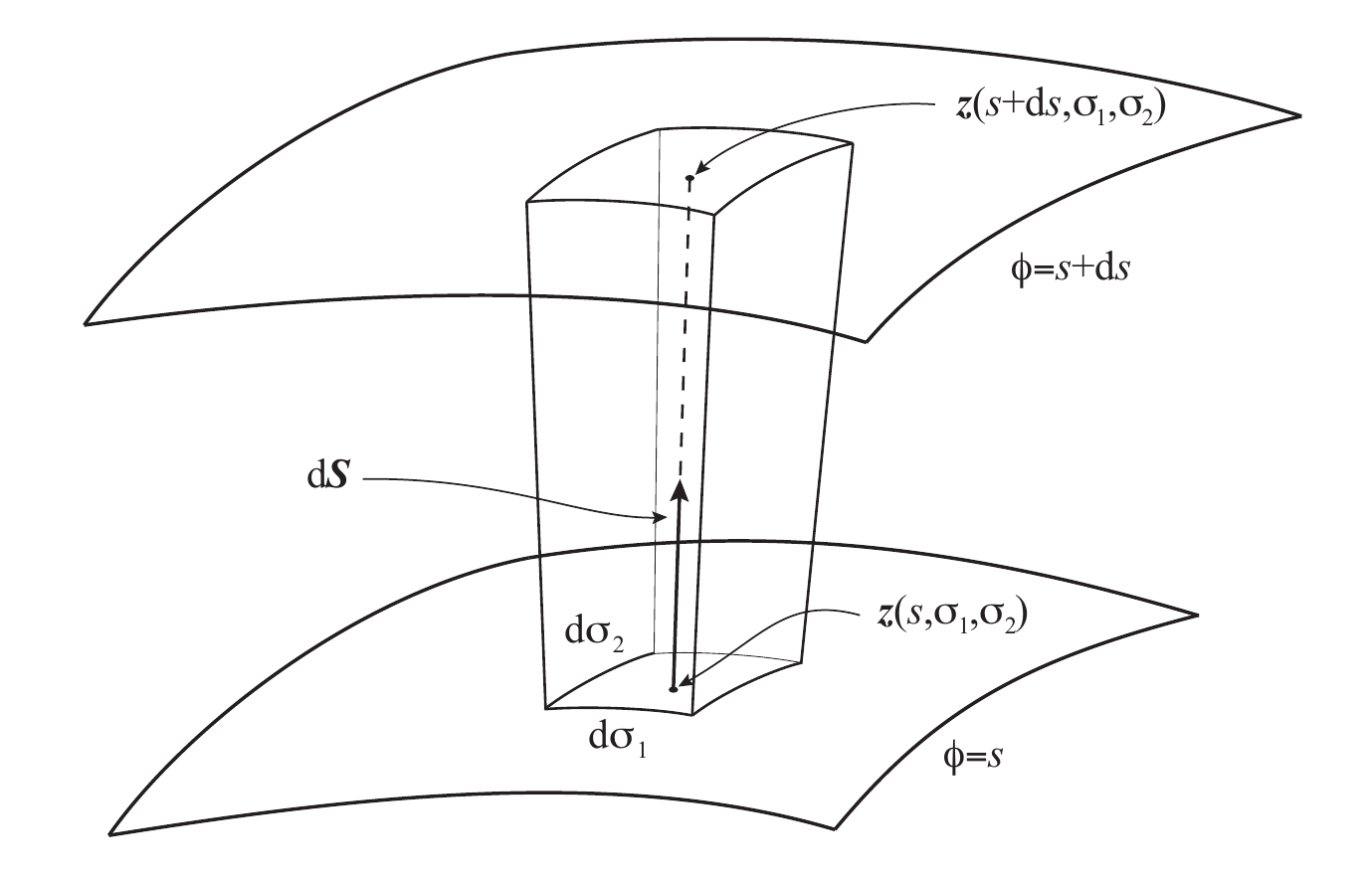}
\caption{An infinitesimal flux tube.
The area element $\dr\bm{S}$ defined in (\ref{RFDefdS}) is proportional 
to both $\bm{\nabla}\phi$ and $\frac{\partial \bm{z}}{\partial s}$.}
\label{RPNahmLaplace}
\end{figure}

It is useful to rewrite the continuum Nahm equations 
in the form
\begin{equation}
-\frac{\partial \bm{z}}{\partial s} = 
 \alpha 
\frac{\partial}{\partial \s^1} \bm{z} 
\times 
\frac{\partial}{\partial \s^2} 
\bm{z}. 
\label{RFcontinuumnahm2}
\end{equation}
This can further be rewritten as
\begin{equation}
-\frac{\partial \bm{z}}
{\partial s}
=
\alpha 
 \frac{\partial}{\partial \s^1} \bm{z} 
\times 
\frac{\partial}{\partial \s^2} 
\bm{z} 
=
\alpha
\frac{\text{d} \bm{S} }{\text{d} \s^1\text{d}\s^2 }.
\label{RFContiNahmVector}
\end{equation}
Here we considered 
an infinitesimal area element
$\text{d} \s^1\text{d}\s^2$ around $(\s^1, \s^2)$
in $\s$-space and the corresponding area element in $z$-space, $\text{d} \bm{S}$, 
and used the fact that 
\begin{equation}
\text{d} \bm{S} 
=
\left( \frac{\partial}{\partial \s^1} \bm{z} 
\times 
\frac{\partial}{\partial \s^2} 
\bm{z} \right)
\,
\text{d} \s^1\text{d}\s^2.
\label{RFDefdS}
\end{equation}
From the definition of the gradient it follows that 
\begin{equation}
\text{d}s 
= \text{d}\phi
= \bm{\nabla} \phi \cdot \text{d} \bm{z},
\end{equation}
and hence
\begin{equation}
1=\bm{\nabla} \phi \cdot \frac{\partial \bm{z}}{\partial s}.
\label{RFNablaPhiDotDZDS}
\end{equation}
Comparing (\ref{RFContiNahmVector}) and (\ref{RFNablaPhiDotDZDS}) we obtain
\begin{equation}
- \bm{\nabla} \phi \cdot \text{d} \bm{S}
= \frac{1}{\alpha} \text{d} \s^1\text{d}\s^2.
\label{RFContiNahmFlux}
\end{equation} 
Applying this formula to the volume element depicted in figure \ref{RPNahmLaplace},
which is a tube whose sides are constructed from
the electric flux lines associated with $\phi$ and the 
bottom and top plates are the area elements corresponding
to $\text{d} \s^1 \text{d}\s^2$ at $\phi=s$ and $\phi=s+\text{d}s$,
we see that
\begin{equation}
\int \bm{\nabla} \phi \cdot \text{d}\bm{S} = 0
\end{equation}
since the sides of the tube, 
where $\bm{\nabla} \phi \cdot \text{d}\bm{S}=0$,  do not contribute 
and the contributions of the top and bottom 
cancel each other.
This implies the three-dimensional Laplace equation.     

One can also derive the continuum Nahm equation 
from the Laplace equation. Starting from 
$\phi(\bm{z})$ satisfying the Laplace equation, 
we construct $\bm{z}(s, \s^1, \s^2)$ 
by solving $\phi(\bm{z}) =s$.
This requires fixing the $\s^1, \s^2$ coordinates.
We first fix  $\s^1,\s^2$ for a given $s$ 
by the requirement that the flux density be constant as in 
(\ref{RFContiNahmFlux}).
The definition of $\s^1, \s^2$ can be extended for 
arbitrary $s$, 
by requiring that
$\frac{\partial \bm{z}}{\partial s}$ be
proportional to $\bm{\nabla}\phi$. 
Once the paramterisation is fixed,
one can deduce
the continuum Nahm equations from the Laplace equation by 
basically backtracking the
above proof.  

\section{Behaviour at the splitting point}
\label{RSASplittingPoint}
In this appendix, we discuss the behaviour of families of equipotential 
surfaces for solutions of the Laplace equation of the type presented in
section~\ref{RSSolutions} near points where the splitting occurs.

The splitting point is a singularity of the continuum Nahm equation,
which is a partial differential equation (PDE).
It is a well-known idea to study such singularities by describing 
the solutions of a PDE as a level set of a given function~\cite{RBPDE}.
The singular point of the PDE is not singular for the function and thus
this approach provides a way to discuss the phenomena described
by the PDE avoiding the singularity.
In the present case the PDE describes the evolution of a surface 
and another representation of the same membrane evolution, also free of
singularities, is provided by the matrix regularisation. The equivalence 
between these two regularisations -- the level set description of the PDE and 
the matrix model formulation -- is implicitly assumed in our analysis. 

In a region sufficiently close to a generic splitting point, 
the potential $\phi$ can be approximated by~\footnote{In
this appendix we parametrise what was referred to as the $z$-space in the 
main text by $(x,y,z)$.}
\begin{equation}
s=
\phi(x,y,z)
=
\frac{1}{2}(x^2+y^2)-z^2,
\label{RFhyperbola}
\end{equation}
which satisfies the Laplace equation,
up to suitable rescaling of the $x,y,z$ variables.
The surface described by (\ref{RFhyperbola}) is a connected hyperboloid, 
a cone or a disconnected hyperboloid, depending on whether $s>0$, $s=0$ 
or $s<0$ respectively.
The $s=0$ surface embedded in the $(x,y,z)$ space is a cone with 
tip at $z=0$. 
The continuum Nahm equations are
\begin{equation}
\dot{x} = - \a \{ y, z\}, \ \ \
\dot{y} = - \a \{ z, x\}, \ \ \
\dot{z} = - \a \{ x, y\},
\label{RFcontinuumnahmxyz}
\end{equation}
where $\alpha>0$ and the Lie bracket is defined 
in~(\ref{RFLieBracket}).
At the tip of the cone the equations 
become singular, 
while the potential (\ref{RFhyperbola}) itself is, 
of course, non-singular. 

The form of the potential~(\ref{RFhyperbola})
can be used to study universal features of the splitting point.
In particular, in order to discuss the relation between the matrix regularisation
and the regularisation via the Laplace equation, 
one can assume the matrix size to be sufficiently large. 
Hence one can focus on the vicinity of the splitting point, \ie 
the tip of the cone, 
and the generic form of the potential~(\ref{RFhyperbola}) is sufficient.

We discuss the relation of the two regularisations in the following way.
Firstly, we rewrite the solution to the Laplace equation near the 
singularity~(\ref{RFhyperbola}) using a hodograph transformation, \ie
exchanging the role of independent and dependent variables, 
to obtain
$x(s, \sigma^1,\sigma^2)$, $y(s, \sigma^1,\sigma^2)$ and 
$z(s, \sigma^1,\sigma^2)$.  As functions of the 
$(s, \sigma^1,\sigma^2)$ these satisfy the continuum Nahm 
equation (\ref{RFcontinuumnahmxyz}), 
see~\cite{RBWard,RBHoppeLaplaceEq} and 
appendix~\ref{RSAGeometricalDerivationCNahmtoLaplace}. 
It turns out that this solution is characterised by a cubic equation.
Then we discretise the functions, $x(\sigma^1,\sigma^2)$, 
$y(\sigma^1,\sigma^2)$ and $z(\sigma^1,\sigma^2)$, using the matrix 
regularisation and check if these satisfy the Nahm equations 
when the matrix size is large enough. 

To construct $x$, $y$ and $z$ satisfying (\ref{RFhyperbola}) 
as functions of $(\sigma^1,\sigma^2)$ we proceed as follows. 
For convenience we introduce a radial coordinate $\rho$,
\begin{equation}
\rho
=
\sqrt{x^2+y^2}
=\sqrt{2}\sqrt{z^2+s} \, ,
\label{RFRelationZRho}
\end{equation}
where in the last equality we have used (\ref{RFhyperbola}).
We fix the $\s$ coordinates by using (\ref{RFContiNahmFlux}).
We compute the infinitesimal area in $\s$-space as
follows. Using 
\begin{equation}
\text{d} (\mbox{Area in $(x,y,z)$-space}) 
= 2\pi \rho \sqrt{1+ \left(\frac{\text{d} \rho}{\text{d} z}\right)^2} \text{d} z
=
2\pi \sqrt{2}\sqrt{3 z^2 +  s} \ 
\text{d}z
\end{equation}
and
\begin{equation}
\left|\bm{\nabla} \phi\right|= 
|(x,y,-2z)|= \sqrt{x^2+y^2+4z^2}=\sqrt{6 z^2 + 2 s} =
\sqrt{2}\sqrt{3 z^2 +  s},
\end{equation}
we obtain
\begin{equation}
\text{d} (\mbox{Area in $\s$-space}) 
= \alpha 4\pi (3 z^2+ s) \text{d} z.
\label{RFareainsigma}
\end{equation}
The infinitesimal area above refers to the portion of the membrane
between $z$ and $z+\text{d}z$.

We parametrise the $\s$-space  
by $\xi$ and $\varphi$
with range $(-\infty, \infty)$ and $[0,2\pi)$ respectively. 
We have chosen the parametrisation 
such that $\xi$ increases as $z$ does and 
$\varphi$ is the standard polar angle. 

Integrating both sides of (\ref{RFareainsigma}) we get a cubic equation,
\begin{equation}
\a (z^3+sz) = \frac{\xi}{2} \, ,
\label{RFRelationZXiS}
\end{equation}
from which we obtain $z$ (and $\rho$ via~(\ref{RFRelationZRho})) 
as a function of $\xi$ and $s$. 
When there exist three solutions in (\ref{RFRelationZXiS}), 
we choose a branch: for positive (negative) $\xi$, 
a positive (negative) root of the equation 
is employed.
For $s>0$ there is only one root for every $\xi\in(-\infty,\infty)$.
For $s<0$ there is a discontinuity at $\xi=0$, such that $z=\sqrt{-s}$ and 
$z=-\sqrt{-s}$ for $\xi\to0$ from above and $\xi\to0$ from below, respectively. 
Therefore, the $\s$-space is cylindrical for $s>0$ (corresponding to a 
connected hyperboloid), whereas it consists of two infinite planes 
for $s<0$ (corresponding to a disconnected hyperboloid) and a cone for 
$s=0$. It is interesting that a family of hyperboloids
(described for each $s$ by a quadratic equation)
is parametrised in this way by a cubic equation.

The $\xi, \varphi$ coordinates thus defined can be identified as
the $(\s^1, \s^2)$ coordinates directly so that 
we have
\begin{equation}
\{ f, g \}= 
\frac{\partial f}{\partial \xi}
\frac{\partial g }{\partial \varphi}
-
\frac{\partial g}{\partial \xi}
\frac{\partial f}{\partial \varphi}
\end{equation}
and
\begin{equation}
\int \text{d} \s^1 \text{d} \s^2
=
\int \text{d} \xi \text{d} \varphi.
\end{equation}
One can study  the singularity at the splitting point, using the coordinates 
$\xi, \varphi$. A natural measure for the singularity is the ``energy density''
\begin{equation}
\veps=
\frac{1}{2}
\left(
\{x ,y \}^2
+
\{y , z\}^2
+
\{z , x \}^2
\right).
\end{equation}
We use the relation between the continuum Nahm equation 
and the Laplace equation to obtain
\begin{equation}
\veps
=
\frac{1}{\alpha^2 |\bm{\nabla} \phi|^2}
=
\frac{1}{\alpha^2 \left(x^2+y^2+4z^2\right) }
=
\frac{1}{\alpha^2 \left(6z^2 +s\right)}.
\end{equation}
Thus the ``energy density'' diverges when $s=0$ at $z=0$ (\ie $\xi=0$) as
\begin{equation}
\veps \sim \frac{1}{z^2} \sim \xi^{-2/3} \, .
\end{equation}
We note however that the divergence is weak enough so that the integral of
the ``energy density'' 
\begin{equation}
\int
\veps
\text{d} \xi \text{d}\varphi
\end{equation}
is finite. 

The coordinates, $x$, $y$ and $z$, are functions of 
$(s, \xi,\varphi)$; from~(\ref{RFRelationZXiS}) 
$z(s, \xi, \varphi)=z(s, \xi)$. 
The variables $x$ and $y$ can be conveniently described by using 
\begin{equation}
w
=
x+iy
=
\rho e^{i \varphi}, 
\ \ \
w^\dagger
=
x-iy
=
\rho e^{-i\varphi}.
\label{RFwwdagger}
\end{equation}
It is easy to check that the functions $z$ and $w$ satisfy the continuum 
Nahm equations~(\ref{RFcontinuumnahmxyz}),
\begin{equation}
\dot{z}
= -\frac{i}{2} \a \{ w, w^\dagger \},
\ \ \
\dot{w}
= i \a \{ w, z \},
\label{RFcontinuumnahmzw}
\end{equation}
where we have introduced $\dot{z}=\partial_s z$ and 
$z'=\partial_{\xi} z$. 
The above equations are equivalent to
\begin{equation}
\dot{z}= -\a \rho' \rho,
\ \ \
\dot{\rho}=\a \rho z'.
\label{RFcontinuumnahmzrho}
\end{equation}
Acting with $\partial_s$ and $\partial_\xi$ 
on both sides of the cubic equation~(\ref{RFRelationZXiS})
we have
\begin{equation}
\dot{z}= -\frac{ z}{3 z^2 +s}, \ \ \ 
z'= \frac{1}{2\a(3z^2+s)}.
\end{equation}
Using (\ref{RFRelationZRho}), 
these imply
\begin{equation}
\dot{\rho}=
\frac{
    \sqrt{z^2+s}}
    {
    \sqrt{2}(3 z^2 + s)}, \ \ \ 
\rho'=
\frac{
    z}
    {
    \sqrt{2}\a\sqrt{z^2+s}
    (3z^2+s)}.
\end{equation}
Substituting into the continuum Nahm equations~(\ref{RFcontinuumnahmzrho}) 
we see that the equations 
are satisfied.

We then discretise the functions, $z$ and $w$, 
via the matrix regularisation and we check if they satisfy the 
Nahm equations when the matrix size is large enough. 
Here we write the Nahm equations in the form
\begin{equation}
\dot{X}= 
- \alpha \frac{1}{iC} [Y, Z] 
\ \ \
\dot{Y}= 
- \alpha \frac{1}{iC} [Z, X],  
\ \ \
\dot{Z}= 
- \alpha \frac{1}{iC} [X, Y],  
\end{equation}
or equivalently,
\begin{equation}
\dot{Z}= - \frac{\a}{2 C}[W, W^\dagger], 
\ \ \
\dot{W}=  \frac{\a}{C} [W,Z],
\label{RFdiscnahm2}
\end{equation}
where
\begin{equation}
W=X+iY, \ \ \ W^{\dagger}=X-iY.
\end{equation}
Here $X$, $Y$, $Z$ and $W$ are infinite dimensional matrices
(corresponding to the non-compact $\s$-space).
If the original matrix size $J$ is sufficiently large, we have $C \ll 1$.~\footnote{
The constant $C$ is analogous to $\hbar$ in 
quantum mechanics. 
For compact membranes,
$C$ is related to the size of matrices $J$ by $C=[\s]/(2\pi J)$
where $[\s]=\int \text{d}^2 \s$. This can be understood for example from 
(\ref{RFMRCommutator}).  By focussing on the splitting point, 
we have turned the original compact $\s$-space into 
a non-compact $\s$-space, and 
thus the matrix size $J$ should also be taken to infinity in such a way that 
$C$ is fixed.}

Our ansatz for the matrices is as 
follows~\cite{RBShimadaMR, RBHyakutake}. We take 
$Z$ to be diagonal and assume that $W$ has only non-zero sub-diagonal 
entries, \ie the only non-zero components in the matrices are 
$Z_{mm}=Z_m$, $W_{m, m+1}$ respectively. 
We use a slightly unusual convention in which the rows and columns
of infinite-size matrices are labelled by half-integer valued indices. The 
index $m$ can take both positive values, $m=1/2, 3/2, 5/2, \ldots$, and 
negative values, $m=-1/2, -3/2, -5/2, \ldots$
Using this ansatz, we propose a candidate for an approximate solution 
to the Nahm equations~(\ref{RFdiscnahm2}). 
For this purpose, we introduce an auxiliary function,
$z[m]$, defined as a root of the cubic equation
\begin{equation}
z^3+ sz = \frac{C}{2\a}m.
\label{RFdisccubic}
\end{equation}
In case this equation has three solutions, we choose a branch in the same 
way as described below (\ref{RFRelationZXiS}). 
Note that in (\ref{RFdisccubic}) we have discretised $\xi$ 
introduced in (\ref{RFRelationZXiS}) as $\xi = Cm$, 
\ie the mesh of $\xi$ is taken to be equally spaced,
following the Bohr-Sommerfeld quantisation \cite{RBShimadaMR}. 
Then the approximate solution to the Nahm equations (\ref{RFdiscnahm2})
is
\begin{equation}
Z_{mm}= z[m],
\end{equation}
\begin{equation}
W_{m, m+1}=\sqrt{2} \sqrt{z [ m+1/2 ]^2 +s} \, .
\end{equation}
Note that this equation is well-defined 
even for $W_{-1/2, +1/2}$ for $s<0$; 
$z[m]$ has a discontinuity at $m=0$,
but $z[m]^2$ does not. 
Furthermore, we note that  $W_{-1/2, +1/2}=0$ here.
For $s<0$ there are two disconnected portions, \ie
the two sheets of the hyperboloid. 
The values, $m=-1/2$ and $m=1/2$, 
correspond to these two parts and we do not expect non-zero matrix 
elements between them.

We show that these matrices satisfy the Nahm 
equations~(\ref{RFdiscnahm2}) up to terms which are sub-leading for 
$C \ll 1$. Indeed,
\begin{align}
[W,W^{\dagger}]_{mm} 
& = W_{m,m+1}^2 - W_{m-1,m}^2 \notag \\
& \approx 2\left(z\left[ m+1/2 \right] - z\left[ m-1/2 \right] \right)
\left(z\left[ m+1/2 \right] + z\left[ m-1/2 \right] \right) \notag \\
& \approx
4z[m] \frac{\partial}{\partial m} z[m] 
\approx
-\frac{2C}{\alpha} \dot{Z}_{mm}.
\end{align}
Similarly, since
\begin{align}
[W,Z]_{m,m+1}
& = \left(z[m+1]-z[m] \right)W_{m,m+1} \notag \\
& \approx \sqrt{2}\sqrt{z[m]^2+s} \ 
\frac{\partial}{\partial m}z[m] 
= \frac{C}{\alpha}\frac{\sqrt{z[m]^2+s}}{\sqrt{2}(3z[m]^2+s)}
\end{align}
and
\begin{align}
\dot{W}_{m,m+1}
& = \frac{2z\left[ m+1/2 \right] \dot{z}\left[ m-1/2 \right] + 1}{W_{m,m+1}} 
\approx
\frac{\sqrt{z[m]^2+s}}{\sqrt{2}(3z[m]^2+s)},
\end{align}
we have
\begin{equation}
[W,Z]_{m,m+1}
\approx
\frac{C}{\alpha}\dot{W}_{m,m+1}.
\end{equation}
The existence of the approximate solution suggests that
there exist an exact solution with the same qualitative behaviour.
It would be also interesting to find the exact solution of the 
Nahm equations within the tridiagonal ansatz made above. In that case, it is 
known that the Nahm equations are closely related to those  
describing the so-called Toda lattice. 

\section{Other solutions}
\label{RSAOtherSolutions}

\subsection{BHP solution}
\label{RSAOtherSolutionsBHP}

The BHP solution~(\ref{RFBHPSolution}) describes an instanton 
interpolating between a representation $L^i_{(-\infty)}$ at $t=-\infty$ and the 
$J$-dimensional representation $\ul{1} \oplus \ul{1}\oplus\cdots\oplus\ul{1}$
at $t=+\infty$. We consider the special case in which $L^i_{(-\infty)}$ is the 
irreducible representation $\ul{J}$. This corresponds to a process in which 
a membrane with angular momentum $J$ shrinks to a point-like object. 
Extrapolating the interpretation of the vacuum associated
with a direct sum of irreducible representations as a collection of concentric
membranes, the final state may be thought of as consisting of $J$ 
membranes each carrying one unit of angular momentum. However, we
stress that such an extrapolation goes well beyond the region of applicability
of the approximations used in this paper. Using the proposal 
in~\cite{RBMSVR}, the final state can also be understood as corresponding
to a single M5-brane carrying angular momentum $J$. 
 
It is not difficult to construct a solution to the Laplace equation which 
approximates this instanton configuration. 
To obtain a solution with the right properties, we consider the Coulomb 
potential generated by a positive point charge $J$ at the origin with the 
addition of a negative constant,
\begin{equation}
\phi(\bm{z}) 
= \frac{J}{4\pi |\bm{z}|} - e^{-2t_0/R},
\label{RFpotential2}
\end{equation}
where $\bm{z}=(z^1,z^2,z^3)$.
Using the transformation (\ref{RFContiBPSToContiNahm}) in reverse order, 
one can compute the distance from the origin of the membrane corresponding 
to~(\ref{RFpotential2}) in terms of the variables $\bm{y}=(y^1,y^2,y^3)$. The 
result is
\begin{equation}
|\bm{y}(t)| 
= 
\frac{J}{(2\pi T)R^2} 
\frac{1}{1+e^{2(t-t_0)/R}} \, .
\label{RFmembranelocation2}
\end{equation}
This represents a sphere with time dependent radius, $r(t)=|\bm{y}(t)|$, 
given by the right hand side of~(\ref{RFmembranelocation2}).
For large values of the potential, \ie $|\bm{z}| \approx 0$, corresponding to 
the far past ($t\to-\infty$), (\ref{RFpotential2})-(\ref{RFmembranelocation2})
describe a configuration which approaches a stable sphere of radius 
$r=J/(2\pi T R^2)$. On the other hand, the potential~(\ref{RFpotential2}) 
becomes zero at a finite distance from the origin, 
$|\bm{z}|=J/(4\pi e^{-2t_0/R})$, in the far future ($t \to +\infty$), 
where the radius of the membrane, $|\bm{y}|$, shrinks to zero. 

The explicit form of the continuum counterpart of the BHP solution, in the 
case in which $L^i_{(-\infty)}$ is irreducible, is therefore
\begin{equation}
y^i (t,\s^1,\s^2) 
= r(t) n^i(\s^1,\s^2),
\end{equation}  
where the membrane radius is 
\begin{equation}
r(t) = \frac{J}{(2\pi T)R^2} \frac{1}{1+e^{2(t-t_0)/R}} 
\label{RFContinuumBHPradius}
\end{equation} 
and $n^i(\s^1,\s^2)$ is a unit vector in the radial direction. 

\subsection{Solution with positive and negative point charges}
\label{RSAOtherSolutionsBHPApollonius}

In this section we present a new solution approximately describing 
the instanton interpolating between the irreducible vacuum $\ul{J}$
and the vacuum associated with the representation 
$\ul{J}' \oplus \ul{1} \oplus\cdots\oplus\ul{1}$, where $\ul{1}$ 
appears with multiplicity $(J-J')$.  
From the membrane point of view in this process 
a single membrane with angular momentum $J$ splits 
into a membrane with angular momentum $J'$ 
and a second membrane which shrinks to a point in the final state.

The corresponding solution to the Laplace equation is the sum of 
two Coulomb potentials associated respectively with a positive charge 
$J$ at the origin and a negative charge of magnitude $(J-J')$ at 
a point $\bm{z}=\bm{z}_0 (\ne \bm{0})$,
\begin{equation}
\phi(\bm{z}) 
= 
\frac{J}{4\pi |\bm{z}|} - \frac{J-J'}{4\pi |\bm{z}-\bm{z}_0|} \, ,
\label{RFpotential3}
\end{equation} 
where $J'$ is assumed to be positive. 
The evolution of the membrane profile, corresponding to the equipotential 
surfaces of~(\ref{RFpotential3}), can be described as follows.
The relevant equipotential surfaces are those  with
$s=\phi \in  (+\infty, 0]$ which corresponds to $t\in (-\infty, +\infty)$.
For $t\sim -\infty$, \ie $s=\phi\sim +\infty$, we have a single spherical 
membrane of small radius. As $t$ increases this membrane grows and 
deforms, until it splits into two membranes (both with spherical topology) 
at a value $s>0$ of the potential. One of the membranes continues to grow 
as $t$ increases towards $+\infty$, \ie $s=\phi\to 0$, and approaches the 
behaviour of the equipotential surfaces of a Coulomb solution with charge 
$J'$. The other membrane at $\phi\sim 0$ is a sphere of finite radius. 
(This is a well-known fact used for the electrostatic potential of a spherical 
conductor.) The first membrane, due to the 
transformation~(\ref{RFContiBPSToContiNahm}),
corresponds to a static spherical membrane in the original $\bm{y}$
coordinates (see section \ref{RSSSolutionCoulomb}). In the same 
coordinates the second membrane shrinks to a point.
Therefore, one finds indeed that the potential (\ref{RFpotential3}) describes 
the process in which a spherical membrane with angular momentum $J$ 
at $t=-\infty$ splits into a spherical membrane with angular momentum $J'$
and a point-like object carrying angular momentum $(J-J')$~\footnote{
It is also possible to consider the solution in which $J' <0$. In this case,
as $t$ increases, the equipotential surfaces in the $z$-coordinates
do not split (in the relevant region $\phi>0$) and instead approach a 
sphere of finite radius as $t\to\infty$. Thus this solution corresponds to an 
instanton interpolating between the irreducible representation $\ul{J}$ and 
the trivial representation $\ul{1}\oplus\cdots\oplus\ul{1}$.
}.

A description of the above solution in terms of a suitable Riemann space 
leads to an interesting interpretation for the negative point charge.
We consider a Riemann space consisting of $m+1$ sheets, with $m$ 
identical branch disks connecting the first sheet to the remaining $m$ 
sheets. The insertion of the negative charge can be realised as the limit in 
which $m$ is sent to infinity while the size of the branch disks is sent to 
zero, so as to keep the total flux out of the first sheet fixed.

It is clear that this solution can be generalised. One can put an arbitrary 
number of  negative point charges at arbitrary locations. 
It is also possible to put the negative charges on different sheets of a 
Riemann space. 

As mentioned earlier the $\ul{1}\oplus \cdots \oplus\ul{1}$ solution was 
conjectured~in \cite{RBMSVR} to represent a single M5-brane.
It is interesting to notice that combining this proposal with the above
considerations, we are led to the conclusion that
while M2-branes in our construction correspond
to positive charges (which should be located at the origin of each sheet of 
a Riemann space), negative charges (which can be placed at arbitrary 
locations) are associated with M5-branes. 

\section{Relation between Hobson's and Sommerfeld's solutions}
\label{RSASommerfeldHobson}

The solution found by Sommerfeld~\cite{RBSommerfeld} is
\begin{equation}
\phi(\bm{r}, \bm{r_0}) = \frac{1}{|\bm{r}-\bm{r_0}|}
\left(
\frac{1}{2}
+
\frac{1}{\pi}
\arcsin{\left(
\frac{
    \cos{\frac{\varphi_S-\varphi_{S0}}{2}}
    }
    {
    \cosh{\frac{\alpha_S}{2}}
    }
\right)}
\right),
\label{RFSommerfeldPotential}
\end{equation}
where 
\begin{equation}
\cosh \alpha_S=
\frac{r_S^2 + r_{S 0}^2 + (z_S- z_{S0})^2}{2 r_S r_{S0}},
\label{RFSommerfeldAlpha}
\end{equation}
up to an overall constant.
The notation used in the above formulae is as follows.
We use the standard cylindrical coordinates $(r_S, \varphi_S, z_S)$ 
for the point $\bm{r}$
where the $z$-axis coincides with the straight branch line.
For the source point $\bm{r_0}$ the cylindrical coordinates are 
$(r_{S0}, \varphi_{S0}, z_{S0})$.

The solution is related to 
Hobson's solution by an inversion tranformation, 
where the point of inversion $A$ is taken to be a point on the
circumference of the disk.
It is clear that the circular branch loop in Hobson's solution
is mapped to the straight branch line in Sommerfeld's solution.
The relation between the two solutions was observed 
and numerically verified
in work by Heise~\cite{RBHeise}.
Below we provide an analytic proof.

For the purpose of comparison, 
we present here again Hobson's solution (\ref{RFHobsonspotential}), 
(\ref{RFHobsonscoshalpha}) in an appropriately normalised form,
\begin{equation}
\phi(\bm{r},\bm{r}_0) = 
\frac{1}{|\bm{r}-\bm{r}_0|}  
\left( \frac{1}{2} + \frac{1}{\pi} \arcsin\!\left( 
\frac{
\cos \frac{\theta_H - \theta_{H 0}}{2} 
}
{
\cosh \frac{\alpha_H}{2}
}
\right) \right) \, ,
\label{RFHobsonspotentialAppendix}
\end{equation}
where 
\begin{equation}
\cosh\alpha_H = \cosh \rho_H \cosh \rho_{H0} - 
\cos (\varphi_H-\varphi_{H0}) \sinh \rho_H \sinh \rho_{H 0}
\label{RFHobsonscoshalphaAppendix}
\end{equation}
and $(\rho_H,\theta_H,\v_H)$ are the peripolar coordinates 
defined in section~\ref{RSSAnalyticSolution}.
We note that we have used
\begin{equation}
\cosh{\frac{\alpha_H}{2}}=
\sqrt{\frac{\cosh \alpha_H +1}{2}}.
\end{equation}
In this appendix we use the subscripts $H$ and $S$ to emphasise the 
distinction between the coordinates and the variables used in Hobson's and 
Sommerfeld's solutions.

In order to show that 
the solution (\ref{RFHobsonspotentialAppendix}) maps to 
(\ref{RFSommerfeldPotential}) under an inversion transformation,
we note that the prefactor $1/|\bm{r}-\bm{r}_0|$
in both solutions transforms under  
an inversion transformation in the well-known manner.
Hence to show the equivalence between 
(\ref{RFHobsonspotentialAppendix})
and (\ref{RFSommerfeldPotential}) under the inversion transformation
it is sufficient to
prove the following identities
\begin{equation}
\varphi_S= \th_H,
\label{RFPhiSEqualsThetaH}
\end{equation}
\begin{equation}
\alpha_S= \alpha_H.
\label{RFAlphaSEqualsAlphaH}
\end{equation}
The first identity needs only be satisfied 
up to a sign and constant shifts because
the variables in the formula always appear in combinations such as
$\cos{((\varphi_S-\varphi_{S0})/2)}$.

To prove (\ref{RFPhiSEqualsThetaH})
we first focus on the plane which 
contains the point of inversion $A$ and the center of the 
branch disk and which is orthogonal to the branch disk.
The definition of $\theta_H$ is depicted in the first panel of figure 
\ref{RPperipolar-cartesian}.
We denote by $P', B'$ the points corresponding to $P, B$ by 
an inversion transformation around the point $A$.
By definition we have $\widehat{APB} =\th_H$.
We also have 
$\widehat{P'B' A}=\varphi_S$,
since the straight branch line passes 
through the plane orthogonally at the point $B'$.
By elementary geometry, 
it is easy to show that $\widehat{APB}=\widehat{P'B' A}$,
which is (\ref{RFPhiSEqualsThetaH}),
using the definition of the inversion transformation.
It is possible furthermore to show that the validity of
(\ref{RFPhiSEqualsThetaH}) can be extended to 
the whole $\R^3$. Indeed, we observe that
the constant $\th_H$ surface in Hobson's coordinate is a part of a sphere 
bounded by the branch loop. 
Since the surface contains the inversion point, it will be mapped to a
half-plane bounded by the straight branch line.
Thus a constant $\th_H$ surface is mapped 
to  a constant $\varphi_S$ surface under the inversion
and therefore it follows that (\ref{RFPhiSEqualsThetaH})
is valid generally.

It finally remains to show (\ref{RFAlphaSEqualsAlphaH}).
As $\alpha_H$ and $\alpha_S$ do not depend on the $\th_H=\varphi_S$ 
variable, one can focus on the special case $\th_H=\pi$.
The quantity $\alpha_H$ will then be defined on a pair of points on the branch 
disk itself and we wish to show that it is equal to $\alpha_S$ for  
the corresponding pair of points on the branch half-plane.
This can be established by noting that $\alpha_S$ and $\alpha_H$ 
are nothing but the geodesic distances in the classic Poincar\'e upper half 
plane and disk models (the latter also referred to as the
conformal disk model) of hyperbolic geometry, 
respectively~\cite{RBHyperbolicGeometry}.
These two models are related to each other by an inversion transformation 
and hence (\ref{RFAlphaSEqualsAlphaH}) follows.

It is interesting to note that the quantity $\alpha_H$ (or $\alpha_S$),
which is an important ingredient in the exact solution,
has an interpretation in terms of some non-trivial geometry
defined on the branch disk (or the branch half-plane).

\section{Boundary conditions at branch disks and reformulation
in terms of integral equations}
\label{RSAConnectingConditionIntegralEq}

In this appendix, 
we formulate  boundary conditions
associated with branch disks and 
present a reformulation of the Laplace equation
on a Riemann space in terms of 
a set of integral equations.
The unknown variables in the integral equations
are functions defined on the branch surfaces.
Here we use the vector $\bm{r}=(x,y,z)$ to 
denote the position of a point in what was referred to as $z$-space in the 
main text. 

For concreteness we focus on the simplest case where there are
two copies of $\R^3$ connected by a single branch disk $D$.
We put a point charge $J$ at the point designated by
a vector $\bm{r_0}$. The potential function will be denoted by 
$\phi_{(1)}(\bm{r})$ in the first space and by $\phi_{(2)}(\bm{r})$ in the 
second space.

\begin{figure}[htb]
\centering
\includegraphics[width=0.9\textwidth]{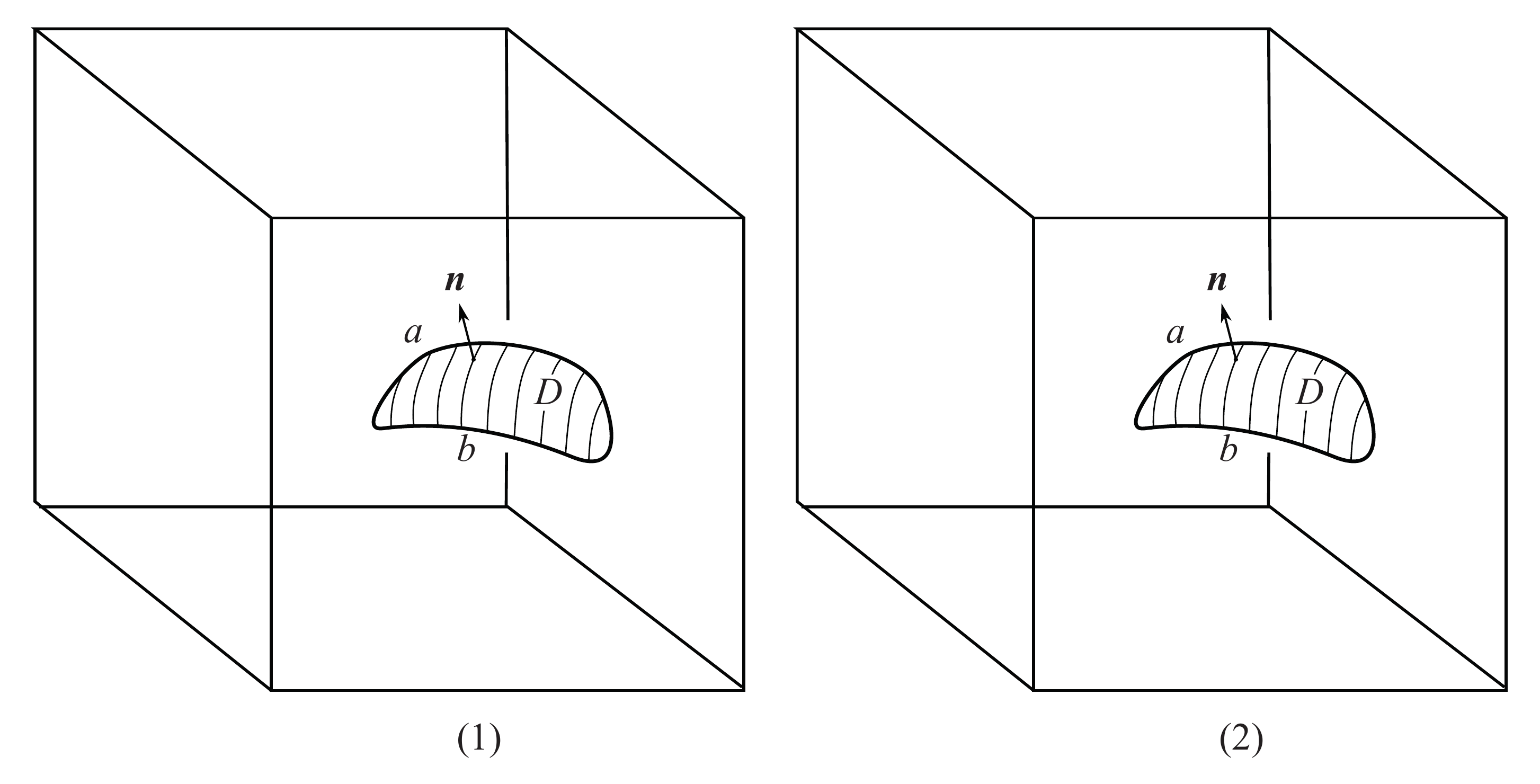}
\caption{Riemann spaces}
\label{RPBoundarCondition}
\end{figure}

As functions in $\R^3$ both these potential functions have a discontinuity at 
the branch disk. In order to take care of this discontinuity, we introduce the 
following notation. We refer to the two sides of the branch disk as $a$ and 
$b$. The values of the potential functions in each space, $\phi_{(i)}$, 
$(i=1,2)$, on the $a$ and $b$ sides of the branch disk will be denoted 
respectively by
\begin{equation}
\phi_{(i)a}(\bm{r})
\quad {\rm and} \quad 
\phi_{(i)b}(\bm{r}).
\end{equation}
In these expressions it is implicitly assumed that the vector $\bm{r}$ 
corresponds to a point on the branch disk which is approached from the
$a$ or $b$ side.

We introduce the normal vector orthogonal to
the branch disk defined at the point on the branch disk,
$\bm{n}$, with orientation such that $\bm{n}$ points away from the disk
on the $a$ side and towards the disk on the $b$ side 
as shown in figure \ref{RPBoundarCondition}.

The 
precise formulation  
of the condition that the first and the second spaces 
are connected by the branch disk is given by the following formulae,
\begin{equation}
\phi_{(1)a}(\bm{r})
=
\phi_{(2)b}(\bm{r}), 
\qquad
\phi_{(2)a}(\bm{r})
=
\phi_{(1)b}(\bm{r}),
\label{RFConnectingCondition1}
\end{equation}
and
\begin{equation}
\partial_n \phi_{(1)a}(\bm{r})
=
\partial_n \phi_{(2)b}(\bm{r}),
\qquad
\partial_n \phi_{(2)a}(\bm{r})
=
\partial_n \phi_{(1)b}(\bm{r}),
\label{RFConnectingCondition2}
\end{equation}
where 
$\partial_n \phi(\bm{r})$ denotes the 
normal derivative of the potential functions at the branch disk,
\begin{equation}
\partial_n \phi(\bm{r})
= \bm{n} \cdot \bm{\nabla} \phi.
\end{equation}
We further require, as a part of the boundary conditions,
that the $\phi_{(i)}$'s be finite at the boundary of the branch disk (the 
branch loop) and that they go to zero at infinity sufficiently rapidly.

Below we reformulate the Laplace equations
with given boundary conditions in terms 
of integral equations.
This reformulation is known, in the case of more conventional boundary
conditions, to be useful in many respects, including
numerical computations and the study of existence and
uniqueness properties for the solutions to the Laplace 
equation~\cite{RBIntegralEquation}.
We will use Green's theorem, 
\begin{equation}
\int_M 
\left( 
\phi \bm{\nabla}^2 \psi 
- \psi \bm{\nabla}^2 \phi  \right) (\bm{r'}) 
\text{d}^3 \bm{r'}
= 
\int_{\partial M}  
\left( \phi \bm{\nabla} \psi  - \psi  \bm{\nabla} \phi  \right) 
(\bm{r'})
\cdot \text{d}^2 \bm{ S'},
\label{RFGreen}
\end{equation} 
for the special case where 
$\psi(\bm{r'})$ is
\begin{equation}
\psi(\bm{r'}) = - \frac{1}{4\pi | \bm{r}' - \bm{r} |} \, ,
\end{equation}
satisfying
\begin{equation}
\bm{\nabla}^2 \psi(\bm{r'}) = \delta^3( \bm{r}' - \bm{r}).
\end{equation}
The potential on the first sheet,
$\phi_{(1)}$, also has a $\delta$-function source,
\begin{equation}
-\bm{\nabla}^2\phi_{(1)}(\bm{r}')
=
J \delta^3( \bm{r}' - \bm{r_0})
\end{equation}
or equivalently behaves as 
\begin{equation}
\phi_{(1)} (\bm{r'})
\sim
\frac{J}{4\pi |\bm{r}' - \bm{r_0}|},
\end{equation}
when $|\bm{r}' - \bm{r_0}|\sim 0$.

We define the following two important functions on the branch disk,
\begin{align}
&f= \phi_{(1)a} - \phi_{(1)b},
\\
&g= \partial_n \phi_{(1)a} - \partial_n\phi_{(1)b}.
\end{align}
They are the discontinuities of the potential and the 
normal derivative (in the first space).

The functions $f$ and $g$ may be interpreted as 
the dipole moment density and the charge density 
of a fictitious charge distribution placed at the disk,
which, together with the point charge at $\bm{r}_0$, reproduces the 
electrostatic potential $\phi_{(1)}$.
The functions $f$ and $g$ are the two unknown functions in the 
reformulation in terms of integral equations.

We apply Green's theorems to $\phi_{(i)}$, choosing for $M$  
the region in $\R^3$ obtained excluding the branch 
disk~\footnote{It is also possible to avoid the use of the $\delta$-functions. 
In this case one defines the manifold $M$ excluding, in addition to the 
branch disk, the points $\bm{r}$ for both $\phi_{(1)}$ and
$\phi_{(2)}$ and $\bm{r_0}$ for $\phi_{(1)}$. One then requires
Coulomb-like behaviour for $\psi$ near $\bm{r}$ and for $\phi_{(1)}$ near  
$\bm{r_0}$.}. We then obtain 
\begin{equation}
\phi_{(1)}(\bm{r})
=
\frac{J}{4 \pi|\bm{r}-\bm{r_0}|}
+\chi(\bm{r}) ,
\label{RFIntegralRep1}
\end{equation}
\begin{equation}
\phi_{(2)}(\bm{r})
=
-\chi(\bm{r}).
\label{RFIntegralRep2}
\end{equation}
Here we have introduced $\chi(\bm{r})$ defined as
\begin{equation}
\chi(\bm{r}) = 
 \int_{D} 
\left( 
-
f(\bm{r}')
\frac{1}{4\pi }
\frac{(\bm{r}' - \bm{r})}{ | \bm{r}' - \bm{r}  |^3 }
\cdot
\bm{n}
-  
g(\bm{r}')
\frac{1}{4 \pi}
\frac{1}{ | \bm{r} - \bm{r}' |} 
 \right ) 
\text{d}^2 S'
, 
\label{RFDefChi}
\end{equation}
where the integration variable, $\bm{r}'$, is restricted to the 
branch disk. 
Here, the condition that the $\phi_{(i)}$'s themselves
be finite at the branch loop is important. Without this condition, 
it is possible to have an extra term in the above 
formulae originating from the branch loop.
(Consider superposing to the $\phi_{(i)}$'s the electrostatic 
potential generated by a charged wire placed at the branch loop.)

Formulae (\ref{RFIntegralRep1})-(\ref{RFDefChi})
give an integral representation of
the potential on the two sheets of the Riemann space,
in terms of the discontinuities $f$ and $g$ at the branch disk.

Substituting the integral representations back into the connecting 
conditions (\ref{RFConnectingCondition1})-(\ref{RFConnectingCondition2}), 
one obtains
\begin{equation}
\chi_{a}(\bm{r})+
\chi_{b}(\bm{r})
=
- \frac{J}{4\pi |\bm{r}-\bm{r}_0|},
\label{RFIntegralEq1}
\end{equation}
\begin{equation}
\partial_n \chi_{a}(\bm{r})+
\partial_n \chi_{b}(\bm{r})
=
-\frac{1}{4\pi} 
\partial_n
\frac{J}{|\bm{r}-\bm{r}_0|}.
\label{RFIntegralEq2}
\end{equation}
Here $\bm{r}$ is assumed to correspond to points on the branch disk.
The  function $\chi$ also has a discontinuity 
across the disk and we denote by $\chi_{a}$ and $\chi_{b}$
the values of $\chi$ on side $a$ and $b$ respectively. 
Formulae (\ref{RFDefChi})-(\ref{RFIntegralEq2}) 
are the integral equations for the functions
$f$ and $g$, which are defined on the disk.

Focussing on the special case in which the disk $D$ lies in the $xy$-plane,
one can rewrite the integral equations in a more explicit form as
\begin{align}
&\int_D 
\frac{1}{\sqrt{(x-x')^2 + (y-y')^2}}
g(x',y')
\text{d}x'\text{d}y'
 \notag \\
& \ \ \ =  
\frac{1}{2} \frac{J}{ |\bm{r}-\bm{r}_0| }
=\frac{1}{2}
\frac{J}{\sqrt{
(x-x_0)^2+(y-y_0)^2+z_0^2
}}
,
\end{align}
\begin{align}
&\int_D 
\left(  
\frac{-1}{\sqrt{(x- x')^2 + (y-y')^2 +\varepsilon^2}^3}
+ 
\frac{3 \varepsilon^2}{ \sqrt{( (x-x')^2 + (y-y')^2 + \varepsilon^2}^{5} }   \right)
f(x', y') 
\text{d}x' \text{d}y'
\notag \\ 
& \ \ \  =
\frac{1}{2} 
\partial_z \frac{J}{ |\bm{r}-\bm{r}_0| }
=
J
\frac{z_0}{
\sqrt{(x-x_0)^2+(y-y_0)^2+z_0^2}^3
}.
\end{align}
They are Fredholm integral equations of the first type,
the second integral equation having a singular kernel.
We note that the first integral equation is
equivalent to the integral equation for the electrostatic potential 
associated with a conducting disk.

\section{Equations for general pp-wave matrix model}
\label{RSAGeneralisedFormulae}
In the main text we have given most of the formulae for the pp-wave matrix 
model which is relevant for the large $J$ sector of the ABJM duality in the 
special case $k=1$. In this appendix 
we give generalisations of some of the important equations
to the general pp-wave matrix model 
written in terms of a mass parameter, $\mu$, and for $k\neq 1$.
To make the use of the generalised formulae below more straightforward 
we have numbered each equation with the same equation number as the 
corresponding one in the main text, with the addition of a suffix ``a''. So for
example equation~(\ref{RFRFeuclideanactiona}) below is the generalisation 
of~(\ref{RFeuclideanaction}) in section~\ref{RSBPSInstantons}. 

Below we consider matrices of size
\begin{equation} 
K=\frac{J}{k} \label{RFDefK}.
\end{equation}
We note that the angular momentum quantum number
$J$ is a multiple of $k$ on $S^7/\Z_k$.
The radius of $S^7$, $R$, is related to the 
radius of the M-theory circle, $R_{11}$, by
\begin{equation}
R_{11}=\frac{R}{k}. \label{RFDefR11}
\end{equation}
For the pp-wave approximation
of AdS$_4 \times S^7/\Z_k$, the mass parameter $\mu$ below 
should be set to the value,
\begin{equation}
\mu =\frac{6}{R}. \label{RFMuAndR}
\end{equation}

All formulae in the rest of this appendix are written in terms of the parameters 
$K$, $R_{11}$ and $\mu$. In order to generalise the formulae in the
main text to the $k\neq 1$ case, it is sufficient to apply (\ref{RFDefR11}) and 
(\ref{RFMuAndR}) to the equations below and remember that the matrix size 
$K$ is given by  (\ref{RFDefK}).

Using the replacements $X^m \to -X^m$ and $Y^i \to - Y^i$, 
changing the sign of the mass parameter, $\mu \to -\mu$, and setting 
$2 \pi T=1$ brings the following equations to the form used in the literature on 
the pp-wave matrix model, \eg  \cite{RBDSVR1}.
Note that the redefinition of $\mu$ and $Y$ has the effect of exchanging 
instantons and anti-instantons, which can be seen 
from~(\ref{RFBPSEqa}).   

\begin{align}
S_E = \int & \dr t \; \text{tr}  
\biggl\{ \frac{1}{2R_{11}} \left( \frac{D Y^i}{D  t} \right)^2
+\frac{1}{2R_{11}} \left( \frac{D X^m}{D  t} \right)^2 \nonumber \\
&+(2\pi T)^2 \frac{R_{11}}{4}
\left((i[X^m, X^n])^2 + 2 (i[X^m,Y^i])^2
\right) \nonumber \\
& + \frac{1}{2R_{11}} \left( \frac{\mu}{6} \right)^2 (X^m)^2
+ (2\pi T)^2 \frac{R_{11}}{2}
\left( \frac{i}{2} \epsilon^{ijk} [Y^j, Y^k]
+ \frac{\mu}{3(2\pi T)R_{11}}  Y^i
\right)^{\!\!2} \nn \\
&+\frac{1}{2} \Psi^T \frac{D \Psi}{D t}
+ 2\pi T \frac{R_{11}}{2}  \left(\Psi^T \gamma^m[ X^m, \Psi]
+\Psi^T \gamma^i[ Y^i, \Psi] \right)
- i \frac{\mu}{8} \, \Psi^T \gamma^{123} \Psi
\biggl\} . 
\label{RFRFeuclideanactiona}
\tag{\ref{RFeuclideanaction}a}
\end{align}
\begin{align}
& Y_0^i = \frac{\mu}{3(2\pi T)R_{11}} L^i \, , \quad {\rm with} \quad 
[L^i,L^j] = i \epsilon^{ijk}L^k \,  .
\label{RFMMVacuaa}
\tag{\ref{RFMMVacua}a}
\end{align}
\begin{equation}
r = \frac{\mu\sqrt{K^2-1}}{6(2\pi T)R_{11}} 
\approx \frac{\mu K}{6(2\pi T) R_{11}} \, .
\label{RFradiusa}
\tag{\ref{RFradius}a}
\end{equation}
\begin{align}
S_E 
=
\frac{1}{2 R_{11}}  \int \dr t\: \text{tr} & \biggl[ 
\left( \frac{\text{d} X^m}{\text{d}  t} \right)^2
+ \left( \frac{\mu}{6} \right)^2 \left( X^m \right)^2
+\frac{(2\pi T)^2R^2_{11}}{2} 
\left\{ 
(i[X^m,X^n])^2 +(i[X^m,Y^i])^2  
\right\} 
\notag \\
&+ \left( \frac{\text{d} Y^i}{ \text{d} t} 
\pm \frac{\mu}{3} Y^i \pm i (2\pi T) 
\frac{R_{11}}{2} \epsilon^{ijk} [Y^j,Y^k]   \right)^2 \notag \\
&\mp \frac{\text{d}}{\text{d} t} 
\left(  \frac{\mu}{3}  Y^iY^i + i (2\pi T) 
\frac{R_{11}}{3} \epsilon^{ijk}Y^i[Y^j,Y^k]  \right) 
+ {\rm fermions} \biggl] \, .
\label{RFactionsumofsquaresa}
\tag{\ref{RFactionsumofsquares}a}
\end{align}
\begin{equation}
S_E \ge \left.\mp\fr{2R_{11}}\, W[Y]\right|_{-\infty}^{+\infty}
= \mp \frac{\mu}{18 R_{11}} \left( \frac{\mu}{3(2\pi T) R_{11}} \right)^{\!2} 
\tr\!\left( L^i_{(+\infty)} L^i_{(+\infty)} - L^i_{(-\infty)} L^i_{(-\infty)} \right) \, .
\label{RFbounda}
\tag{\ref{RFbound}a}
\end{equation}
\begin{equation}
W[Y] = \tr\left( \frac{\mu}{3}  Y^iY^i + i (2\pi T) \frac{R_{11}}{3} 
\epsilon^{ijk}Y^i[Y^j,Y^k]\right)  \, .
\label{RFsuperpotentiala}
\tag{\ref{RFsuperpotential}a}
\end{equation}
\begin{equation}
\frac{\text{d} Y^i}{ \text{d}t} 
\pm \frac{\mu}{3} Y^i \pm i (2\pi T) 
\frac{R_{11}}{2} \epsilon^{ijk} [Y^j,Y^k]  =0 \, .
\label{RFBPSEqa}
\tag{\ref{RFBPSEq}a}
\end{equation}
\begin{equation}
W[Y_0] 
= \frac{\mu}{9}\left( \frac{\mu}{3(2\pi T)R_{11}} \right)^{\!2}  
\tr\!\left(L^iL^i\right) 
= \frac{\mu}{9}\left( \frac{\mu}{3(2\pi T)R_{11}} \right)^2 \frac{1}{4} 
\sum^{n}_{i=1} \tr \left[ \left( K^2_i -1 \right) \Id_{K_i \times K_i}  \right] \, ,
\label{RFW0a}
\tag{\ref{RFW0}a}
\end{equation}
where $K_j=J_j/k$, $j=1,2,\ldots,n$, denote the (integer) size of the blocks
in the case in which the matrices $L^i$ are generators of a reducible SU(2) 
representation. The $K_j$'s satisfy $\sum_j K_j =K$.

\begin{equation}
Z^i =
C\,e^{(\mu t)/3} Y^i, 
\ \ \ 
s=e^{-(\mu t)/3}\, .
\label{RFBHPToNahma}
\tag{\ref{RFBHPToNahm}a}
\end{equation}
\begin{equation}
\frac{\text{d} Z^i}{ \text{d} s} 
=  
i(2\pi T)\frac{3R_{11}}{2 C \mu }  \epsilon^{ijk}[Z^j,Z^k] \, .
\label{RFNahma}
\tag{\ref{RFNahm}a}
\end{equation}

\begin{align}
\rho(f g) &\approx \frac{1}{2} \Big(\rho(f) \rho(g) + \rho(g) \rho(f)\Big) \, .
\tag{\ref{RFMRMultiplication}a}
\label{RFMRMultiplicationa} \\
\rho\left(\{f, g\}\right) &\approx \frac{2\pi K}{i \tA} 
\left[ \rho(f), \rho(g) \right] \, .
\tag{\ref{RFMRCommutator}a}
\label{RFMRCommutatora} \\
\frac{1}{\tA}\int f \, \dr^2 \s &\approx \frac{1}{K} \tr \big(\rho(f)\big) \, .
\tag{\ref{RFMRTr}a}
\label{RFMRTra}
\end{align}

\begin{equation}
\frac{\partial y^i}{\partial t} 
\pm \frac{\mu}{3}  y^i 
\mp \frac{(2\pi T)R_{11}}{4\pi}\frac{[\s]}{K} \epsilon^{ijk} \{y^j,y^k \} 
=0 \, .
\label{RFContinuumBPSEqa}
\tag{\ref{RFContinuumBPSEq}a}
\end{equation}
\begin{equation}
z^i=
\frac{(2\pi T)}{4\pi }\left( \frac{6}{\mu} \right)^2 e^{ (\mu t)/3} y^i, \ \ \ 
s =  
e^{- (\mu t)/3} \, . 
\label{RFchangevariable2a}
\tag{\ref{RFContiBPSToContiNahm}a}
\end{equation}
\begin{equation}
\frac{\partial z^i}{ \partial s} 
= - \frac{\mu R_{11}}{12}  \frac{[\s]}{K} 
\epsilon^{ijk} \{ z^j,z^k \} \, .
\label{RFContinuumnahma}
\tag{\ref{RFContinuumnahm}a}
\end{equation}

\end{document}